\documentclass[journal,draftclsnofoot,onecolumn,12pt]{IEEEtran}
\usepackage{cite}
\usepackage{graphicx}
\usepackage{amsmath}
\usepackage{amssymb}
\usepackage{algorithmicx}
\usepackage{algorithm}
\usepackage{color}
\usepackage{pdfcomment}

\usepackage{caption}

\allowdisplaybreaks

\def\sinc{\mathrm{sinc}}

\begin{document}

\title{Wireless Communications with Reconfigurable Intelligent Surface: Path Loss Modeling and Experimental Measurement}

\author{Wankai Tang, Ming Zheng Chen, Xiangyu Chen, Jun Yan Dai, Yu Han,\\Marco Di Renzo, Yong Zeng, Shi Jin, Qiang Cheng, and Tie Jun Cui

\thanks{W. Tang, X. Chen, Y. Han, Y. Zeng, and S. Jin are with the National Mobile Communications Research
Laboratory, Southeast University, Nanjing, China.}
\thanks{M. Z. Chen, J. Y. Dai, Q. Cheng, and T. J. Cui are with the State Key Laboratory of Millimeter Waves, Southeast University, Nanjing, China.}
\thanks{M. Di Renzo is with Universit\'e Paris-Saclay, CNRS, CentraleSup\'elec, Laboratoire des Signaux et Syst\`emes, Paris, France.}
}

\maketitle
\vspace{-1.6cm}
\begin{abstract}
Reconfigurable intelligent surfaces (RISs) comprised of tunable unit cells have recently drawn significant attention due to their superior capability in manipulating electromagnetic waves. In particular, RIS-assisted wireless communications have the great potential to achieve significant performance improvement and coverage enhancement in a cost-effective and energy-efficient manner, by properly programming the reflection coefficients of the unit cells of RISs. In this paper, free-space path loss models for RIS-assisted wireless communications are developed for different scenarios by studying the physics and electromagnetic nature of RISs. The proposed models, which are first validated through extensive simulation results, reveal the relationships between the free-space path loss of RIS-assisted wireless communications and the distances from the transmitter/receiver to the RIS, the size of the RIS, the near-field/far-field effects of the RIS, and the radiation patterns of antennas and unit cells. In addition, three fabricated RISs (metasurfaces) are utilized to further corroborate the theoretical findings through experimental measurements conducted in a microwave anechoic chamber. The measurement results match well with the modeling results, thus validating the proposed free-space path loss models for RIS, which may pave the way for further theoretical studies and practical applications in this field.
\end{abstract}
\vspace{-0.4cm}
\begin{IEEEkeywords}
\vspace{-0.4cm}
Path loss, reconfigurable intelligent surface, metasurface, intelligent reflecting surface, large intelligent surface, wireless propagation measurements.
\end{IEEEkeywords}

\IEEEpeerreviewmaketitle

\section{Introduction}
It is foreseen that the commercial service of the fifth-generation (5G) of mobile communications will be launched on a worldwide scale from 2020. Massive multiple-input multiple-output (MIMO) base stations (BSs) with full-digital transceivers are being commercially deployed and tested in several countries \cite{HW}. Meanwhile, the success of millimeter wave (mmWave) trials and testbeds across the world ensures that mmWave wireless communications will be realized by 2020 \cite{mmwave-band}. As the two key technologies of the 5G physical layer, massive MIMO and mmWave will significantly increase the network capacity and resolve the spectrum shortage issue in current cellular communication systems \cite{mmwave}. On the other hand, the International Telecommunication Union (ITU) predicted that the exponential growth of the global mobile data traffic will continue, which will reach 5 zettabytes (ZB, 1 ZB = $10^{21}$ Bytes) per month by 2030 \cite{6G}. The ever-increasing traffic demands are driven by emerging data-intensive applications, such as virtual reality, augmented reality, holographic projection, autonomous vehicle, tactile Internet, to name a few, which can hardly be offered by 5G. Therefore, there has been increasing research interest from both academia and industry towards the sixth-generation (6G) of mobile communications.

Looking forward to future 6G wireless communication technologies, one possibility is to further extend the spectrum to terahertz (THz) band \cite{terahertz} and the antenna array scale to ultra-massive MIMO (UM-MIMO) \cite{ummimo}\cite{activeLIS}, which can further boost the spatial diversity and expand the available spectrum resources. Besides, artificial intelligence (AI), orbital angular momentum (OAM) multiplexing, visible-light communications (VLC), blockchain-based spectrum sharing, quantum computing, etc., to name just a few, are being actively discussed as potential enabling technologies for 6G mobile communication systems \cite{6Gtech}. Furthermore, reconfigurable intelligent surfaces (RISs)\footnote{Also known as intelligent reflecting surfaces (IRSs) and passive large intelligent surfaces (LISs) in the literature.}, which are an emerging technology to manipulate electromagnetic waves, have drawn significant attention due to their capability in tailoring electromagnetic waves across a wide frequency range, from microwave to visible light. RIS technology is enabled by metasurfaces comprised of sub-wavelength unit cells with tunable electromagnetic responses, such as amplitude, phase, polarization and frequency, which can be controlled by external signals in a real-time reconfigurable manner during the light-matter interaction \cite{MetaInfo,MetaNsr,MetaNature}. Their programmable electromagnetic properties make RISs especially appealing for wireless communications.

Recently, we have witnessed fast growing research effort on wireless communication using RISs. RISs have been utilized to realize new wireless transceiver architectures in \cite{MetaCom,MetaEL,MetaMag,MetaAMT}, which may bring a paradigm-shift on the transceiver design and reduce the hardware cost of future wireless communication systems. Moreover, RISs can artificially shape the electromagnetic wave propagation environment \cite{MetaAI}. In current wireless communication systems, the wireless environment, i.e., the physical objects that influence the propagation of the electromagnetic waves, is uncontrollable. During the research and design processes of wireless communication systems, one can only design transceivers, signal processing algorithms, and transmission protocols to adapt to the radio environment. RISs, on the other hand, may in principle make the wireless environment controllable and programmable, thus bringing unprecedented new opportunities for enhancing the performance of wireless communication systems \cite{MetaSmart}\cite{MetaSoftware}.

Against this background, various RIS-assisted wireless communication systems have recently been studied, by providing cost-effective and energy-efficient solutions that offer performance improvements and coverage enhancement in wireless networks \cite{RIS1}. For example, \cite{IRS1} showed that, by jointly optimizing active and passive beamforming in an RIS-assisted wireless network, the energy consumption and the coverage performance can be significantly improved. The joint beamforming design was also investigated for improving the physical layer security \cite{IRS3}\cite{IRS4}. The authors of \cite{RIS3,RIS4,IRS5} optimized the achievable rate of RIS-assisted systems in different communication scenarios. The energy efficiency of RIS-based downlink multi-user MISO systems was maximized in \cite{RIS2}\cite{LIS3}.
Energy harvesting performance, ergodic spectral efficiency, symbol error probability, and outage probability of RIS-assisted wireless communications were derived and optimized in \cite{IRS7,LIS1,LIS2,LIS4}. The comparison with relays was carried out in \cite{Relaycompare}. The impact of hardware imperfections, like discrete phase-shift and phase-dependent amplitude, were studied in \cite{LIS1}\cite{IRS2}\cite{IRS6}. Channel estimation for RIS-assisted systems was discussed in \cite{IRSR7,IRSR8,IRSR9}. The impact of channel estimation and feedback overhead was recently analyzed in \cite{IRSR10}.

However, the major limitation of existing research on RIS-assisted wireless communications is the lack of tractable and reliable physical and electromagnetic models for the RISs. Most of the existing research works are based on simplistic mathematical models that regard the RIS as a diagonal matrix with phase shift values. The responses of the RISs to the radio waves have not yet been extensively studied from the physics and electromagnetic point of view, which may lead to relatively simplified algorithm designs and performance predictions. To the best of the authors' knowledge, in particular, there is no experimentally validated path loss model for RIS-assisted wireless communications, even in the simple free-space propagation environment. The path loss models used in most existing works do not consider physical factors such as the size of the RISs, and the near-field/far-field effects of the RISs. Therefore, channel measurements are urgently needed, which may provide authentic information on the path loss of RIS-assisted wireless communications. Motivated by these conditions, we develop free-space path loss models for RIS-assisted wireless communications in different scenarios based on the physics and electromagnetic nature of the RISs. We validate them by using numerical simulations, and we also conduct channel measurements by using fabricated RISs in an anechoic chamber to practically corroborate our findings. In particular, our main contributions are summarized as follows:

1) We introduce a system model for RIS-assisted wireless communications from the perspective of electromagnetic theory, which takes into account important physical factors like the physical dimension of the RISs, and the radiation pattern of the unit cells. A general formula is derived to characterize the free-space path loss of RIS-assisted wireless communications.

2) Based on the derived general formula, we propose three free-space path loss models for RIS-assisted wireless communications, which capture three relevant scenarios and unveil the relationships between the free-space path loss of RIS and the distances from the transmitter/receiver to the RIS, the size of the RIS, the near-field/far-field effects of the RIS, as well as the radiation patterns of antennas and unit cells.

3) We report the world's first experimental results that assess and validate the path loss of RIS-assisted wireless communications. Three different RISs (metasurfaces) of various size are utilized in the measurements. The measurement results are shown to be in good agreement with the modeling results. Besides, the power consumption of the RISs and the effect of the incident angle are also measured and discussed.
\vspace{-0.25cm}
\section{Preliminaries}
\vspace{-0.2cm}
This section reviews several basic concepts, which will be used in the free-space path loss modeling of RIS-assisted wireless communications in the next section.
\vspace{-0.59cm}
\subsection{Far Field and Near Field}\label{FieldofRIS}
\vspace{-0.18cm}
When a transmitter is sufficiently far away from an antenna array, the spherical wave generated by the transmitter can be approximately regarded as a plane wave at the antenna array side. Specifically, when the maximum phase difference of the received signal on the antenna array does not exceed $\frac{\pi }{8}$, the transmitter is in the far field of the antenna array \cite{Book0}. Based on this assumption, the boundary of the far field and the near field of the antenna array is defined as $L=\frac{{2{D^2}}}{\lambda}$, where $L$, $D$ and $\lambda$ denote the distance between the transmitter and the center of the antenna array, the largest dimension of the antenna array and the wave length of the signal, respectively \cite{Book0}. It's worth noting that when the transmitter is replaced by a receiver, the above definition remains the same due to the reciprocity of the antenna array. We will use the same definition of the far field and near field for RISs in the next section. When the distance between the transmitter/receiver and the center of the RIS is less than $\frac{{2{D^2}}}{\lambda }$, the transmitter/receiver are considered to be in the near field\footnote{We assume that the distance between the transmitter/receiver and each unit cell of the RIS is larger than $5\lambda$ throughout this paper\cite{Book1}, which is regarded as the lower bound of the near field of the RIS in this paper. This assumption ensures that for each unit cell of the RIS, we can consider only the unit cell's far field component: the incident/reflected power of each unit cell follows the inverse square law\cite{TSE}. As a result, the near field range of the RIS is defined as between $5\lambda$ and $\frac{{2{D^2}}}{\lambda}$ in this paper. It is worth noting that the RIS used for assisting wireless communication usually satisfy that $\frac{{2{D^2}}}{\lambda}$ is greater than $5\lambda$, ie., $D > 1.58\lambda$. If an RIS is too small to support $\frac{{2{D^2}}}{\lambda} > 5\lambda$, then this RIS only has far-field region where the distance $> 5\lambda$.} of the RIS. Otherwise, they are in the far field of the RIS.
\vspace{-0.4cm}
\subsection{Power Radiation Pattern and Gain}\label{Radiationpattern}
\begin{figure}[H]
	\centering
	\includegraphics[height=1.5in]{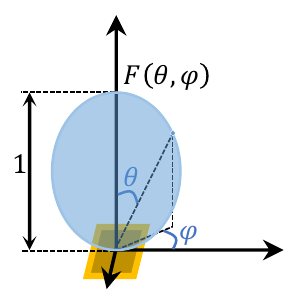}
    \vspace{-0.5cm}
	\caption{. Diagram of normalized power radiation pattern.}
	\label{radiationpattern}
\end{figure}

\vspace{-0.24cm}
The power radiation pattern defines the variation of the power radiated or received by an antenna as a function of the direction away from the antenna, which allows us to visualize where the antenna transmits or receives the maximum power. The normalized power radiation pattern can be written as a function $F\left( {\theta ,\varphi } \right)$ in the spherical coordinate system as shown in Fig. \ref{radiationpattern}, where $\theta$ and $\varphi$ are the elevation and azimuth angles from the antenna to a certain transmitting/receiving direction. An example of normalized power radiation pattern is the following\cite{Book1}
\begin{equation}\label{s1}
F\left( {\theta ,\varphi } \right) = \left\{
\begin{array}{rcl}
{{\cos ^3}\theta} & & {\theta  \in \left[ {0,\frac{\pi }{2}} \right], \varphi  \in \left[ {0,2\pi } \right]}\\
0 & & {\theta  \in \left( {\frac{\pi }{2},\pi } \right], \varphi  \in \left[ {0,2\pi } \right]}
\end{array} \right.
\end{equation}

The specific normalized power radiation pattern represented by (\ref{s1}) is only a function of the elevation angle $\theta$ and is maximized when $\theta=0$, which indicates that the corresponding antenna has maximum gain towards the direction of $\theta=0$. The term antenna gain describes how much power is transmitted or received in the direction of peak radiation relative to that of an isotropic antenna, which, by assuming 100$\%$ antenna efficiency, can be written as  \cite{Book1}
\begin{equation}\label{s2}
Gain = \frac{{4\pi }}{{\int\limits_{\varphi  = 0}^{2\pi } {\int\limits_{\theta  = 0}^\pi  {{{ {F\left( {\theta ,\varphi } \right)} }}\sin \theta d\theta d\varphi } } }}.
\end{equation}

Consider the antenna with power radiation pattern described by (\ref{s1}) as an example, its gain is equal to 8 (9.03 dBi) under the assumption of 100$\%$ radiation efficiency, which means that the power transmitted or received in the direction of peak radiation ($\theta=0$) is 9.03 dB higher than that of an isotropic antenna. In the next section, we will use the definition of normalized power radiation pattern not only for the antennas of the transmitter and receiver, but also for the unit cells of RIS in the free-space path loss modeling of RIS-assisted wireless communications\footnote{We use (\ref{s1}) to match the normalized power radiation pattern of the unit cell of the RIS in this paper, because simulated -3dB main lobe width of the unit cell of three utilized RISs in Section IV and V is about $75^{\circ}$ $(\cos^3 (\frac{75^{\circ}}{2}) = 0.5)$. The form ${\cos^\alpha}$ can be used to match the normalized power radiation pattern of different unit cell and antenna designs with appropriate $\alpha$\cite{Book1}.}.
\vspace{-0.7cm}
\vspace{-1.4cm}
\section{Free-Space Path Loss Modeling of RIS-Assisted Wireless Communications}\label{Modeling}
\vspace{-0.2cm}
In this section, we describe the overall system model and develop the free-space path loss model for RIS-assisted wireless communications in different scenarios.
\vspace{-0.6cm}
\subsection{System Model}\label{SystemModel}
\begin{figure}[H]
	\centering
	\includegraphics[height=2.6in]{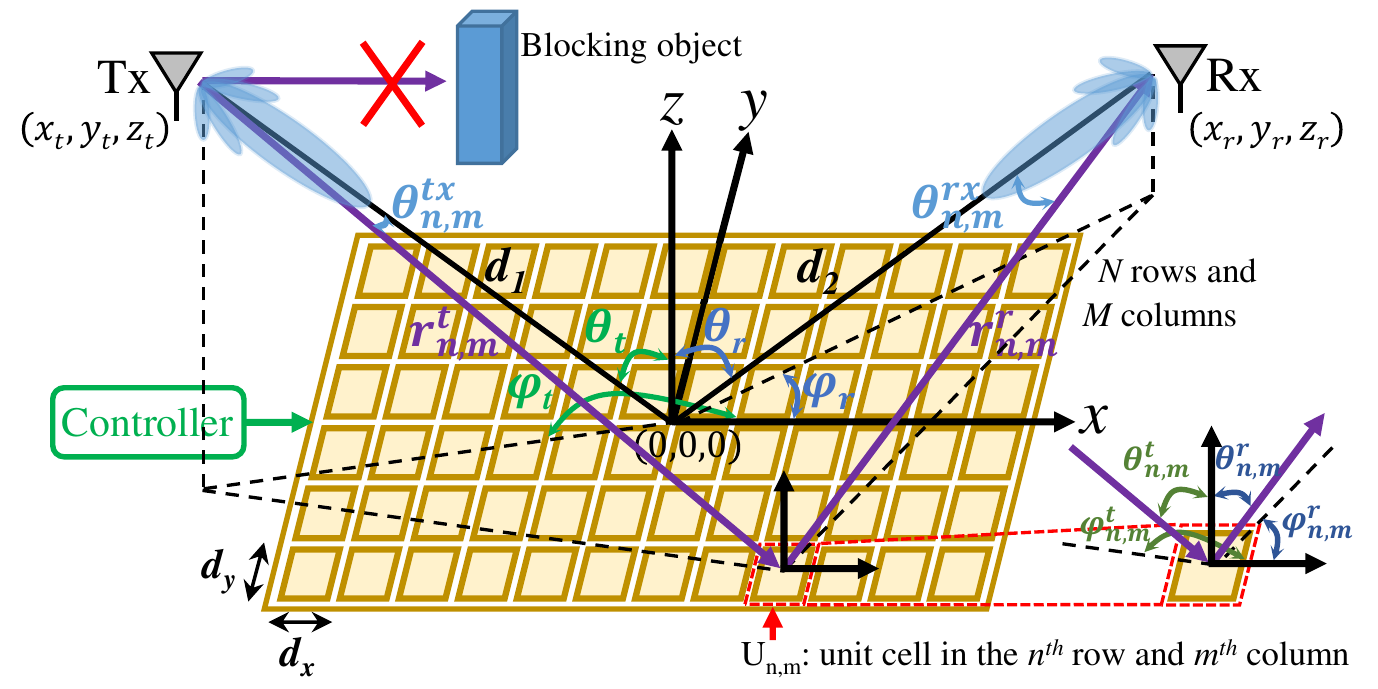}
    \vspace{-0.4cm}
	\caption{. RIS-assisted wireless communication without the direct path between the transmitter and the receiver.}
	\label{systemmodel}
\end{figure}

\vspace{-0.2cm}
We consider a general RIS-assisted single-input single-output (SISO) wireless communication system as shown in Fig. \ref{systemmodel}. Since this paper aims to study the free-space path loss modeling of RIS-assisted wireless communications, the direct link between the transmitter and the receiver is ignored in the rest of the paper\footnote{If the direct link exists, the received signal becomes the
sum of the signal reflected from the RIS and the signal transmitted from the direct link. Since the channel model of the direct link has already been well studied\cite{Book2}, this work focuses on the RIS-assisted link and aims to provide its free-space path loss models in different scenarios.}. The RIS is placed in the x-y plane of a Cartesian coordinate system, and the geometric center of the RIS is aligned with the origin of the coordinate system.
Let $N$ and $M$ denote the number of rows and columns of the regularly arranged unit cells of the RIS, respectively. The size of each unit cell along the x axis is $d_{x}$ and that along the y axis is $d_{y}$, which are usually of subwavelength scale within the range of $\frac{\lambda}{10}$ and $\frac{\lambda}{2}$. $F\left( {\theta ,\varphi } \right)$ is the normalized power radiation pattern of the unit cell, which reveals the dependence of the incident/reflected power density of the unit cell on the incident/reflected angle. $G$ is the gain of the unit cell, which is defined by (\ref{s2}) and is only related to the normalized power radiation pattern $F\left( {\theta ,\varphi } \right)$ of the unit cell. $U_{n,m}$ denotes the unit cell in the $n^{th}$ row and $m^{th}$ column with the programmable reflection coefficient $\Gamma _{n,m}$. The center position of $U_{n,m}$ is $((m-\frac{1}{2})d_{x}, (n-\frac{1}{2})d_{y}, 0)$, where $m \in \left[ {1 - \frac{M}{2},\frac{M}{2}} \right]$ and $n \in \left[ {1 - \frac{N}{2},\frac{N}{2}} \right]$, assuming that both of $N$ and $M$ are even numbers. We use the symbols $d_{1}$, $d_{2}$, $\theta_{t}$, $\varphi_{t}$, $\theta_{r}$ and $\varphi_{r}$ to represent the distance between the transmitter and the center of the RIS, the distance between the receiver and the center of the RIS, the elevation angle and the azimuth angle from the center of the RIS to the transmitter, the elevation angle and the azimuth angle from the center of the RIS to the receiver, respectively. Let $r_{n,m}^t$, $r_{n,m}^r$, $\theta_{n,m}^t$, $\varphi_{n,m}^t$, $\theta_{n,m}^r$ and $\varphi_{n,m}^r$ represent the distance between the transmitter and $U_{n,m}$, the distance between the receiver and $U_{n,m}$, the elevation angle and the azimuth angle from $U_{n,m}$ to the transmitter, the elevation angle and the azimuth angle from $U_{n,m}$ to the receiver, respectively.
\vspace{-0.45cm}
\vspace{-0.7cm}
As shown in Fig. \ref{systemmodel}, the transmitter emits a signal to the RIS with power $P_{t}$ through an antenna with normalized power radiation pattern ${F^{tx}}(\theta,\varphi)$ and antenna gain $G_t$. The signal is reflected by the RIS and received by the receiver with normalized power radiation pattern ${F^{rx}}(\theta,\varphi)$ and antenna gain $G_r$. Let $\theta_{n,m}^{tx}$, $\varphi_{n,m}^{tx}$, $\theta_{n,m}^{rx}$ and $\varphi_{n,m}^{rx}$ represent the elevation angle and the azimuth angle from the transmitting antenna to the unit cell $U_{n,m}$, and the elevation angle and the azimuth angle from the receiving antenna to the unit cell $U_{n,m}$, respectively. We assume that the polarization of the transmitter and receiver are always properly matched, even after the transmitted signal is reflected by the RIS.

The following key result presents the connection between the received signal power of the receiver and various parameters described above in RIS-assisted wireless communication systems.

{\textbf{Theorem 1.}} The received signal power in RIS-assisted wireless communications is as follows
\begin{equation}\label{s14}
{P_r} = {P_t}\frac{{{G_t}{G_r}G{d_x}{d_y}{\lambda ^2}}}{{64{\pi ^3}}}{\left| {\sum\limits_{m = 1 - \frac{M}{2}}^{\frac{M}{2}} {\sum\limits_{n = 1 - \frac{N}{2}}^{\frac{N}{2}} {\frac{{\sqrt {{F_{n,m}^{combine}}}\  {\Gamma _{n,m}}}}{{r_{n,m}^tr_{n,m}^r}}{e^{\frac{{ - j2\pi (r_{n,m}^t + r_{n,m}^r)}}{\lambda }}}} } } \right|^2},
\end{equation}
where ${{F_{n,m}^{combine}}}{=}{F^{tx}}\left( {\theta _{n,m}^{tx},\varphi _{n,m}^{tx}} \right)F\left( {\theta _{n,m}^t,\varphi _{n,m}^t} \right)F\left( {\theta _{n,m}^r,\varphi _{n,m}^r} \right){F^{rx}}\left( {\theta _{n,m}^{rx},\varphi _{n,m}^{rx}} \right)$, which accounts for the effect of the normalized power radiation patterns on the received signal power.

\emph{Proof}: See Appendix A. \hfill $\blacksquare$

Theorem 1 reveals that the received signal power is proportional to the transmitted signal power, the gains of the transmitting/receiving antennas, the gain of the unit cell, the size of the unit cell, the square of the wave length. In addition, Theorem 1 also indicates that the received signal power is related to the normalized power radiation patterns of the transmitting/receiving antennas and unit cells, the reflection coefficients of the unit cells, and the distances between the transmitter/receiver and the unit cells. However, their specific relationships are not that straightforward, which needs further analysis and discussions in the rest of this paper. Theorem 1 also shows that RIS-assisted wireless communication systems are transmitter-receiver reciprocal, since the same signal power can be received if the roles of the transmitter and the receiver are exchanged, which is an especially important property in the uplink and downlink design of time division duplexing (TDD) wireless communication systems.
\vspace{-0.3cm}
\begin{figure}[H]
	\centering
	\includegraphics[height=2.1in]{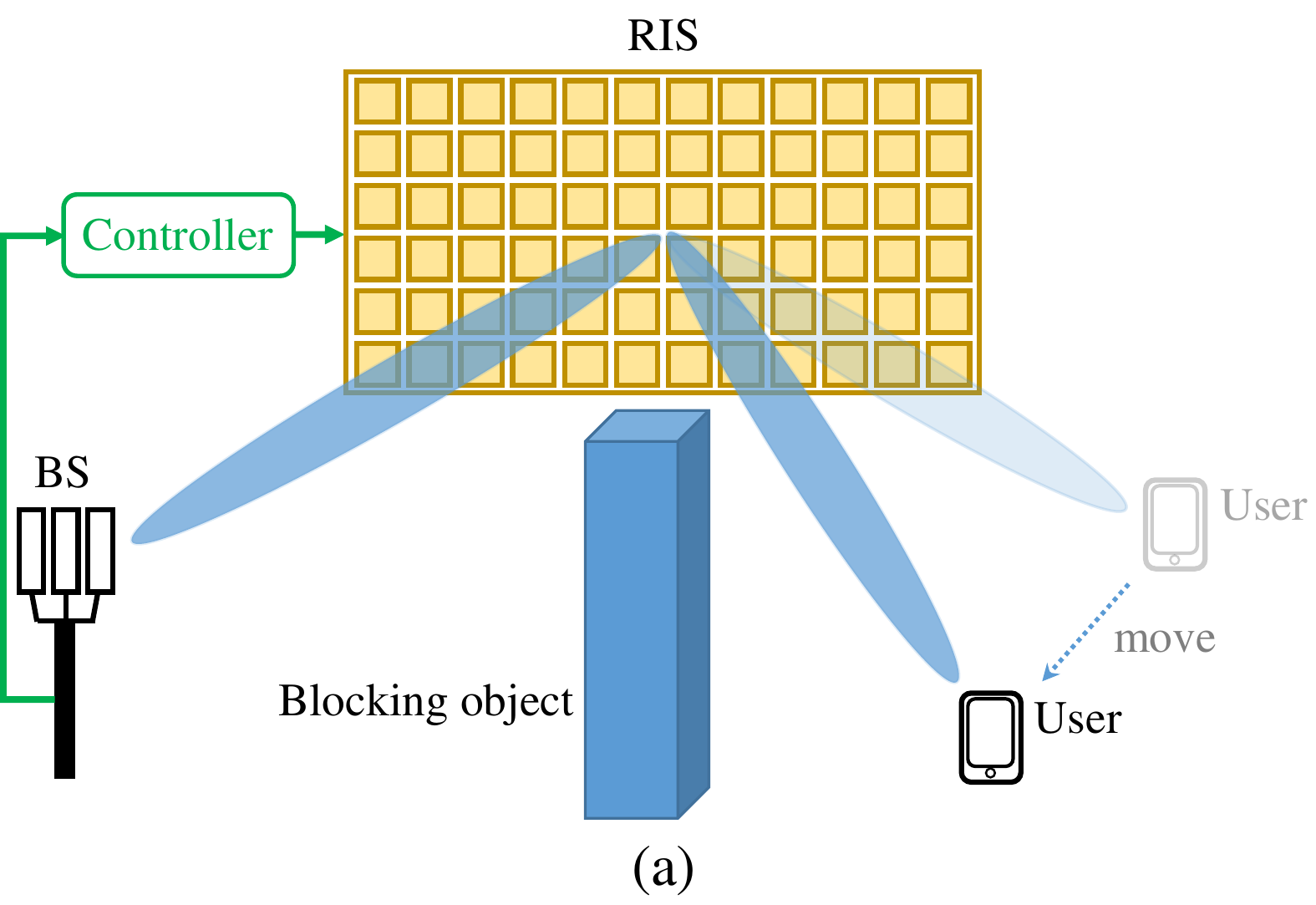}
    \hspace{0.7cm}
    \includegraphics[height=2.1in]{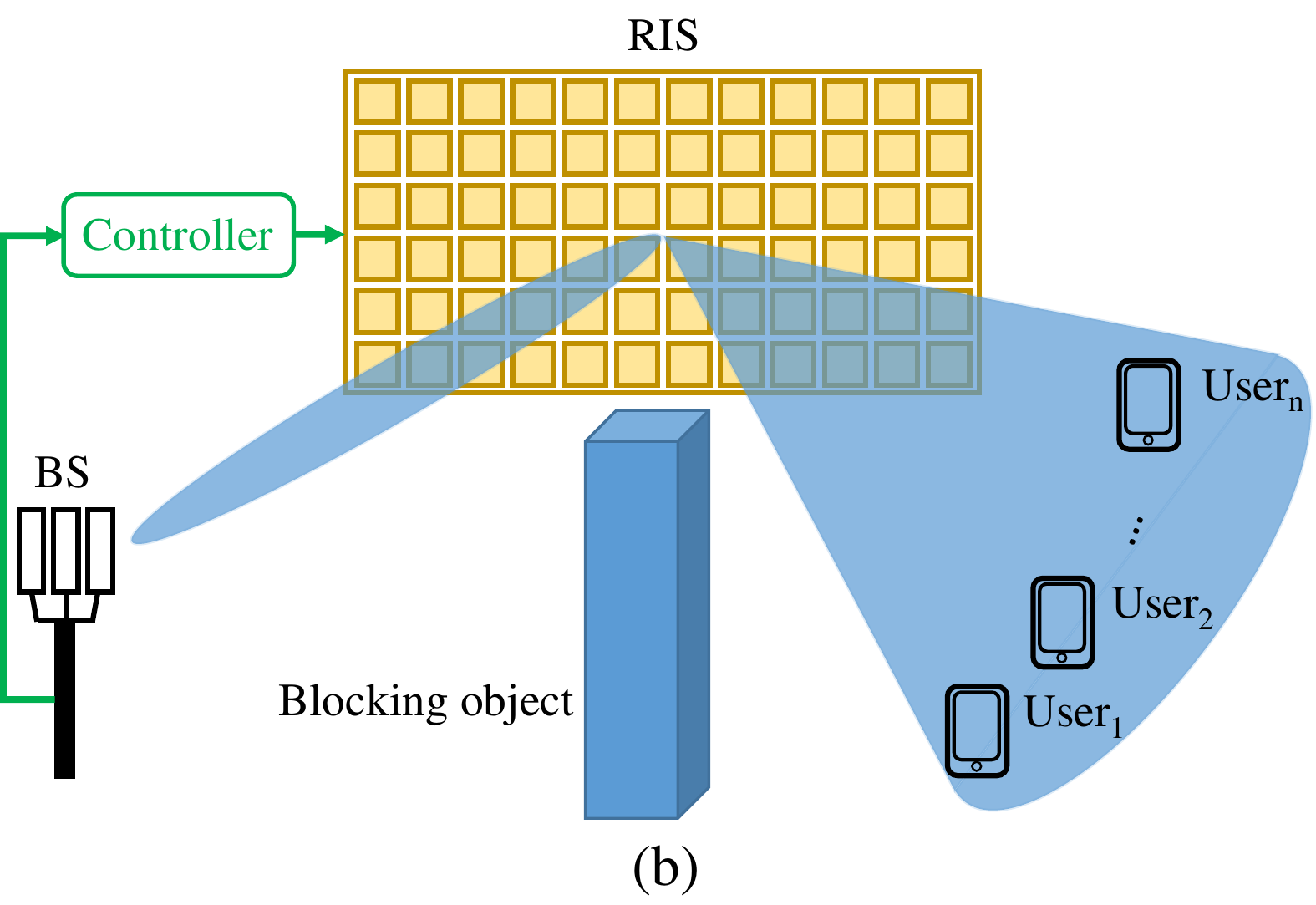}
    \vspace{-1cm}
	\caption{. Two scenarios of RIS-assisted wireless communications. (a) RIS-assisted beamforming. (b) RIS-assisted broadcasting.}
	\label{specificuserarea}
\end{figure}
\vspace{-0.6cm}
In this paper, we develop free-space path loss models for RIS-assisted wireless communications in different scenarios, which can be classified into two categories: RIS-assisted beamforming and RIS-assisted broadcasting as shown in Fig. \ref{specificuserarea}. In the RIS-assisted beamforming scenario, the received signal power is maximized for a single specific user. On the other hand, in the RIS-assisted broadcasting scenario, the signal evenly covers all users in a specific area. We regard (\ref{s14}) as a \textbf{\emph{general formula}}. As each considered scenario depends on the design of the reflection coefficients of the unit cells, as well as the near-field/far-field effects of the RIS, (\ref{s14}) will be further discussed for different cases to draw insights on the free-space path loss of RIS-assisted wireless communications.
\vspace{-0.5cm}
\subsection{Far Field Beamforming Case}\label{FarfieldBeammodel}
The signals reflected by all the unit cells of the RIS to the receiver can be aligned in phase to enhance the received signal power, which makes RISs especially appealing for beamforming applications. Therefore, it is important to study the free-space path loss in RIS-assisted beamforming scenario. The case where both of the transmitter and receiver are in the far field of the RIS is first discussed as follows.

\textbf{Proposition 1.} Assume that the directions of peak radiation of both the transmitting and receiving antennas point to the center of the RIS, and all the unit cells of the RIS share the same reflection coefficient $\Gamma _{n,m}=A{e^{j{\phi}}}$. The received signal power in the far field case can be written as
\begin{equation}\label{s15}
\begin{aligned}
{P_r} = &{P_t}\frac{{{G_t}{G_r}G{M^2}{N^2}{d_x}{d_y}{\lambda ^2}F({\theta _t},{\varphi _t})F({\theta _r},{\varphi _r}){A^2}}}{{64{\pi ^3}{d_1}^2{d_2}^2}}  \\
&\times{\left| {\frac{{\sinc\left( {\frac{{M\pi }}{\lambda }(\sin {\theta _t}\cos {\varphi _t}{+}\sin {\theta _r}\cos {\varphi _r}){d_x}} \right)}}{{\sinc(\frac{\pi }{\lambda }(\sin {\theta _t}\cos {\varphi _t}{+}\sin {\theta _r}\cos {\varphi _r}){d_x})}}\frac{{\sinc\left( {\frac{{N\pi }}{\lambda }(\sin {\theta _t}\sin {\varphi _t}{+}\sin {\theta _r}\sin {\varphi _r}){d_y}} \right)}}{{\sinc\left( {\frac{\pi }{\lambda }(\sin {\theta _t}\sin {\varphi _t}{+}\sin {\theta _r}\sin {\varphi _r}){d_y}} \right)}}} \right|^2}.
\end{aligned}
\end{equation}

If $\theta_{r}  = {\theta _{t}}$ and $\varphi_{r}  = {\varphi _{t}} + {\pi}$, (\ref{s15}) is maximized as
\begin{equation}\label{s16}
P_r^{\max } = \frac{{{G_t}{G_r}G{M^2}{N^2}{d_x}{d_y}{\lambda ^2}F({\theta _t},{\varphi _t})F({\theta _r},{\varphi _r}){A^2}}}{{64{\pi ^3}{d_1}^2{d_2}^2}}{P_t}.
\end{equation}

The path loss corresponding to (\ref{s16}) is
\begin{equation}\label{s23}
PL_{farfield}^{beam} = \frac{{64{\pi ^3}{{({d_1}{d_2})}^2}}}{{{G_t}{G_r}G{M^2}{N^2}{d_x}{d_y}{\lambda ^2}F({\theta _t},{\varphi _t})F({\theta _r},{\varphi _r}){A^2}}}.
\end{equation}

\emph{Proof}: See Appendix B. \hfill $\blacksquare$

Proposition 1 is more insightful compared with the general formula (\ref{s14}) in Theorem 1 and equation  (\ref{s15}) is referred to as the \textbf{\emph{far-field formula}}. Proposition 1 reveals that the free-space path loss of RIS-assisted wireless communications is proportional to $({d_1}{d_2})^2$ in the far field case. The free-space path loss is also related to the unit cells' normalized power radiation pattern $F\left( {\theta ,\varphi } \right)$, which is fixed once the RIS is designed and fabricated. Furthermore, when the reflection coefficients of all the unit cells are the same, the RIS performs specular reflection, that is, the incident signal is mainly reflected towards the mirror direction ($\theta_{r}  = {\theta _{t}}$ and $\varphi_{r}  = {\varphi _{t}} + {\pi}$).

The above analysis assumes that all the unit cells share the same reflection coefficient. The design of ${{\Gamma _{n,m}}}$ is discussed in the following proposition, which enables the RIS to beamform the reflected signal to the receiver in any desired direction $({\theta _{des}},{\varphi _{des}})$ through intelligent reflection\footnote{The intelligent reflection in this paper refers to intelligently controlling the reflection nature of the RIS by properly programming the reflection coefficients of the unit cells of the RIS. The specular reflection in this paper includes two cases: far-field specular beamforming and near-field specular broadcasting, and the reflection coefficients of the unit cells are identical.}.

\textbf{Proposition 2.} Assume that the directions of peak radiation of both the transmitting and receiving antennas point to the center of the RIS, and the reflection coefficients of all the unit cells share the same amplitude value $A$ and different phase shift ${\phi _{n,m}}$ ($\Gamma _{n,m}=A{e^{j{\phi _{n,m}}}}$). The received signal power through intelligent reflection in the far field case can be written as
\begin{equation}\label{s17}
\begin{aligned}
{P_r} = &{P_t}\frac{{{G_t}{G_r}G{M^2}{N^2}{d_x}{d_y}{\lambda ^2}F({\theta _t},{\varphi _t})F({\theta _r},{\varphi _r}){A^2}}}{{64{\pi ^3}{d_1}^2{d_2}^2}}  \\
\times&{\left| {\frac{{\sinc\left( {\frac{{M\pi }}{\lambda }(\sin {\theta _t}\cos {\varphi _t}{+}\sin {\theta _r}\cos {\varphi _r}{+}{\delta _1}){d_x}} \right)}}{{\sinc(\frac{\pi }{\lambda }(\sin {\theta _t}\cos {\varphi _t}{+}\sin {\theta _r}\cos {\varphi _r}{+}{\delta _1}){d_x})}}\frac{{\sinc\left( {\frac{{N\pi }}{\lambda }(\sin {\theta _t}\sin {\varphi _t}{+}\sin {\theta _r}\sin {\varphi _r}{+}{\delta _2}){d_y}} \right)}}{{\sinc\left( {\frac{\pi }{\lambda }(\sin {\theta _t}\sin {\varphi _t}{+}\sin {\theta _r}\sin {\varphi _r}{+}{\delta _2}){d_y}} \right)}}} \right|^2},
\end{aligned}
\end{equation}
where ${\delta _1}\left( {m{-}\frac{1}{2}} \right){d_x}{+}{\delta _2}\left( {n{-}\frac{1}{2}} \right){d_y}{=}\frac{{\lambda {\phi _{n,m}}}}{{2\pi }}$.
When ${\theta _r}{=}{\theta _{des}}$ and ${\varphi _r}{=}{\varphi _{des}}$, (\ref{s17}) is maximized as
\begin{equation}\label{ss17}
P_{{r}}^{\max } = {P_t}\frac{{{G_t}{G_r}G{M^2}{N^2}{d_x}{d_y}{\lambda ^2}F({\theta _t},{\varphi _t})F({\theta _{des}},{\varphi _{des}}){A^2}}}{{64{\pi ^3}{d_1}^2{d_2}^2}},
\end{equation}
when ${\delta _1} =  - \sin {\theta _t}\cos {\varphi _t} - \sin {\theta _{des}}\cos {\varphi _{des}}$ and ${\delta _2} =  - \sin {\theta _t}\sin {\varphi _t} - \sin {\theta _{des}}\sin {\varphi _{des}}$. ${{\phi _{n,m}}}$ is assumed to be continuous for free-space path loss modeling in this paper\cite{MetaCom}\cite{MetaEL}, and the corresponding design of ${{\phi _{n,m}}}$ is
\begin{equation}\label{s18}
\begin{aligned}
\hspace{-1.8cm}
{\phi _{n,m}} = &\bmod (\frac{{2\pi }}{\lambda }({\delta _1}\left( {m{-}\frac{1}{2}} \right){d_x}{+}{\delta _2}\left( {n{-}\frac{1}{2}} \right){d_y}),2\pi )\\
\hspace{-1.8cm}
 = &\bmod ( - \frac{{2\pi }}{\lambda }((\sin {\theta _t}\cos {\varphi _t}{+}\sin {\theta _{des}}\cos {\varphi _{des}})\left( {m{-}\frac{1}{2}} \right){d_x}{+}\\
 &(\sin {\theta _t}\sin {\varphi _t}{+}\sin {\theta _{des}}\sin {\varphi _{des}})\left( {n{-}\frac{1}{2}} \right){d_y}),2\pi ).
\end{aligned}
\end{equation}

\emph{Proof}: See Appendix C. \hfill $\blacksquare$

According to the phase shift design described by (\ref{s18}), the incident signal from direction $\left( {{\theta _t},{\varphi _t}} \right)$ to the RIS can be manipulated and reflected towards any desired direction $\left( {{\theta _{des}},{\varphi _{des}}} \right)$ in the far field case, which is referred to as intelligent reflection. Besides, (\ref{s15}) and (\ref{s17}) also indicate that the free-space path loss of RIS-assisted far-field beamforming has the same scaling law as for specular reflection and intelligent reflection. Moreover, as for the far-field beamforming case, a large-size RIS outperforms a small-size RIS under the same condition based on (\ref{s23}).
\vspace{-0.4cm}
\subsection{Near Field Beamforming Case}\label{NearfieldBeammodel}
\vspace{-0.1cm}
In the previous subsection, the path loss analyses for the RIS-assisted beamforming assume that both of the transmitter and the receiver are in the far field of the RIS. The case study where the transmitter and/or the receiver are in the near field of the RIS is discussed in this subsection, which enables the RIS to focus the reflected signal to the receiver by properly design the reflection coefficients ${{\Gamma _{n,m}}}$.

\textbf{Proposition 3.} Assume that the reflection coefficients of all the unit cells share the same amplitude value $A$ and different phase shifts ${\phi _{n,m}}$ and that the RIS-assisted wireless communication system operates in the near field case. The received signal power is written as follows
\vspace{-0.0cm}
\begin{equation}\label{s20}
\begin{aligned}
{P_r} = {P_t}\frac{{{G_t}{G_r}G{d_x}{d_y}{\lambda ^2}{A^2}}}{{64{\pi ^3}}} {\left| {\sum\limits_{m = 1 - \frac{M}{2}}^{\frac{M}{2}} {\sum\limits_{n = 1 - \frac{N}{2}}^{\frac{N}{2}} {\sqrt {{F_{n,m}^{combine}}}\ \frac{{e^{\frac{{ - j(2\pi (r_{n,m}^t + r_{n,m}^r) - \lambda {\phi _{n,m}})}}{\lambda }}}}{{r_{n,m}^tr_{n,m}^r}}} } } \right|^2}.
\end{aligned}
\end{equation}
For the desired receiver at any position $\left( {{x_{r}},{y_{r}},{z_{r}}} \right)$, (\ref{s20}) is maximized as
\begin{equation}\label{s21}
\begin{aligned}
P_r^{\max } = {P_t}\frac{{{G_t}{G_r}G{d_x}{d_y}{\lambda ^2}{A^2}}}{{64{\pi ^3}}} {\left| {\sum\limits_{m = 1 - \frac{M}{2}}^{\frac{M}{2}} {\sum\limits_{n = 1 - \frac{N}{2}}^{\frac{N}{2}} {\frac{\sqrt {{F_{n,m}^{combine}}}}{{r_{n,m}^tr_{n,m}^r}}} } } \right|^2},
\end{aligned}
\end{equation}
when
\begin{equation}\label{s22}
{\phi _{n,m}} = \bmod (\frac{{2\pi (r_{n,m}^t + r_{n,m}^r)}}{\lambda },2\pi ).
\end{equation}

The path loss corresponding to (\ref{s21}) is
\begin{equation}\label{s24}
PL_{nearfield}^{beam} = \frac{{64{\pi ^3}}}{{{G_t}{G_r}G{d_x}{d_y}{\lambda ^2}{A^2}{{\left| {\sum\limits_{m = 1 - \frac{M}{2}}^{\frac{M}{2}} {\sum\limits_{n = 1 - \frac{N}{2}}^{\frac{N}{2}} { \frac{\sqrt {{F_{n,m}^{combine}}}}{{r_{n,m}^tr_{n,m}^r}}} } } \right|}^2}}}.
\end{equation}

\emph{Proof}: See Appendix D. \hfill $\blacksquare$

Proposition 3 gives the phase gradient design of the RIS for the maximization of the received signal power in the near field case and equation (\ref{s21}) is referred to as the \textbf{\emph{near-field beamforming formula}}. Proposition 3 also formulates the free-space path loss of RIS-assisted near field beamforming case. However, its dependence on $d_1$ and $d_2$ cannot be directly inferred from (\ref{s24}), which will be further discussed based on the simulation results in the next section.
\subsection{Near Field Broadcasting Case}\label{NearfieldBroadmodel}
The studies conducted in the previous subsections assume that the RIS is employed for beamforming, and, then, the received power is maximized for one specific user. The scenario where the RIS is utilized for broadcasting is discussed in this subsection. When the transmitter is in the near field of the RIS, it means that the transmitter is relatively near to the RIS, thus the electromagnetic wave transmitted to the RIS can be regarded as spherical wave as discussed in Section II. If the RIS is electrically large (i.e., both of its length $Md_x$ and width $Nd_y$ are at least 10 times larger than the wavelength $\lambda$), then the incident spherical wave will form a circular and divergent phase gradient on the RIS surface, which is due to the different transmission distances from the transmitter to each unit cell of the RIS.

\begin{figure}[H]
	\centering
	\includegraphics[height=2.7in]{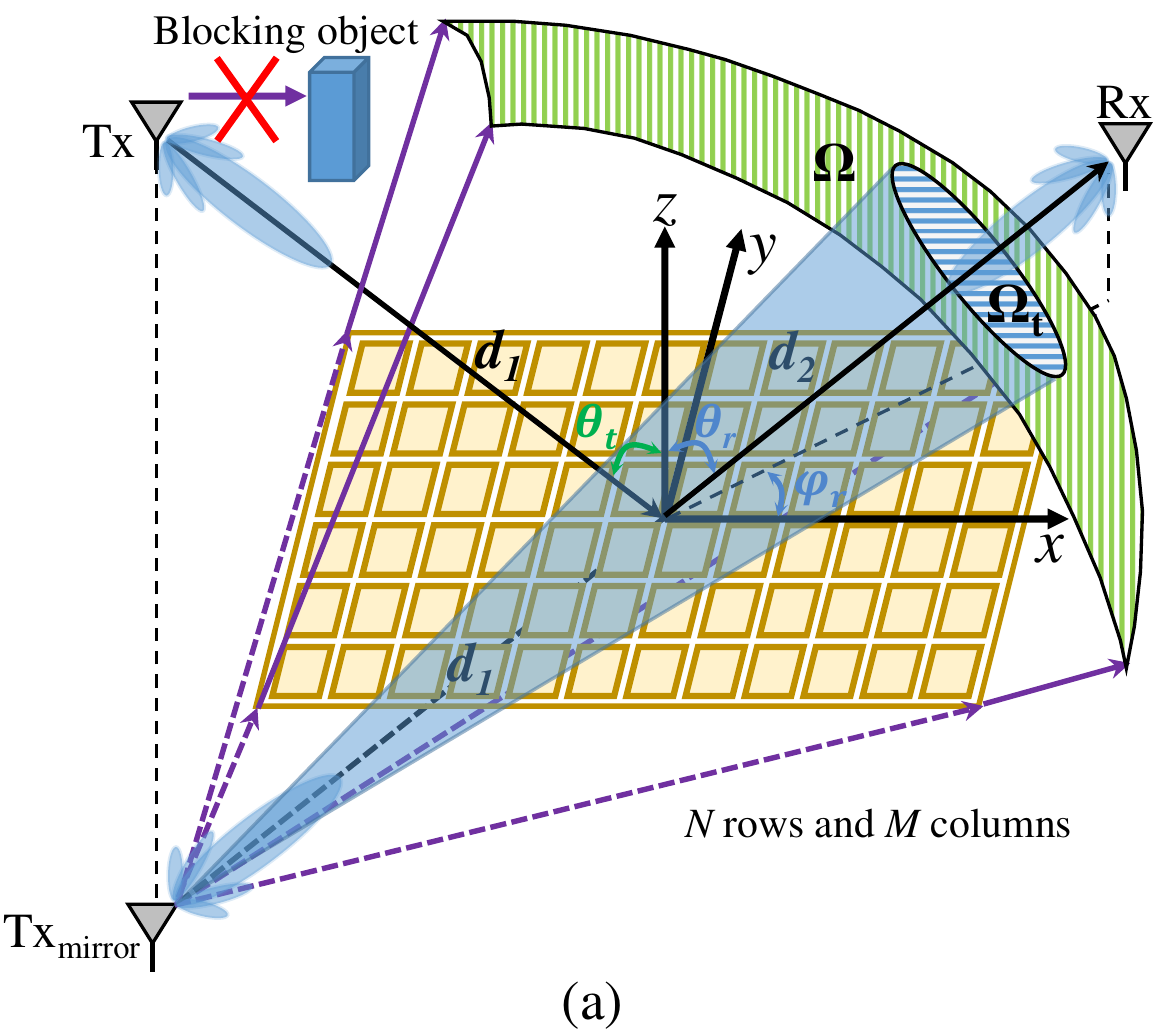}
    \hspace{0.0cm}
    \includegraphics[height=2.7in]{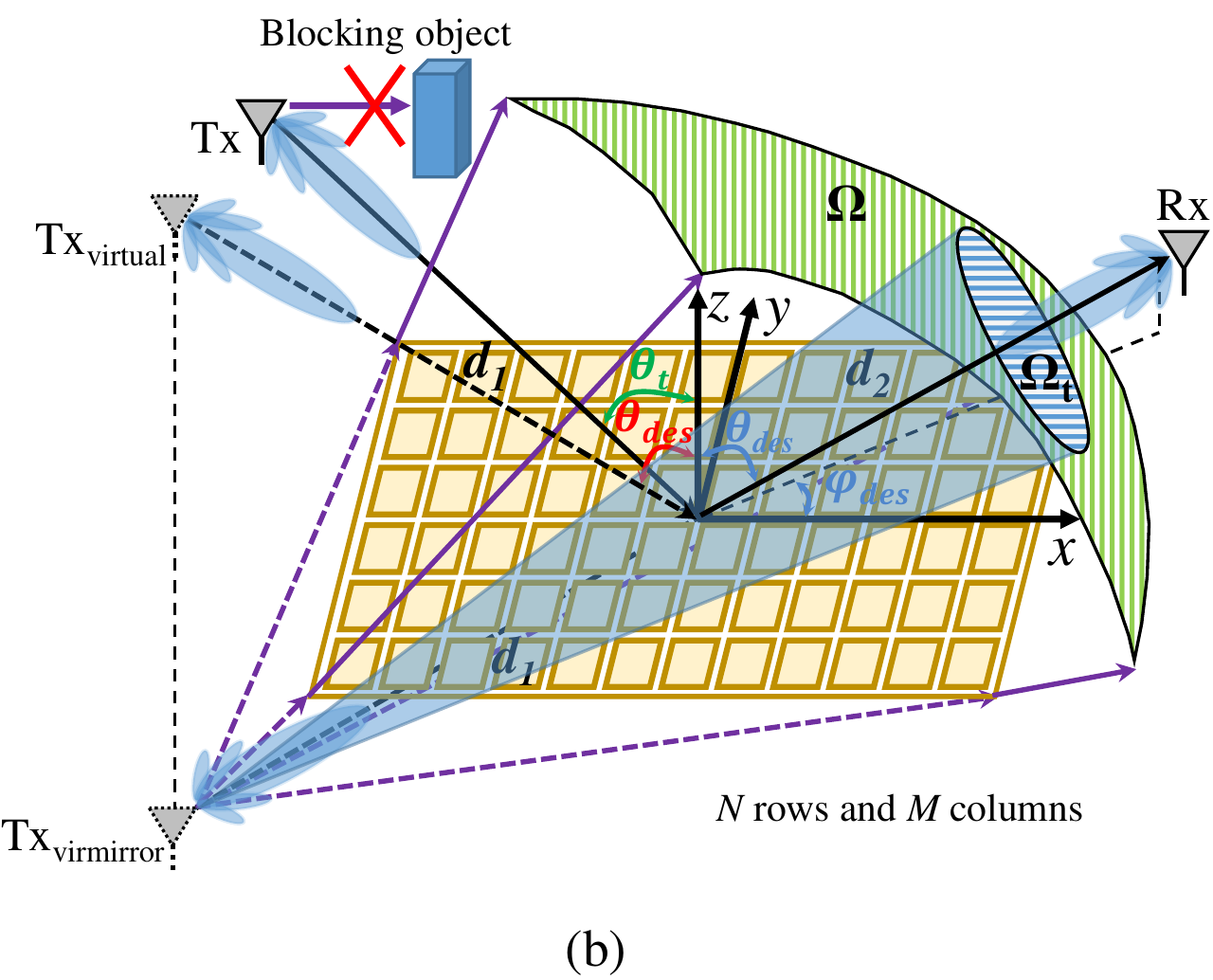}
    \vspace{-1cm}
	\caption{. Near field broadcasting case of RIS-assisted wireless communication in the view of geometrical optics. (a)  Broadcasting through specular reflection. (b) Broadcasting through intelligent reflection.}
	\label{nearfieldsystemmodel}
\end{figure}

\textbf{Proposition 4.} Assume that all the unit cells share the same reflection coefficient $\Gamma _{n,m}=A{e^{j{\phi}}}$, the received signal power through the reflection of an RIS of large electrical size in the near field broadcasting case can be approximately formulated as follows
\begin{equation}\label{s19}
P_r \approx \left\{
\begin{array}{rcl}
{\frac{{{G_t}{A_r}{{A} ^2}}}{{4\pi {{({d_1} + {d_2})}^2}}}{P_t} = \frac{{{G_t}{G_r}{\lambda ^2}{A^2}}}{{16{\pi ^2}{{({d_1} + {d_2})}^2}}}P_t}, & & {{({\theta _r},{\varphi _r}) \in \Omega \cap {\Omega _t}}}\\
0, & & {({\theta _r},{\varphi _r}) \notin \Omega \cap {\Omega _t}}
\end{array} \right.
\end{equation}
where ${A_r} = \frac{{{G_r}{\lambda ^2}}}{{4\pi }}$, and the solid angle $\Omega$ is enclosed by the extension lines from the mirror image of the transmitter ($\mathrm {Tx_{mirror}}$) to the edge of the RIS, as illustrated in Fig. \ref{nearfieldsystemmodel}(a). The solid angle $\Omega_t$ is formed by the main lobe of the transmitting antenna of $\mathrm {Tx_{mirror}}$. If the elevation angle $\theta_r$ and the azimuth angle $\varphi_r$ of the receiver are within the intersection of $\Omega$ and $\Omega_t$, the receiver can receive the reflected signal from the RIS, otherwise it can hardly receive any signal. The RIS performs specular reflection in this near-field broadcasting case, and its free-space path loss is
\begin{equation}\label{s25}
PL_{nearfield}^{broadcast} \approx \frac{{16{\pi ^2}{{({d_1} + {d_2})}^2}}}{{{G_t}{G_r}{\lambda ^2}{A^2}}}.
\end{equation}

\emph{Proof:} In the near field broadcasting case, the free-space path loss of RIS-assisted wireless communications can also be characterized by the general formula (\ref{s14}) in Theorem 1. In addition, a more insightful result can be obtained by using the approximation based on geometric optics, which is typically employed in ray tracing. Since all the unit cells share the same reflection coefficient, the RIS
performs specular reflection in the view of geometrical optics.
As shown in Fig. \ref{nearfieldsystemmodel}(a), the mirror image of the transmitter, namely $\mathrm {Tx_{mirror}}$, can be obtained by taking the RIS as the symmetric plane. The signal transmission process is equivalent to that the signal is transmitted from $\mathrm {Tx_{mirror}}$ and received by the receiver after travelling distance $(d_{1}+d_{2})$, thus we can get Proposition 4 according to the conventional free-space path loss model \cite{Book2}.
\vspace{-0.4cm}
\vspace{-0.9cm}
The above analysis assumes that all the unit cells share the same reflection coefficient. In the following, we introduce a design method for the phase shifts of the unit cells, which enables the RIS to broadcast the reflected signal to the desired direction through intelligent reflection.

\textbf{Proposition 5.} Assume that the reflection coefficients of all the unit cells share the same amplitude value $A$ and different phase shift ${\phi _{n,m}}$, and the RIS aims to broadcast the reflected signal to the desired direction $({\theta _{des}},{\varphi _{des}})$ (i.e., the desired center direction of $\Omega_t$) in the near field broadcasting case, an effective design method for the phase shift ${\phi _{n,m}}$ is the following
\begin{equation}\label{sr1}
{\phi _{n,m}} = \,\bmod \,(\frac{{2\pi }}{\lambda }(r_{n,m}^t - r_{n,m}^{virt}),2\pi ),
\end{equation}
where $r_{n,m}^t$ represents the distance between the transmitter and the unit cell $U_{n,m}$, $r_{n,m}^{virt}$ represents the distance between the virtual transmitter $Tx_{virtual}$ and the unit cell $U_{n,m}$. The position of the virtual transmitter is $ (-{d_1}\sin \theta_{des} \cos \varphi_{des} ,-{d_1}\sin \theta_{des} \sin \varphi_{des} ,{d_1}\cos \theta_{des} )$  as shown in Fig. \ref{nearfieldsystemmodel}{(b)}.

\emph{Proof:} If all the unit cells of the RIS share the same phase shift, i.e., if the RIS is a equiphase surface, it will perform specular reflection in the near-field broadcasting case according to geometric optics. Therefore, as shown in Fig. \ref{nearfieldsystemmodel}{(b)}, if a virtual transmitter $Tx_{virtual}$ is placed in the mirror direction $({\theta _{des}},{\varphi _{des}}+ {\pi})$ of the desired direction $({\theta _{des}},{\varphi _{des}})$ and its distance to the center of RIS is $d_1$, the RIS will broadcast the reflected signal to the desired direction $({\theta _{des}},{\varphi _{des}})$. Since the real transmitter $Tx$ usually doesn't coincide with the imagined virtual transmitter in position, phase compensation can be used to make the real transmitter to behave as the virtual transmitter. As $-\frac{{2\pi }}{\lambda }r_{n,m}^t$ and $-\frac{{2\pi }}{\lambda }r_{n,m}^{virt}$ are the phase alterations caused by free-space propagation from the real transmitter and the virtual transmitter to the unit cell $U_{n,m}$, respectively, the phase shifts of the unit cells can be designed as in (\ref{sr1}) to compensate for the
phase difference between them. With this design, the reflected signal is broadcasted to the desired direction. In addition, the signal transmission process is equivalent to that of a signal is transmitted from $Tx_{virmirror}$
(the mirror image of the virtual transmitter) and received by the receiver after travelling distance ($d_1+ d_2$) as shown in Fig. \ref{nearfieldsystemmodel}{(b)}, thus (\ref{s25}) also holds for the intelligent near field broadcasting.

Propositions 4 and 5 indicate that when the transmitter is in the near field of an RIS with large electrical size and the phase shifts ${\phi _{n,m}}$ are designed as in (\ref{sr1}), the free-space path loss of RIS-assisted wireless communications is proportional to ${{({d_1} + {d_2})}^2}$, instead of ${d_1}^2{d_2}^2$ as for the far-field case. Similarly, when the receiver is in the near field of the RIS, the same result holds according to the transmitter-receiver reciprocity property revealed by Theorem 1. Formula (\ref{s19}) is referred to as the \textbf{\emph{near-field broadcasting formula}}, which describes the relationship between the received and transmitted signal power of RIS-assisted wireless communications in the near field broadcasting case, and is more insightful compared with the general formula (\ref{s14}) in Theorem 1.

\vspace{-0.5cm}
\subsection{Path Loss Model Summary}\label{Modelsummary}
Based on the modeling results above, the free-space path loss models of RIS-assisted wireless communications in different scenarios\footnote{The far field broadcasting case is not discussed in this paper, mainly because it is not that suitable for wireless communications.  In the far-field case,  the path loss of every individual link between the transmitter and receiver provided by each unit cell of the RIS is very large ($\propto {\left( {{d_1}{d_2}} \right)^2}$), so it is necessary to perform beamforming to guarantee the quality of RIS-assisted link, rather than broadcasting. In fact, the ``far field broadcasting'' case is usually used in the research field of metasurface-based electromagnetic diffusion\cite{Diffusion1}\cite{Diffusion2}, which is suitable for the application of radar cross-section (RCS) reduction.} are summarized in Table \ref{pathlosssummary}.
\vspace{-0.3cm}
\newcommand{\tabincell}[2]{\begin{tabular}{@{}#1@{}}#2\end{tabular}}
\begin{table}[H]
\centering
\footnotesize
\caption{Free-Space Path Loss Models of RIS-assisted Wireless Communications in Different Scenarios.}\label{pathlosssummary}
\vspace{-0.2cm}
\begin{tabular}{|l|l|c|l|}
\hline
\textbf{Applications} & \textbf{Near/Far Field Conditions}  & \textbf{Path Loss} & \textbf{Notes}\\
\hline
Beamforming & \tabincell{l}{Transmitter and receiver are both in the \\far field of RIS} & (\ref{s23}) & \tabincell{l}{Apply to both electrically small and large RISs\\${\phi _{n,m}}$ are designed as in (\ref{s18})}\\
\hline
Beamforming & \tabincell{l}{Transmitter and receiver are both or only \\one of them is in the near field of RIS} & (\ref{s24}) & \tabincell{l}{Apply to both electrically small and large RISs\\${\phi _{n,m}}$ are designed as in (\ref{s22})}\\
\hline
Broadcasting & \tabincell{l}{Transmitter and receiver are both or only \\one of them is in the near field of RIS} & (\ref{s25}) & \tabincell{l}{Only apply to electrically large RIS\\${\phi _{n,m}}$ are designed as in (\ref{sr1})}\\
\hline
\end{tabular}
\end{table}
\vspace{-0.6cm}
\section{Validation of Path Loss Models via Numerical Simulations}\label{SimulationResults}
Numerical simulations are conducted based on the general formula (\ref{s14}) to verify the far-field formula (\ref{s15}), the near-field beamforming formula (\ref{s21}), the near-field broadcasting formula (\ref{s19}), and their corresponding path loss formulas (\ref{s23}), (\ref{s24}) and (\ref{s25}), respectively. Three different RISs and two different antennas (corresponding to those used in our measurements, described in Section V) are utilized in the simulations, with their parameters specified in Table \ref{parasummary}.
\vspace{-0.6cm}
\begin{table}[H]
\centering
\footnotesize
\caption{The Parameters of Three Different RISs and Two Different Antennas Utilized in Simulations.}\label{parasummary}
\vspace{-0.2cm}
\begin{tabular}{|l|l|}
\hline
\textbf{Name} & \textbf{Parameters}  \\
\hline
Large RIS1 & \tabincell{l}{$N = 100$, $M = 102$, $d_{x}=d_{y}=0.01\ m$, $\left| {{\Gamma _{n,m}}} \right| = A  = 0.9$, $F(\theta ,\varphi )$ is defined by (\ref{s1}),\\ operating frequency $f{=}10.5\ GHz$, $\lambda{=}c/f{=}0.0286\ m$, $\frac{{2{D^2}}}{\lambda } = \frac{{2MN{d_x}{d_y}}}{\lambda } = 71.4\ m$.} \\
\hline
Large RIS2& \tabincell{l}{$N = 50$, $M = 34$, $d_{x}=d_{y}=0.01\ m$, $\left| {{\Gamma _{n,m}}} \right| = A  = 0.9$, $F(\theta ,\varphi )$ is defined by (\ref{s1}),\\ operating frequency $f = 10.5\ GHz$, $\lambda  = c/f = 0.0286\ m$, $\frac{{2{D^2}}}{\lambda } = \frac{{2MN{d_x}{d_y}}}{\lambda } = 11.9\ m$.}  \\
\hline
Small RIS & \tabincell{l}{$N = 8$, $M = 32$, $d_{x}=d_{y}=0.012\ m$, $\left| {{\Gamma _{n,m}}} \right| = A  = 0.7$, $F(\theta ,\varphi )$ is defined by (\ref{s1}),\\ operating frequency $f = 4.25\ GHz$, $\lambda  = c/f = 0.07\ m$, $\frac{{2{D^2}}}{\lambda } = \frac{{2MN{d_x}{d_y}}}{\lambda } = 1\ m$.}  \\
\hline
X-band horn antenna & \tabincell{l}{$F^{tx}(\theta ,\varphi ) = F^{rx}(\theta ,\varphi ) = \cos^{62} \theta $ when $\theta  \in \left[ {0,\frac{\pi }{2}} \right]$, $F^{tx}(\theta ,\varphi ) = F^{rx}(\theta ,\varphi ) = 0$ when $\theta  \in \left( {\frac{\pi }{2},\pi } \right]$.\\ $G_t = G_r = 21\ dB$ according to (\ref{s2}) under the assumption that the radiation efficiency is $100\%$.\\ The operating frequency is $f = 10.5\ GHz$.}  \\
\hline
C-band horn antenna & \tabincell{l}{$F^{tx}(\theta ,\varphi ) = F^{rx}(\theta ,\varphi ) = \cos^{13} \theta $ when $\theta  \in \left[ {0,\frac{\pi }{2}} \right]$, $F^{tx}(\theta ,\varphi ) = F^{rx}(\theta ,\varphi ) = 0$ when $\theta  \in \left( {\frac{\pi }{2},\pi } \right]$.\\ $G_t = G_r = 14.5\ dB$ according to (\ref{s2}) under the assumption that the radiation efficiency is $100\%$.\\ The operating frequency is $f = 4.25\ GHz$.}  \\
\hline
\end{tabular}
\end{table}
\vspace{-0.8cm}
The electrical sizes of the RISs are ${\text{35}}\lambda{\times}\text{35.7}\lambda$ for the large RIS1, ${\text{17}}{\text{.5}}\lambda{ \times}{\text{11}}{\text{.9}}\lambda$ for the large RIS2, and ${\text{1}}{\text{.36}}\lambda{\times}{\text{5}}{\text{.44}}\lambda$ for the small RIS, respectively.
\vspace{-0.6cm}
\subsection{ RIS-assisted Beamforming}\label{Coveragespecificuser}
\vspace{-0.1cm}
\subsubsection{Specular Reflection in the Far Field Beamforming Case}\label{Specularreflection}
\begin{figure}[H]
	\centering
    \includegraphics[height=2.1in]{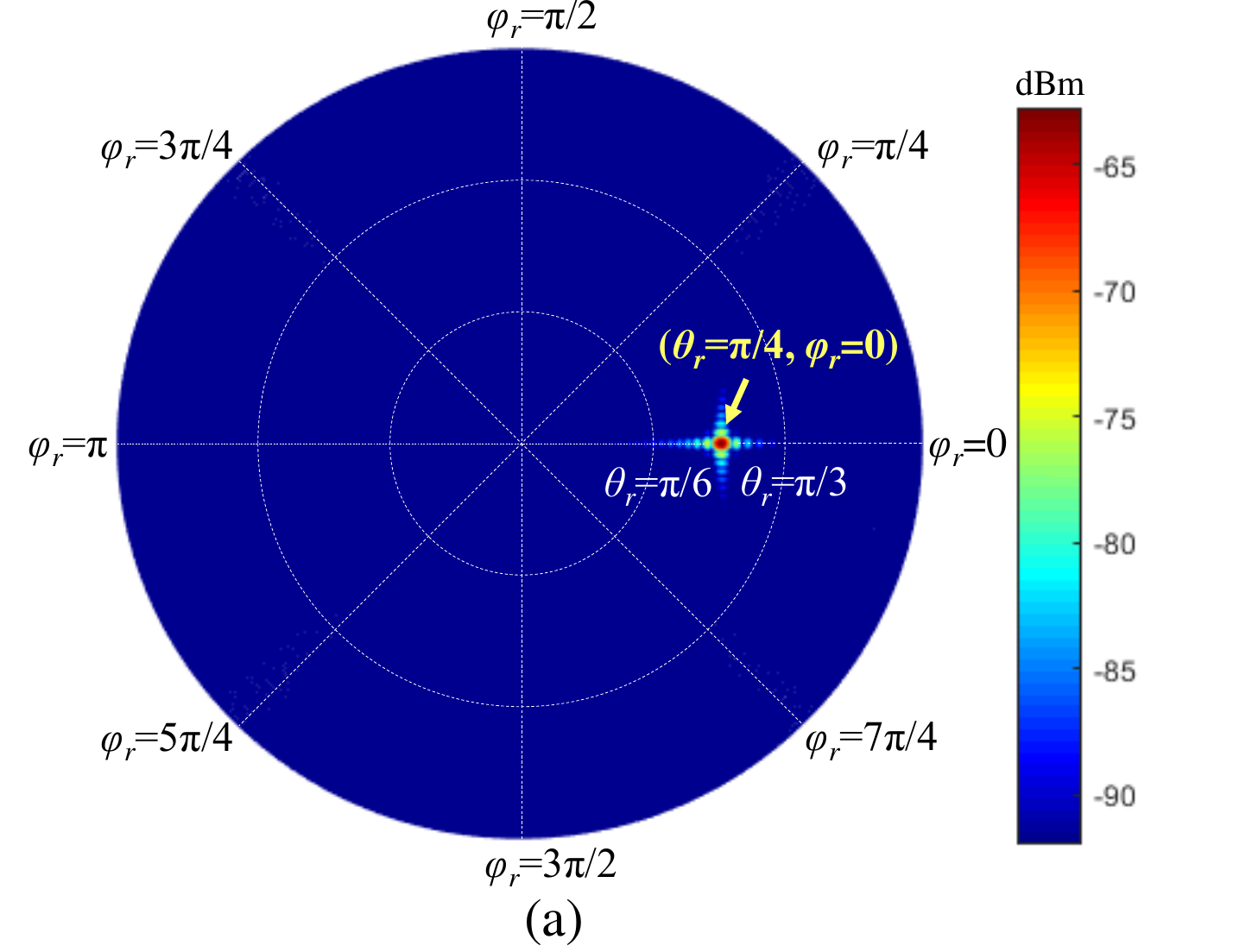}
    \hspace{1cm}
	\includegraphics[height=2.2in]{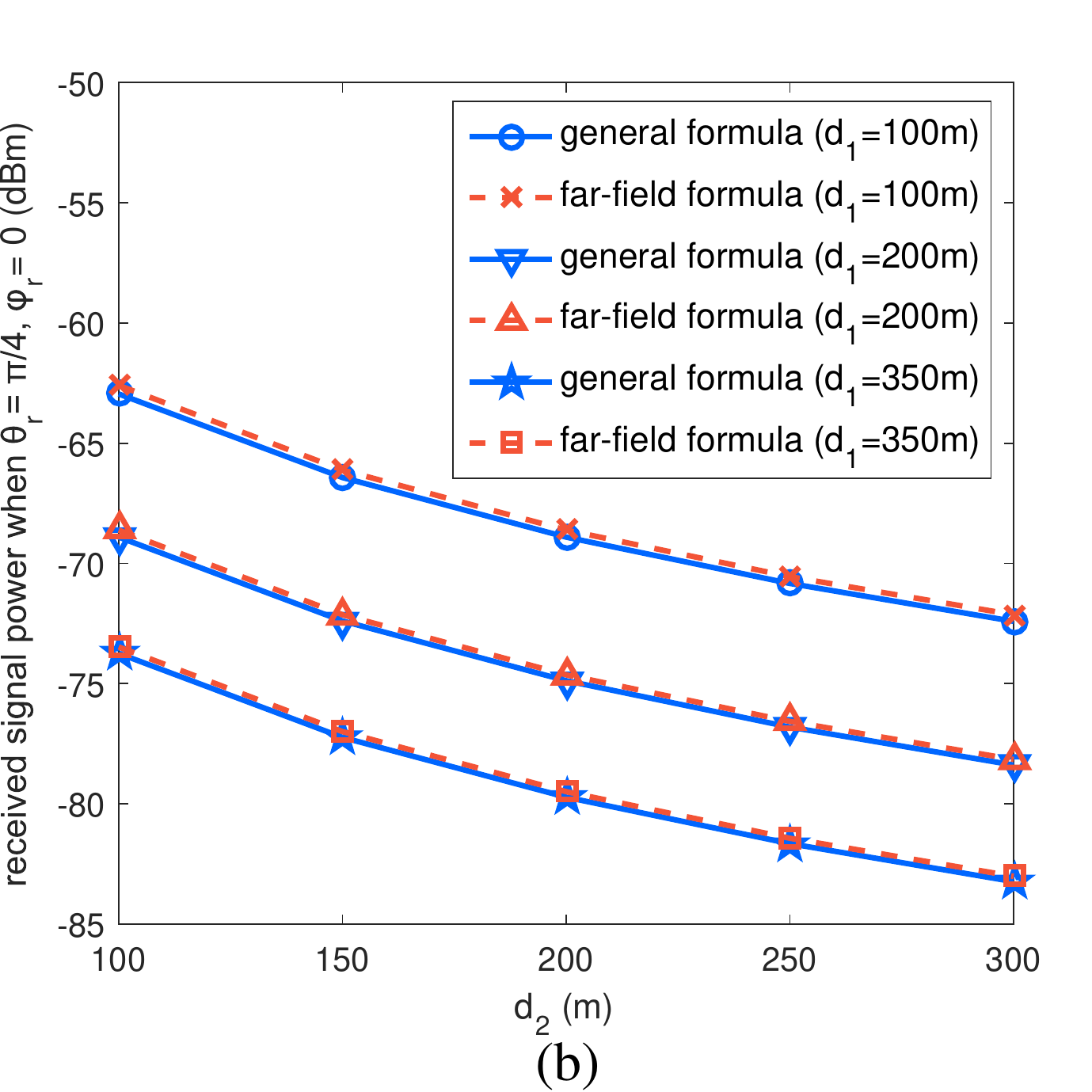}
    \vspace{-0.5cm}
	\caption{. Simulation results of RIS-assisted beamforming through specular reflection of large RIS1 in the far field case. (a) Received signal power distribution when $d_1$ = $d_2$ = 100 m. (b) Received signal power in the specular direction versus $d_1$ and $d_2$.}
	\label{specularfarfield}
\end{figure}

\begin{figure}[H]
	\centering
	\includegraphics[height=2.1in]{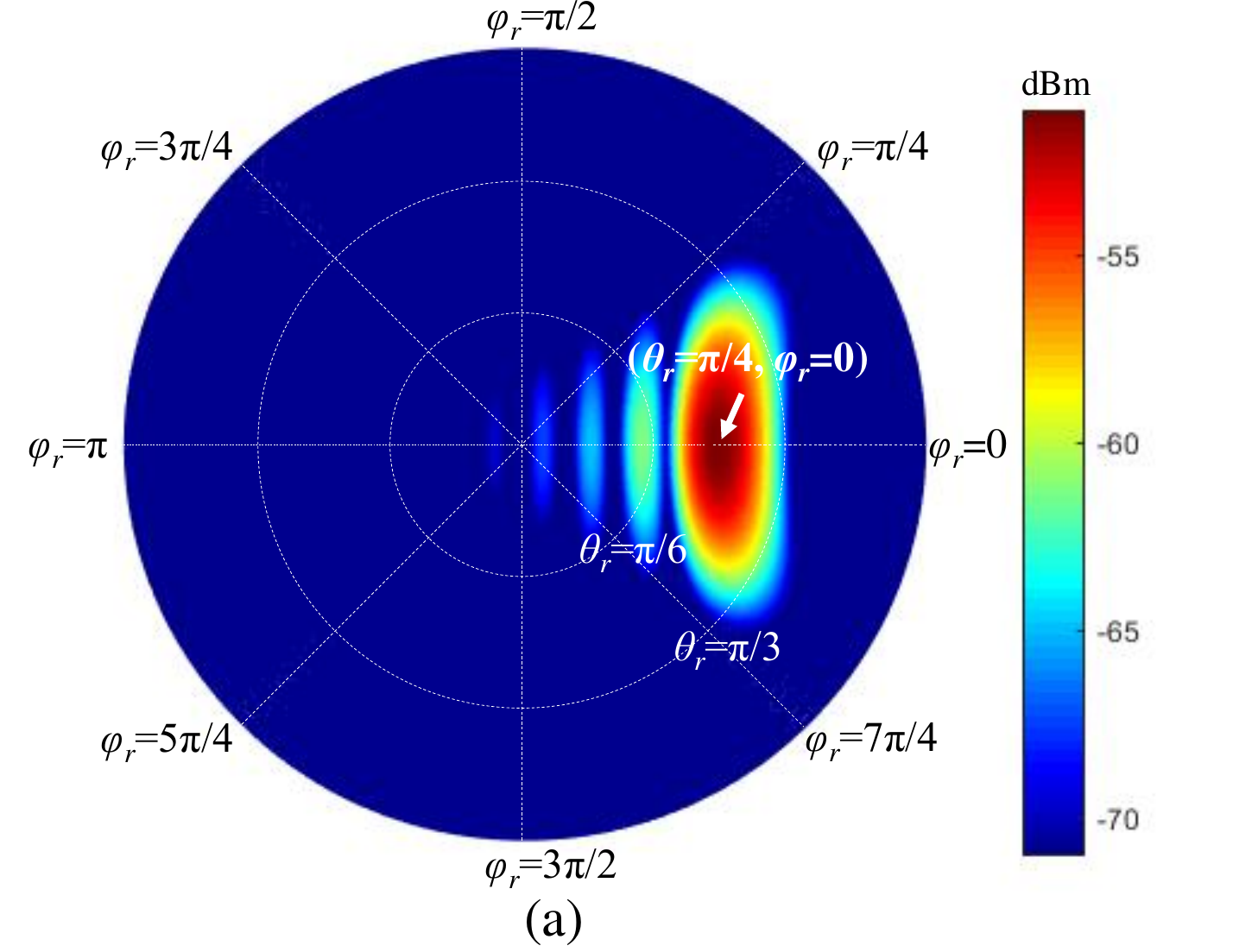}
    \hspace{1cm}
    \includegraphics[height=2.2in]{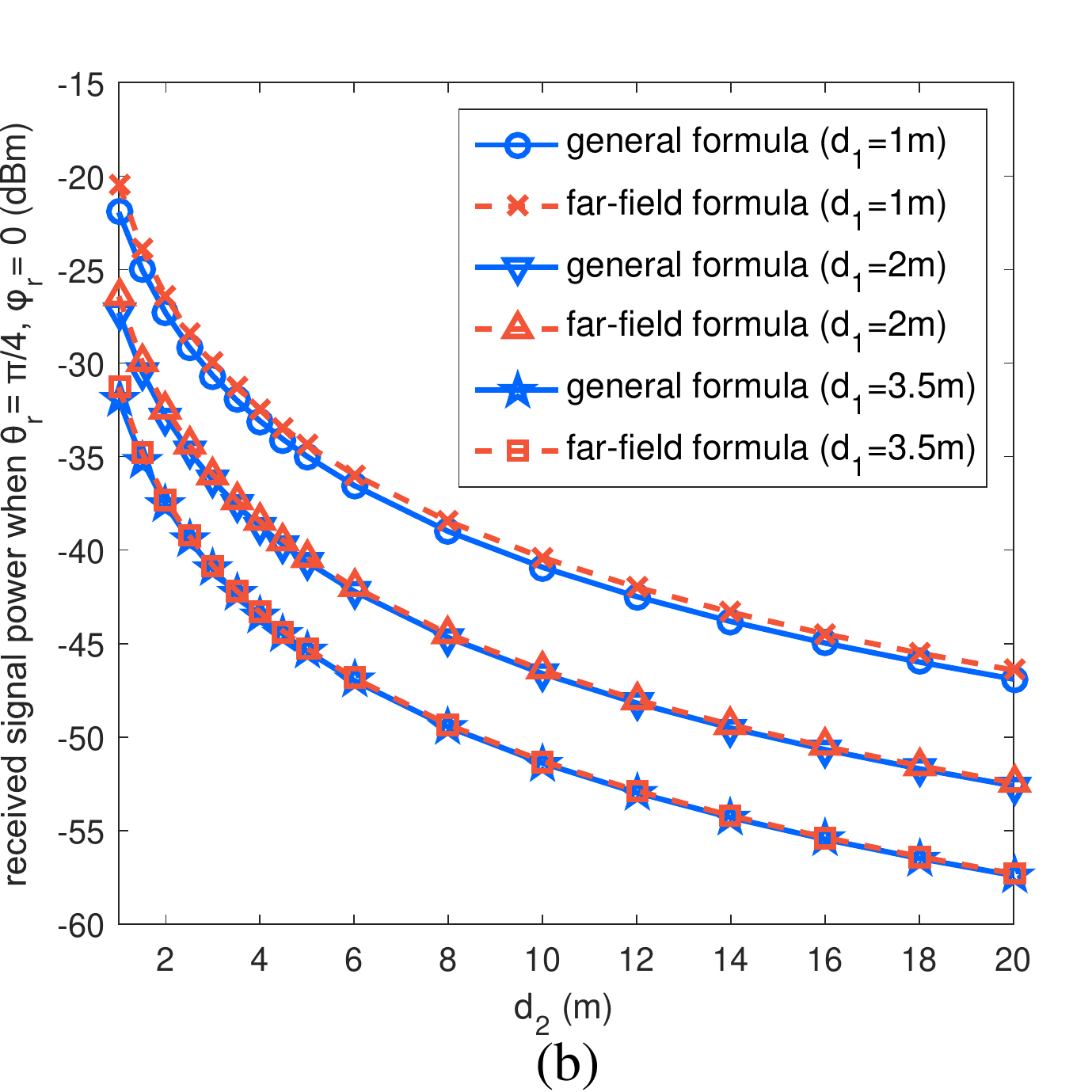}
    \vspace{-0.5cm}
	\caption{. Simulation results of RIS-assisted beamforming through specular reflection of small RIS in the far field case. (a) Received signal power distribution when $d_1$ = 3.5 m and $d_2$ = 10 m. (b) Received signal power in the specular direction versus $d_1$ and $d_2$.}
	\label{specularfarfield2}
\end{figure}

${P_t} = 0\ dBm$, ${\theta _{t}} = \frac{\pi }{4}$, ${\varphi _t} = \pi$, $\angle {\Gamma _{n,m}} = {\phi _{n,m}} = 0$. The large RIS1 and X-band antennas are employed in the simulation. When $d_1$ and $d_2$ are both greater than 71.4 m, the far field condition of RIS1 is fulfilled. The small RIS and C-band antennas are also employed in the simulation under the same conditions. When $d_1$ and $d_2$ are both larger than 1 m, the far field condition of the small RIS is fulfilled. Fig. \ref{specularfarfield}(a) and Fig. \ref{specularfarfield2}(a) show the simulated received signal power distribution in various directions in the large RIS1 case and the small RIS case, respectively. It can be seen that the received signal power is maximized when ${\theta _{r}} = \frac{\pi }{4}$ and ${\varphi _r} = 0$, which indicates that both of the large RIS1 and the small RIS specularly reflect the incident signal. Moreover, the larger the electrical size of the RIS is, the more concentrated the power of the reflected signal. As shown in Fig. \ref{specularfarfield}(b) and Fig. \ref{specularfarfield2}(b), the far-field formula (\ref{s15}) matches well with the general formula (\ref{s14}), which validates that the free-space path loss is compliant with (\ref{s23}) under the scenario of RIS-assisted beamforming through specular reflection of RIS in the far field case.
\vspace{-0.4cm}
\vspace{-1.5cm}
\vspace{-0.85cm}
\subsubsection{Intelligent Reflection in the Far Field Beamforming Case}\label{Specularreflection}
\begin{figure}[H]
	\centering
    \includegraphics[height=2.1in]{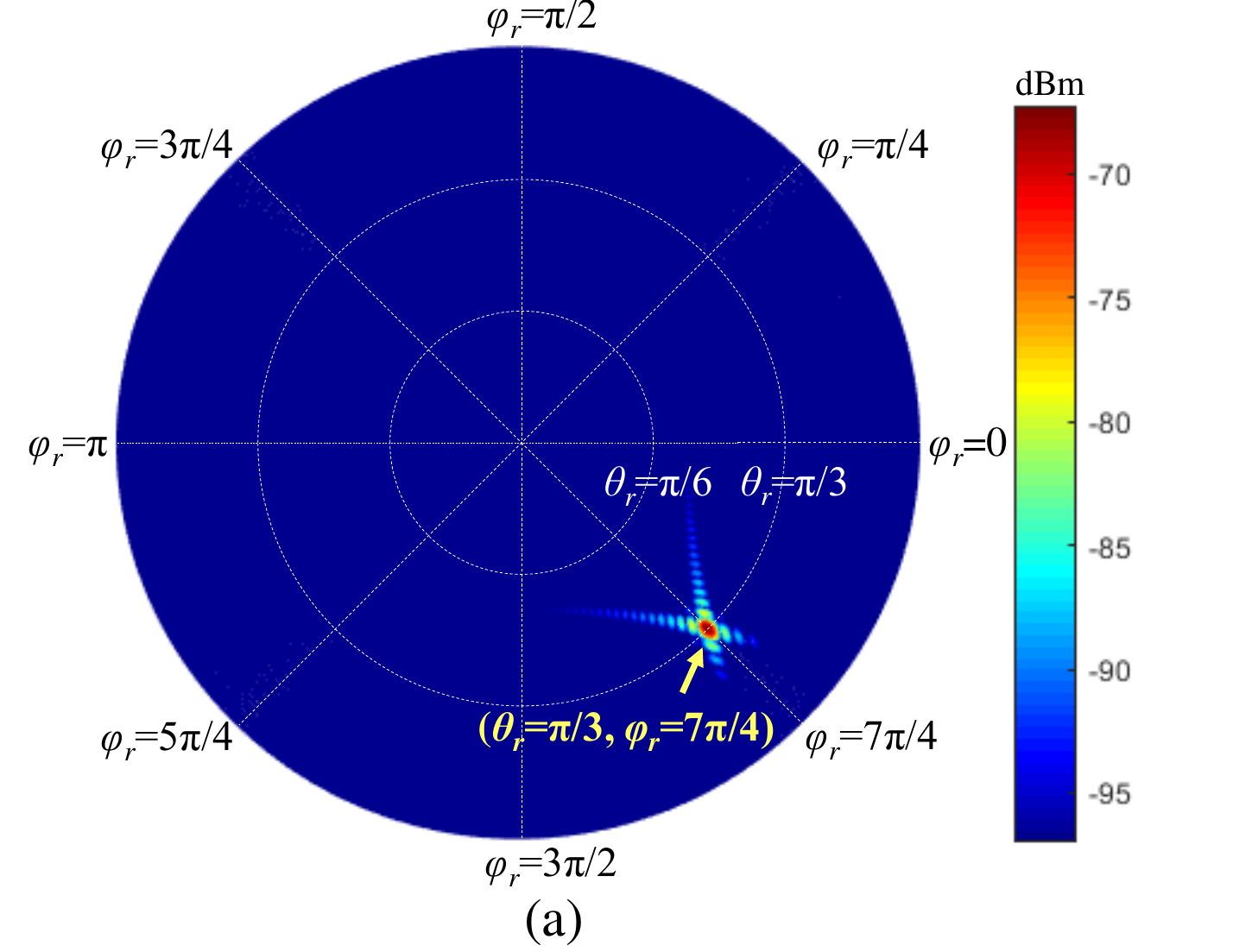}
    \hspace{1cm}
	\includegraphics[height=2.2in]{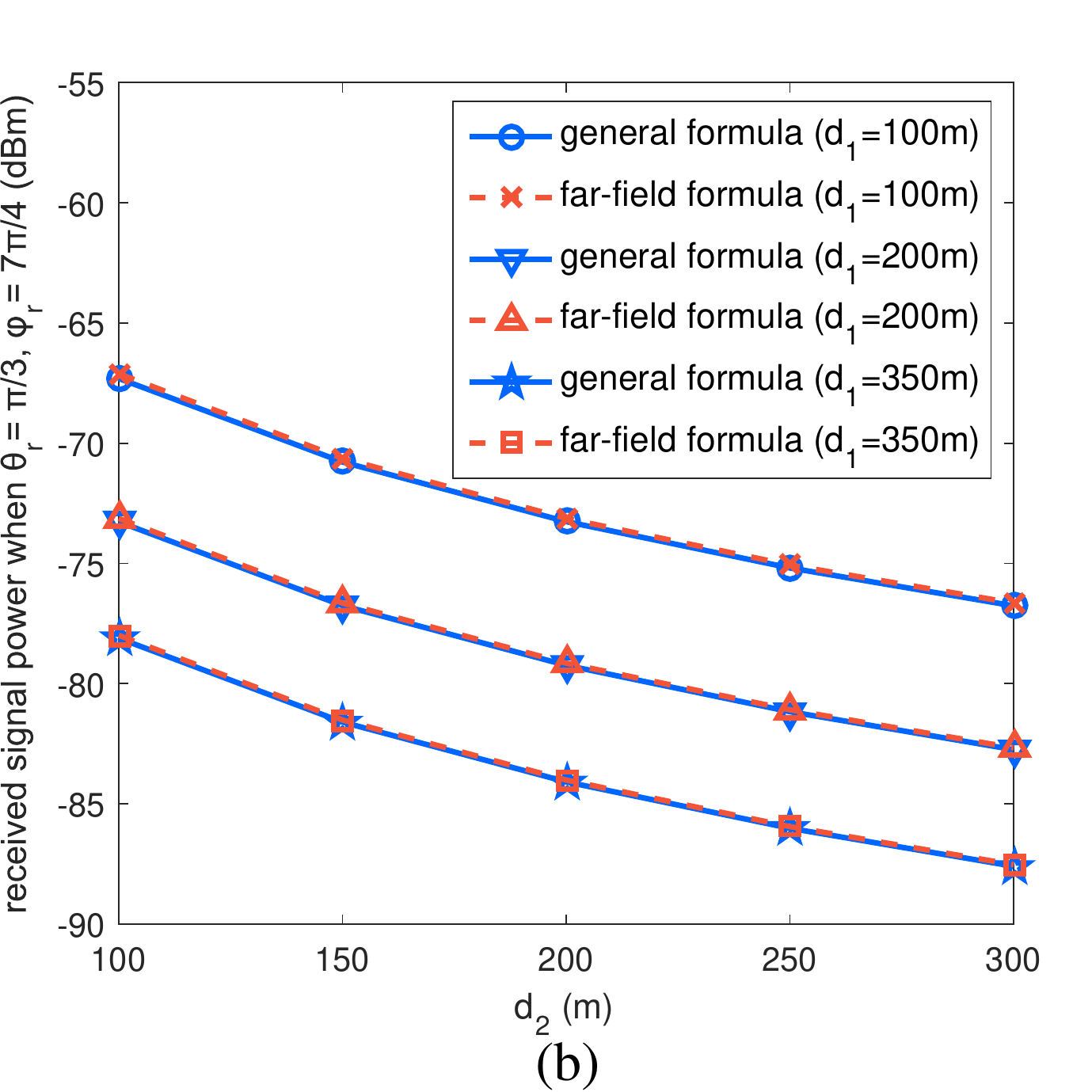}
    \vspace{-0.5cm}
	\caption{. Simulation results of RIS-assisted beamforming through intelligent reflection of large RIS1 in the far field case. (a) Received signal power distribution case when ${\theta _{des}}=\frac{\pi }{3}$, ${\varphi _{des}}=\frac{7\pi }{4}$, $d_1$ = 100 m, $d_2$ = 100 m. (b) Received signal power in the desired direction versus $d_1$ and $d_2$.}
	\label{abnormalfarfield}
\end{figure}

\begin{figure}[H]
	\centering
	\includegraphics[height=2.1in]{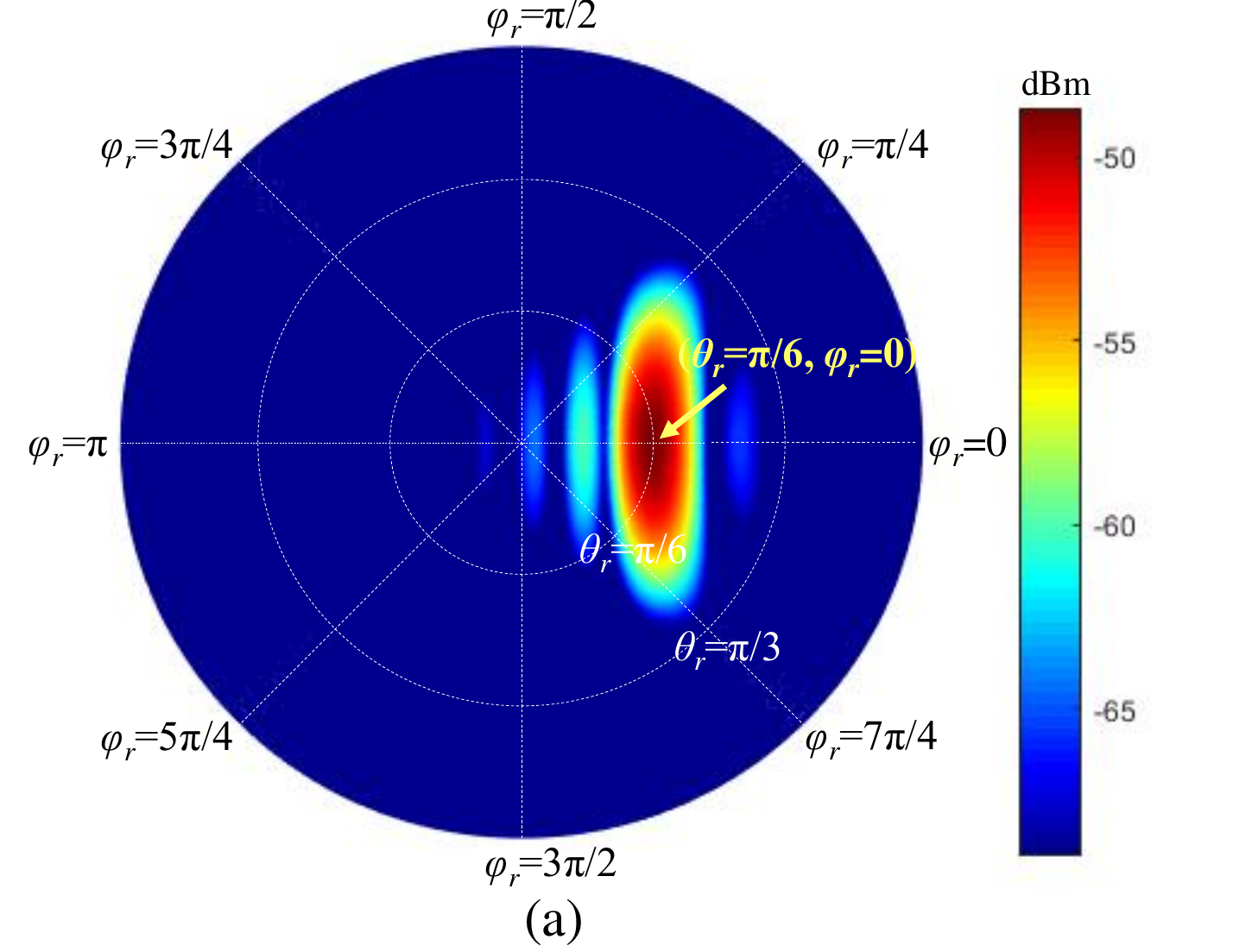}
    \hspace{1cm}
    \includegraphics[height=2.2in]{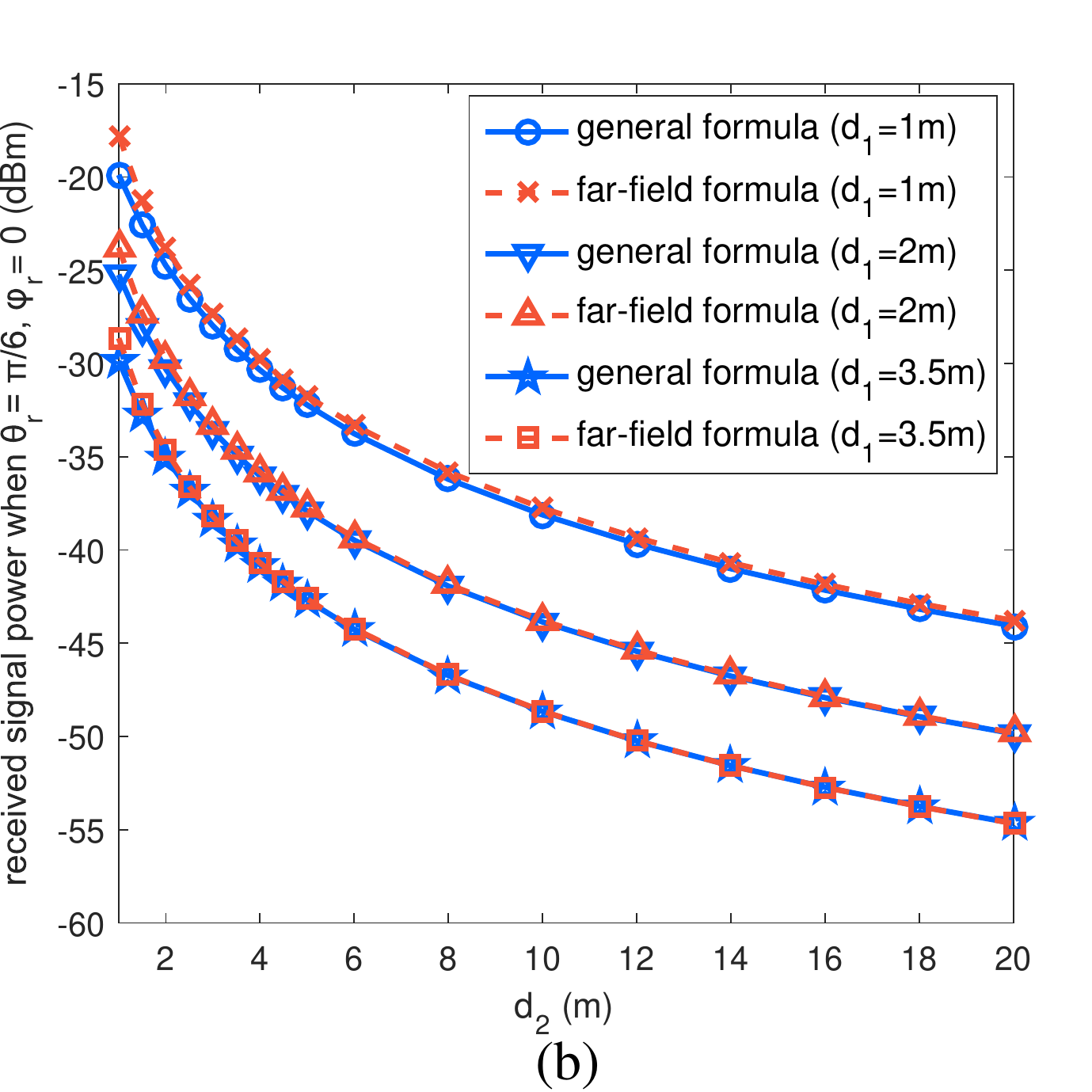}
    \vspace{-0.5cm}
	\caption{. Simulation results of RIS-assisted beamforming through intelligent reflection of small RIS in the far field case. (a) Received signal power distribution when ${\theta _{des}}=\frac{\pi }{6}$, ${\varphi _{des}}=0$, $d_1$ = 3.5 m, $d_2$ = 10 m. (b) Received signal power in the desired direction versus $d_1$ and $d_2$.}
	\label{abnormalfarfield2}
\end{figure}

${P_t} = 0\ dBm$, ${\theta _{t}} = \frac{\pi }{4}$, ${\varphi _t} = \pi$. The large RIS1 and X-band antennas are employed  in the simulation, and the phase shifts ${\phi _{n,m}}$ are designed to steer the reflected signal to a desired direction of ${\theta _{des}} = \frac{\pi }{3}$ and ${\varphi _{des}} = \frac{7\pi }{4}$ according to (\ref{s18}). The small RIS and C-band antennas are also employed in the simulation, and ${\phi _{n,m}}$ are designed to steer the reflected signal to a desired direction of ${\theta _{des}} = \frac{\pi }{6}$ and ${\varphi _{des}} = 0$. Fig. \ref{abnormalfarfield}(a) and Fig. \ref{abnormalfarfield2}(a) show that the received signal power is maximized when ${\theta _{r}} = \frac{\pi }{3}$, ${\varphi _r} = \frac{7\pi }{4}$ and ${\theta _{r}} = \frac{\pi }{6}$, ${\varphi _r} = 0$, respectively, which indicates that both of the large RIS1 and the small RIS successfully reflect the incident signal towards the desired direction. As shown in Fig. \ref{abnormalfarfield}(b) and Fig. \ref{abnormalfarfield2}(b), the far-field formula (\ref{s17}) with phase design matches well with the general formula (\ref{s14}), which validates that the free-space path loss adheres to (\ref{s23}) under the scenario of RIS-assisted beamforming through intelligent reflection of RIS in the far field case.
\vspace{-0.4cm}
\vspace{-1.5cm}
\vspace{-0.85cm}
\subsubsection{Intelligent Reflection in the Near Field Beamforming Case}\label{Specularreflection}
\begin{figure}[H]
	\centering
	\includegraphics[height=2.1in]{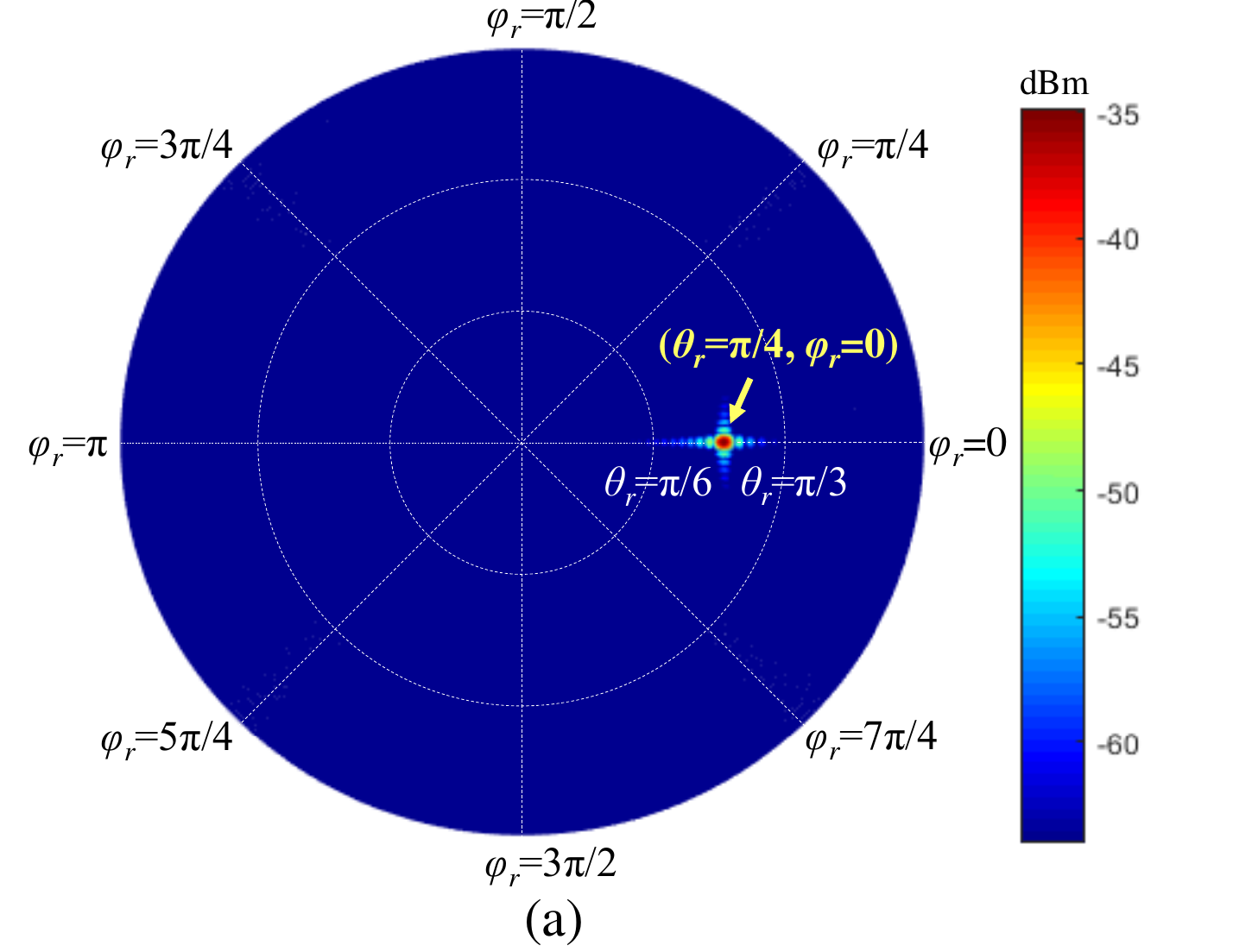}
    \hspace{1cm}
	\includegraphics[height=2.2in]{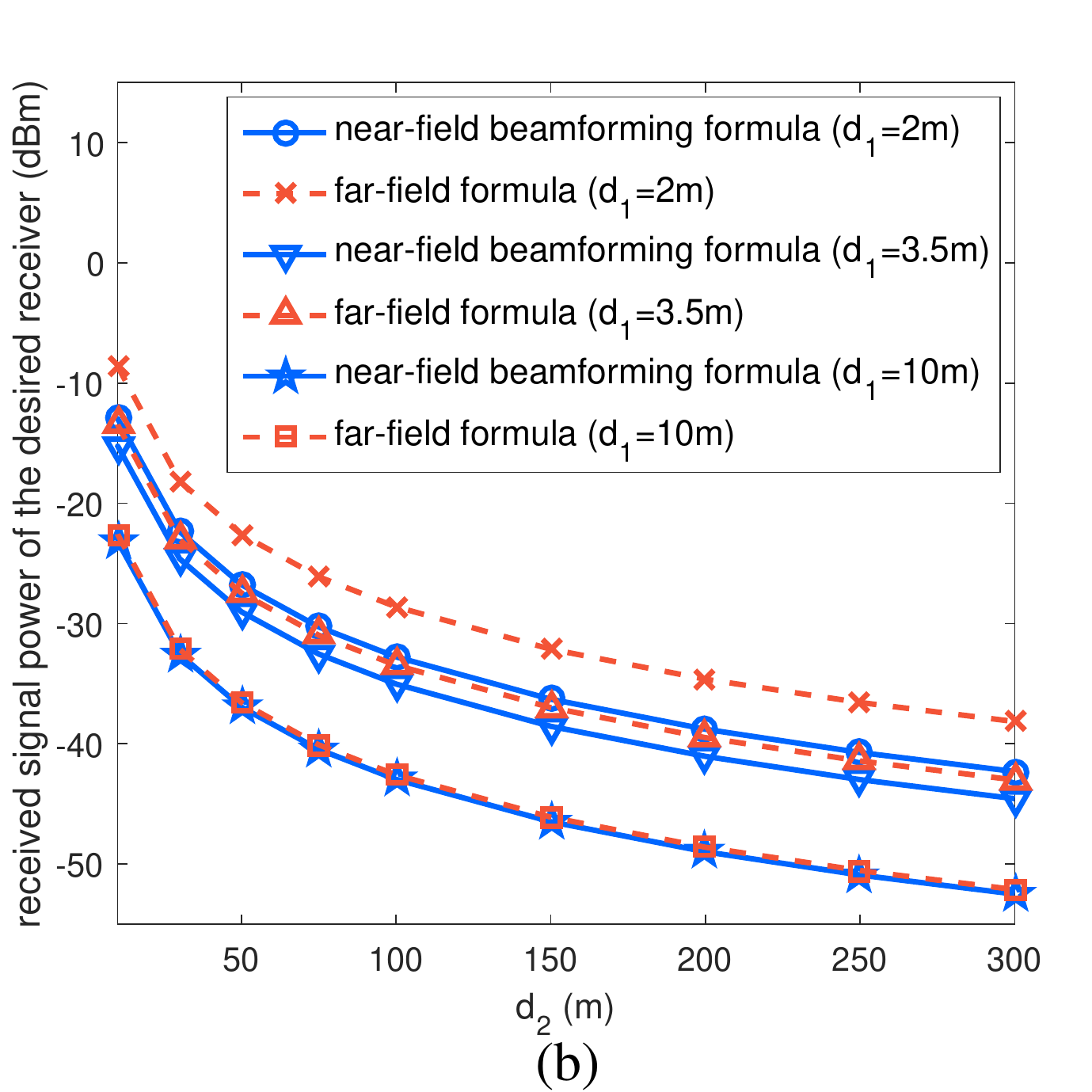}
    \vspace{-0.35cm}
	\caption{. Simulation results of RIS-assisted beamforming through intelligent reflection of large RIS1 in the near field case. (a) Received signal power distribution when $d_1$ = 3.5 m and the desired receiver is at $d_{2} = 100 m$, ${\theta _{r}} = \frac{\pi }{4}$ and ${\varphi _r} = 0$. (b) Received signal power of the desired receiver versus $d_1$ and $d_2$.}
	\label{focussixbignearfield}
\end{figure}

${P_t} = 0\ dBm$, ${\theta _{t}} = \frac{\pi }{4}$, ${\varphi _t} = \pi$. The large RIS1 and X-band antennas are employed  in the simulation, and ${\phi _{n,m}}$ are designed to focus the reflected signal towards the desired receiver at a certain location of $d_{2} = 100 m$, ${\theta _{r}} = \frac{\pi }{4}$ and ${\varphi _r} = 0$ according to (\ref{s22}). The transmitter is in the near field of RIS1. Fig. \ref{focussixbignearfield}(a) shows the power distribution of the simulated received signal when $d_1$ = 3.5 m and $d_2$ = 100 m. It can be observed that the reflected signal has been successfully focused towards the desired receiver ($d_{2} = 100 m$, ${\theta _{r}} = \frac{\pi }{4}$ and ${\varphi _r} = 0$) by applying (\ref{s22}).  Fig. \ref{focussixbignearfield}(b) shows that as $d_{1}$ increases, the near-field beamforming formula (\ref{s21}) fits better with the far-field formula (\ref{s15}), although the transmitter is in the near field of the RIS. This is because, when both $d_{1}$ and $d_{2}$ increase, the distances between the transmitter/receiver and each unit cell become more similar, and the RIS is gradually covered by the main lobe of the transmitting/receiving antenna. Therefore, (\ref{s21}) can be written as
\begin{equation}\label{s26}
P_r^{\max } \approx {P_t}\frac{{{G_t}{G_r}G{d_x}{d_y}{\lambda ^2}{A^2}}}{{64{\pi ^3}}} \frac{{{M^2}{N^2}F({\theta _{\text{t}}},{\varphi _t})F({\theta _r},{\varphi _r})}}{{{d_1}^2d_2^2}} \hfill,
\end{equation}
by applying ${F^{tx}}(\theta _{n,m}^{tx},\varphi _{n,m}^{tx}) \approx 1$, ${F^{rx}}(\theta _{n,m}^{rx},\varphi _{n,m}^{rx}) \approx 1$, $F(\theta _{n,m}^t,\varphi _{n,m}^t) \approx F({\theta _{\text{t}}},{\varphi _t})$, $F(\theta _{n,m}^r,\varphi _{n,m}^r) \approx F({\theta _r},{\varphi _r})$, $r_{n,m}^t \approx {d_1}$, and $r_{n,m}^r \approx {d_2}$ into (\ref{s21}). Formula (\ref{s26}) is exactly the same as (\ref{s16}), which reveals that the free-space path loss gradually approaches (\ref{s23}) under the RIS-assisted beamforming scenario in the near field case when both $d_{1}$ and $d_{2}$ increase. The above conclusion also holds for the small RIS as it is not related to the size of RIS.
\vspace{-0.5cm}
\vspace{-1.5cm}
\subsection{RIS-assisted Broadcasting}\label{Coveragespecificarea}
\vspace{-0.2cm}
\subsubsection{Specular Reflection in the Near Field Broadcasting Case}\label{Specularreflection}
\begin{figure}[H]
	\centering
	\includegraphics[height=2.2in]{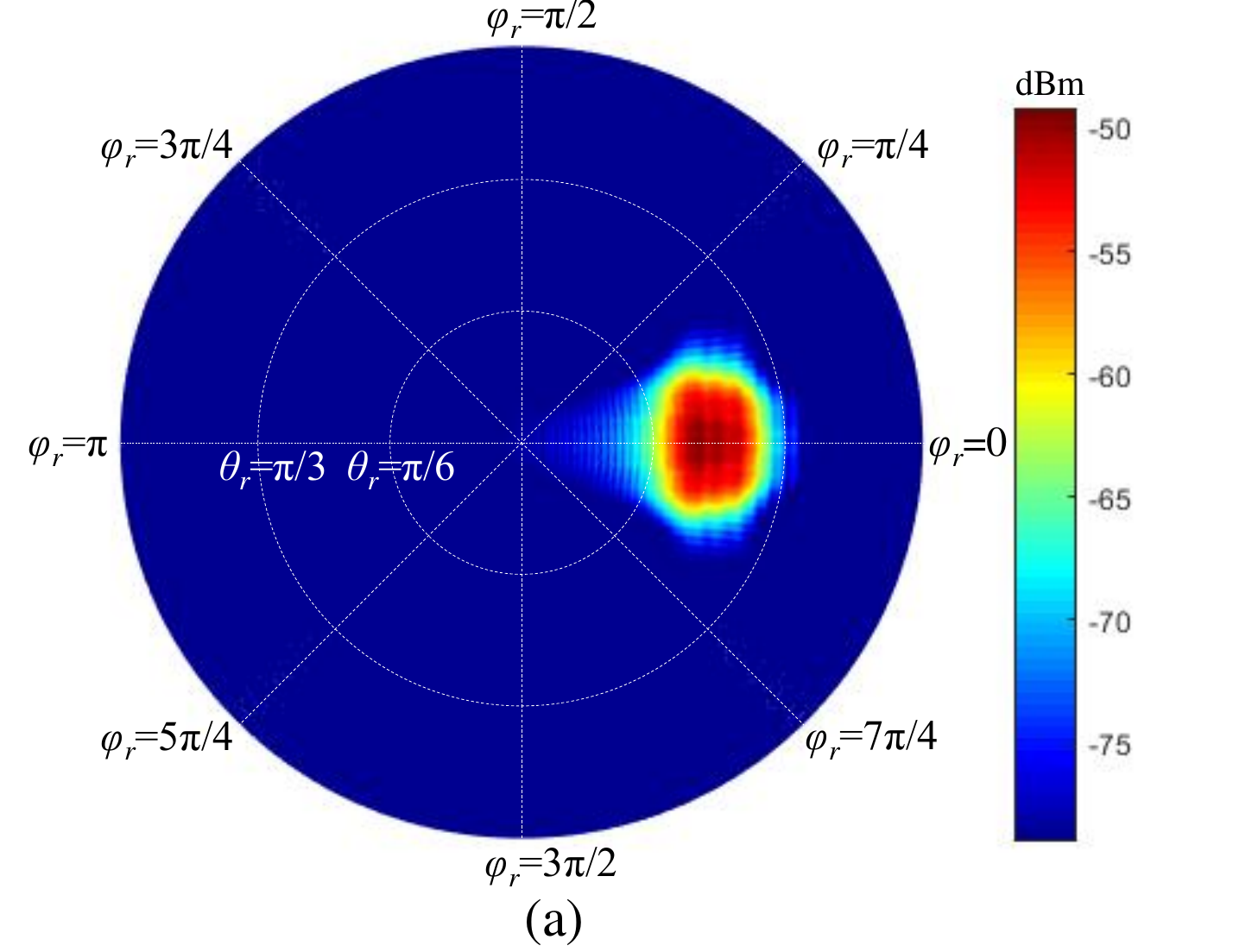}
    \hspace{1cm}
	\includegraphics[height=2.2in]{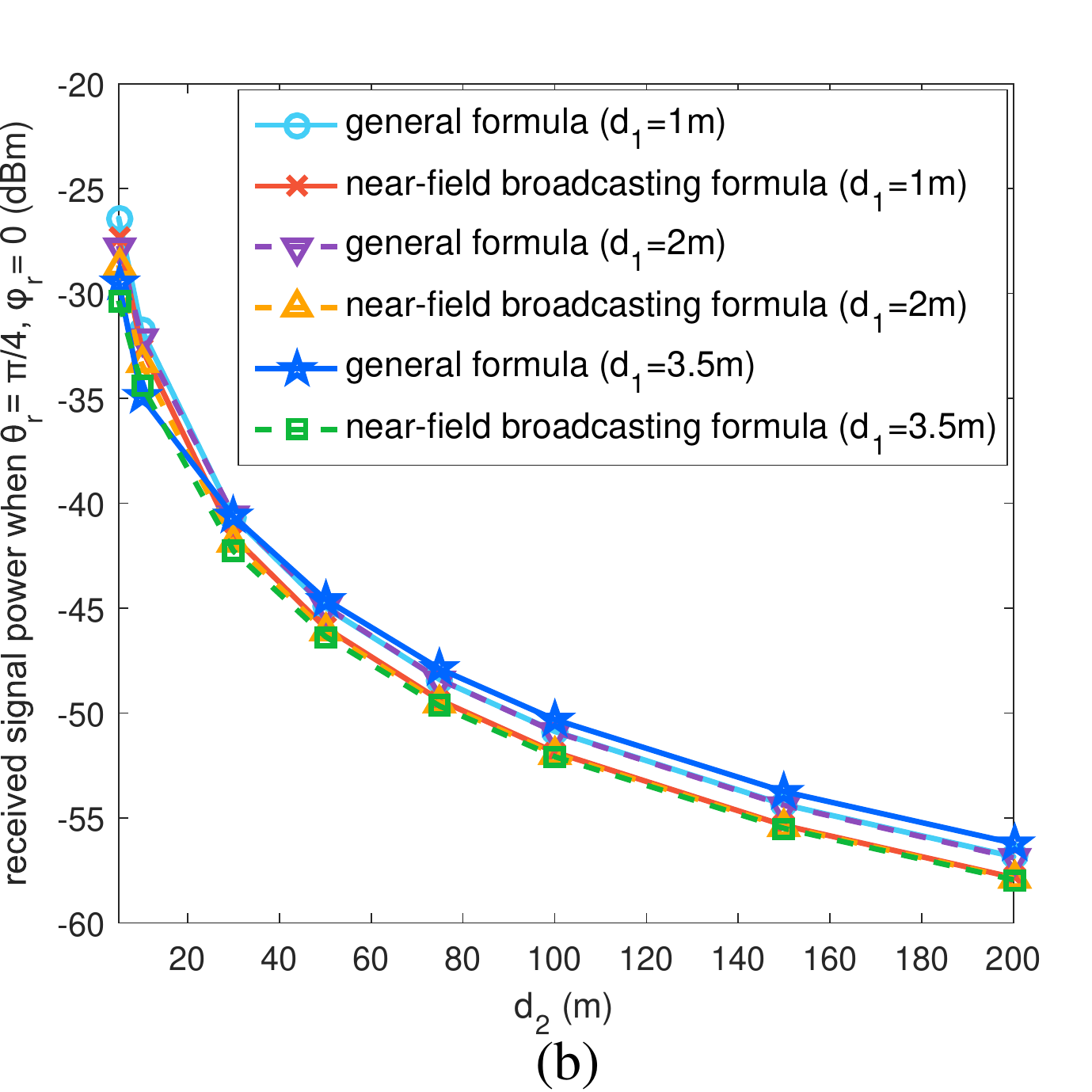}
	\includegraphics[height=2.2in]{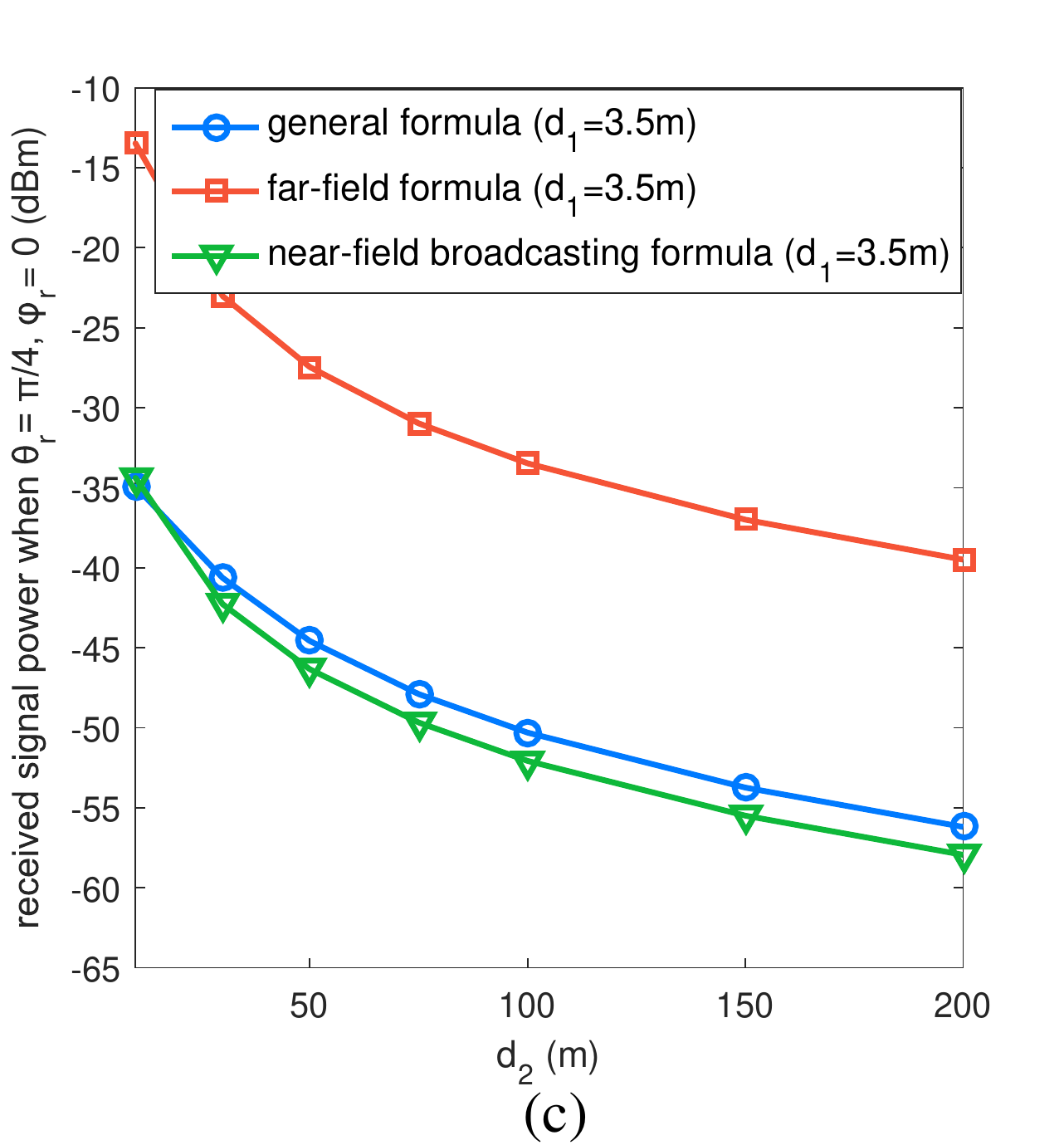}
    \includegraphics[height=2.2in]{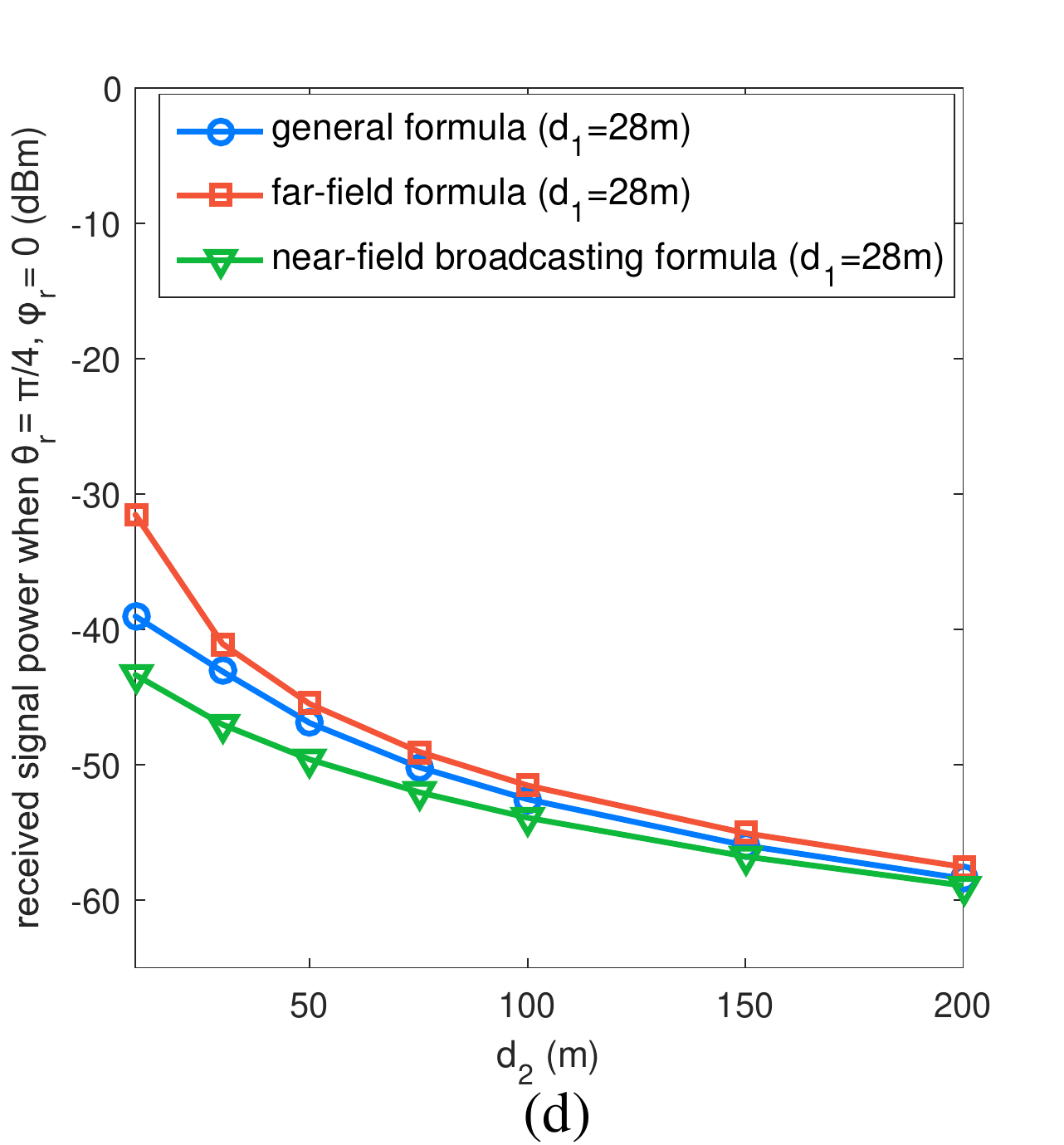}
    \includegraphics[height=2.2in]{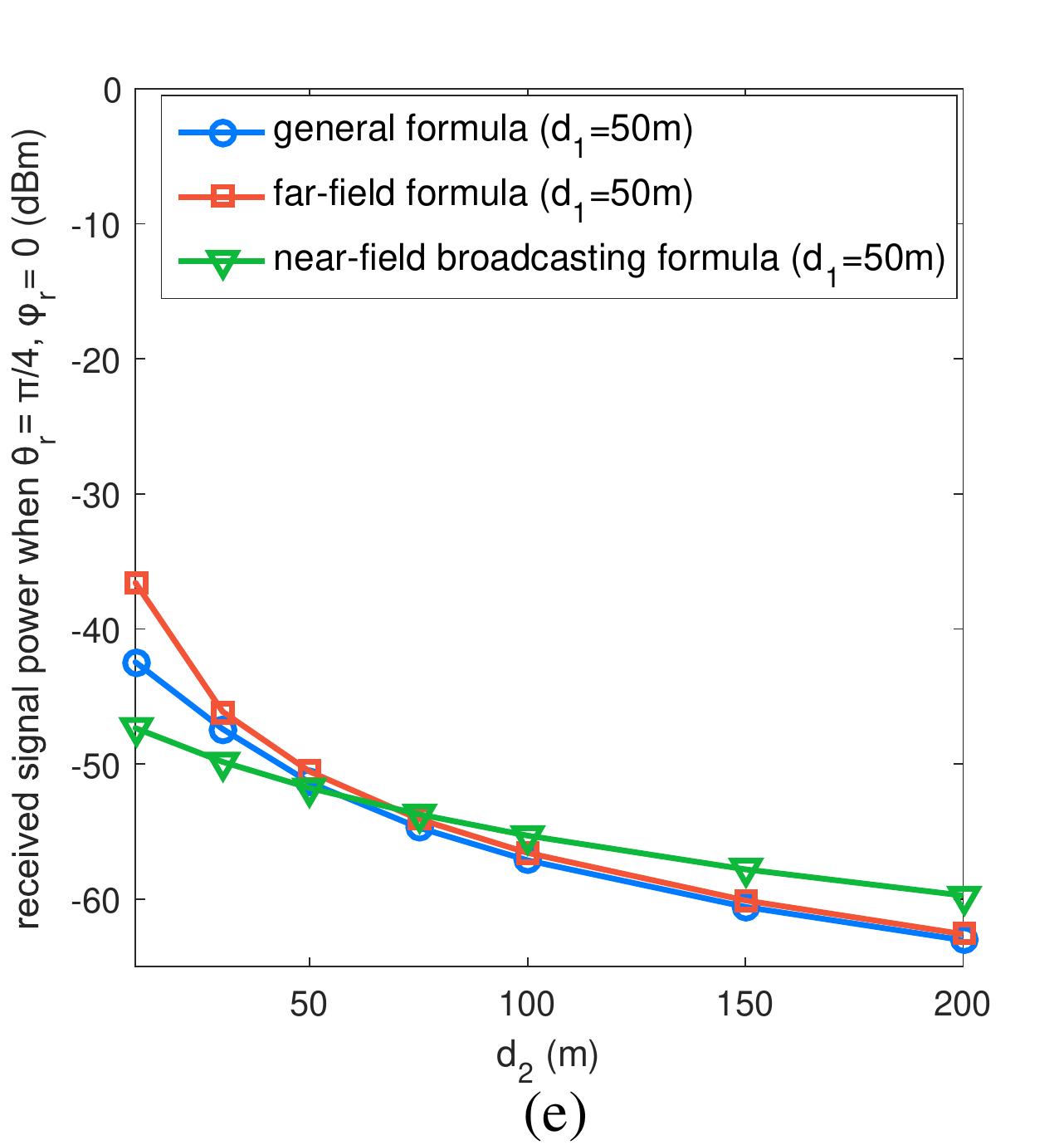}
    \vspace{-0.4cm}
	\caption{. Simulation results of RIS-assisted broadcasting through specular reflection of large RIS1 in the near field case. (a) Received signal power distribution when $d_1$ = 2 m and $d_{2} = 100 m$. (b) Received signal power of the broadcasted receiver versus $d_1$ and $d_2$ when $d_1$ = 1 m, 2 m, and 3.5 m. (c)-(e) Received signal power of the broadcasted receiver versus $d_1$ and $d_2$ when $d_1$ = 3.5 m, 28 m, and 50 m, respectively.}
	\label{sixbignearfield}
\end{figure}

${P_t} = 0\ dBm$, ${\theta _{t}} = \frac{\pi }{4}$, ${\varphi _t} = \pi$, $\angle {\Gamma _{n,m}} = {\phi _{n,m}} = 0$. The large RIS1 and X-band antennas are employed in the simulation. When $d_1$ or $d_2$ is less than 71.4 m, the near field condition of the large RIS1 is fulfilled. Fig. \ref{sixbignearfield}(a) shows the simulated received signal power distribution in various directions when $d_1$ = 2 m and $d_2$ = 100 m. It can be seen that a specific area which is defined as the solid angle $\Omega \cap {\Omega _t}$ in (\ref{s19}) is lit up. As shown in Fig. \ref{sixbignearfield}(b), the near-field broadcasting formula (\ref{s19}) fits well with the general formula (\ref{s14}) when $d_1$ = 1 m, 2 m, and 3.5 m, which reveals that the free-space path loss follows (\ref{s25}) under the scenario of RIS-assisted broadcasting through specular reflection of a large RIS in the near field case. In addition, the general formula curve gradually moves from the near-field broadcasting formula curve to the far-field formula curve as $d_1$ increases as shown in Fig. \ref{sixbignearfield}(c)-(e), which indicates a transition process between the near field and the far field of the RIS. It is worth noting that the curves of the far-field formula and the near-field broadcasting formula almost overlap when $d_1$ = 28 m. Therefore, we redefine the boundary of the far field and the near field of the RIS as $L_{bound}$, by letting (\ref{s23}) equals to (\ref{s25}) as
\begin{equation}\label{s27}
\frac{{{G_t}{G_r}G{M^2}{N^2}{d_x}{d_y}{\lambda ^2}F({\theta _t},{\varphi _t})F({\theta _r},{\varphi _r}){A^2}}}{{64{\pi ^3}L_{_{bound}}^2d_2^2}} = \frac{{{G_t}{G_r}{\lambda ^2}{A^2}}}{{16{\pi ^2}{{({L_{bound}} + {d_2})}^2}}}.
\end{equation}
When $d_{2}$ is much larger than ${L_{bound}}$ (the receiver is in the far field of RIS), (\ref{s27}) is solved as
\begin{equation}\label{s28}
{L_{bound}} \approx MN\sqrt {\frac{{G{d_x}{d_y}F({\theta _t},{\varphi _t})F({\theta _r},{\varphi _r})}}{4\pi}}.
\end{equation}
For the large RIS1 simulation, ${L_{bound}}$ equals to 28.77 m, which is in good agreement with Fig. \ref{sixbignearfield}(d). In the rest of the paper, (\ref{s28}) is referred to as the \textbf{\emph{boundary of the far field and the near field for RIS-assisted wireless communications}}.
\vspace{-0.5cm}
\vspace{-0.8cm}
\subsubsection{Intelligent Reflection in the Near Field Broadcasting Case}\label{Specularreflection}
\begin{figure}[H]
	\centering
    \includegraphics[height=2.25in]{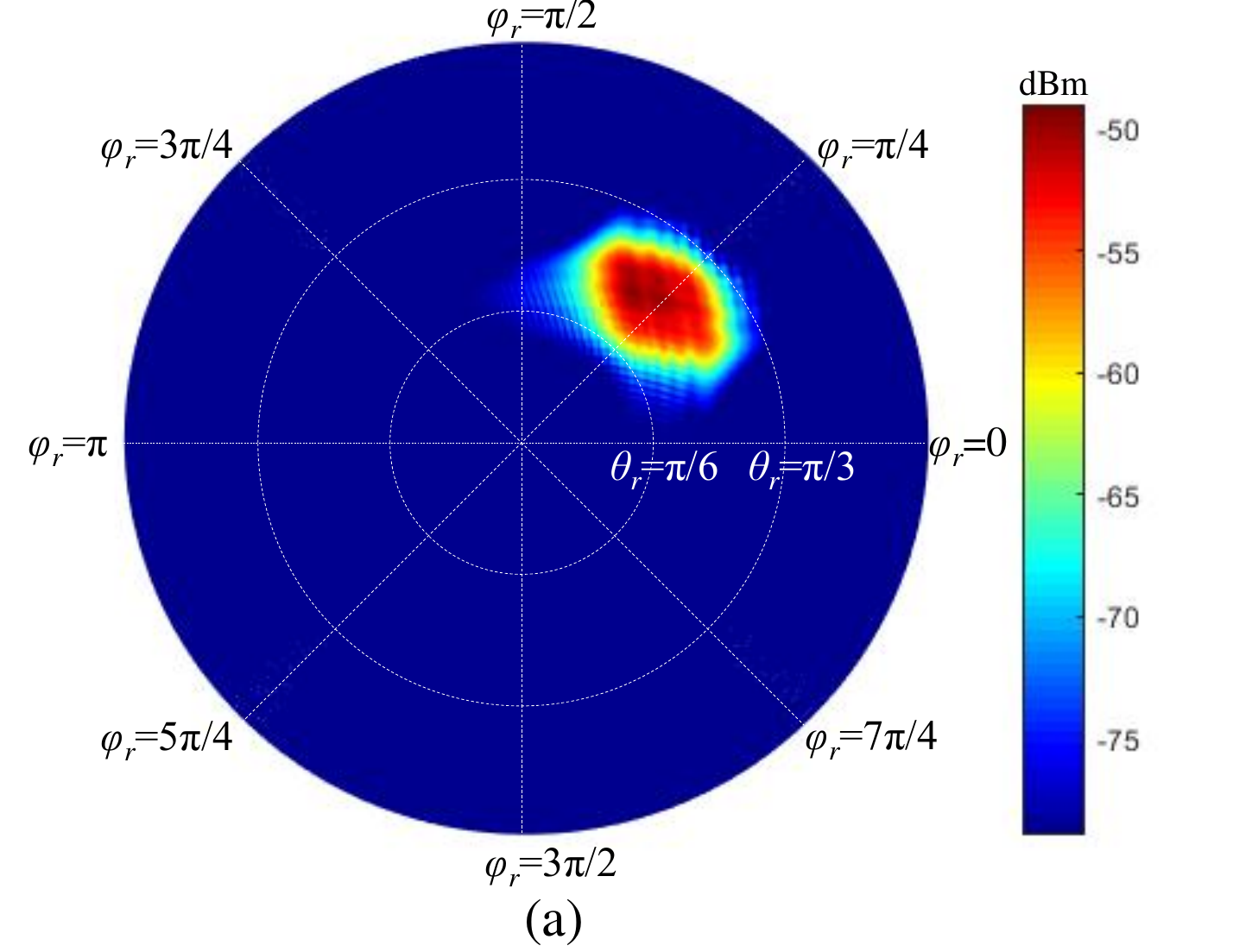}
    \hspace{1cm}
    \includegraphics[height=2.25in]{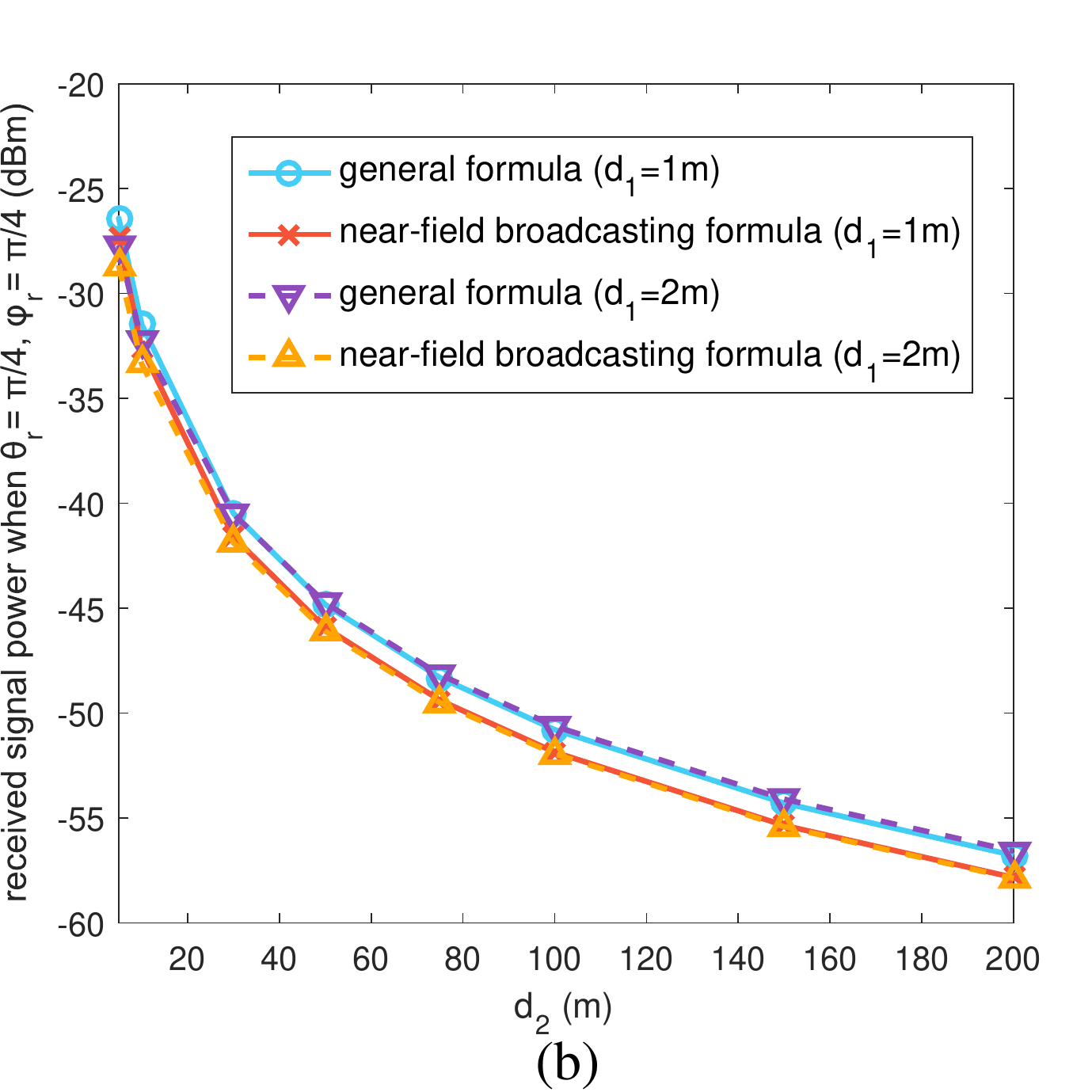}
    \includegraphics[height=2.25in]{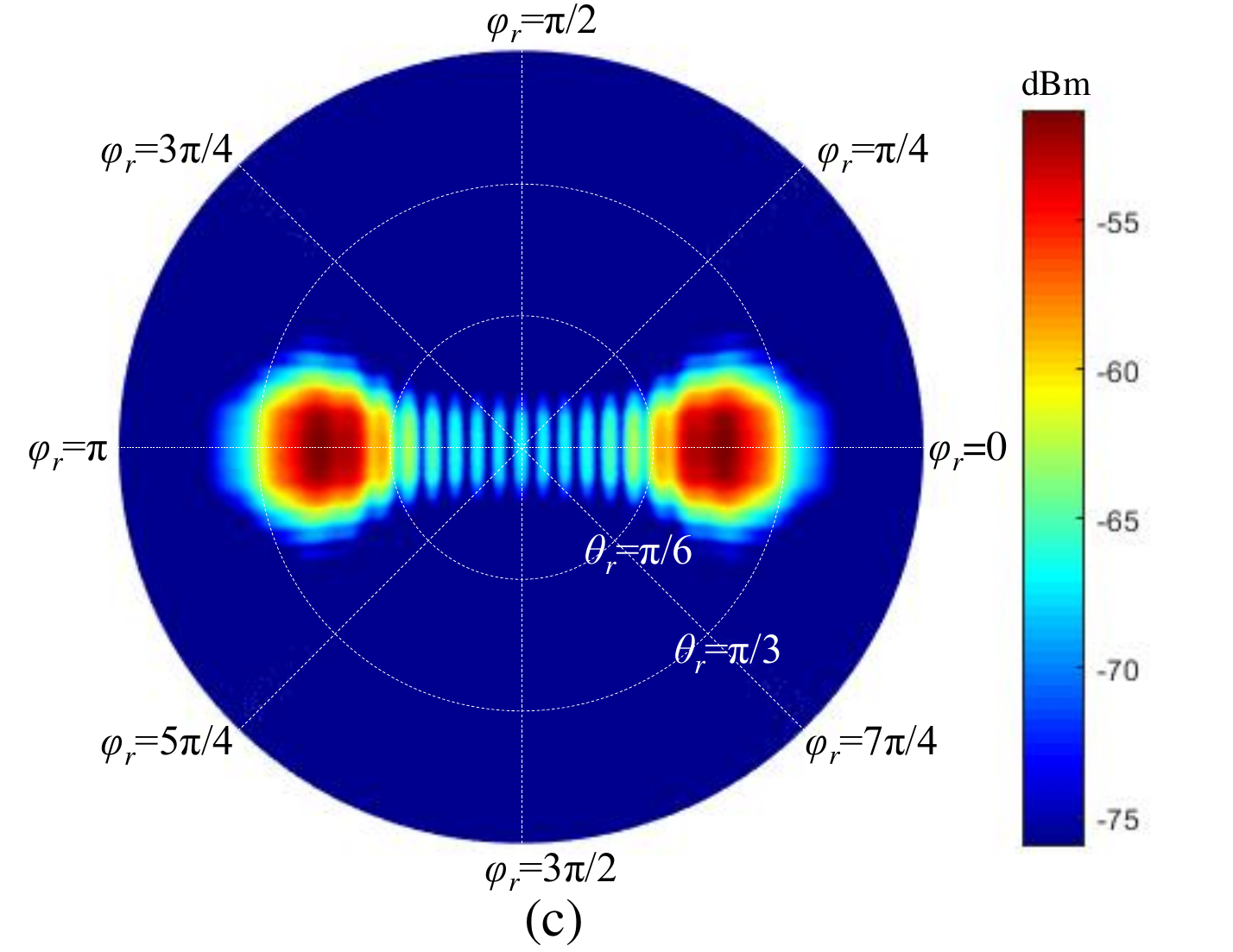}
    \hspace{1cm}
    \includegraphics[height=2.25in]{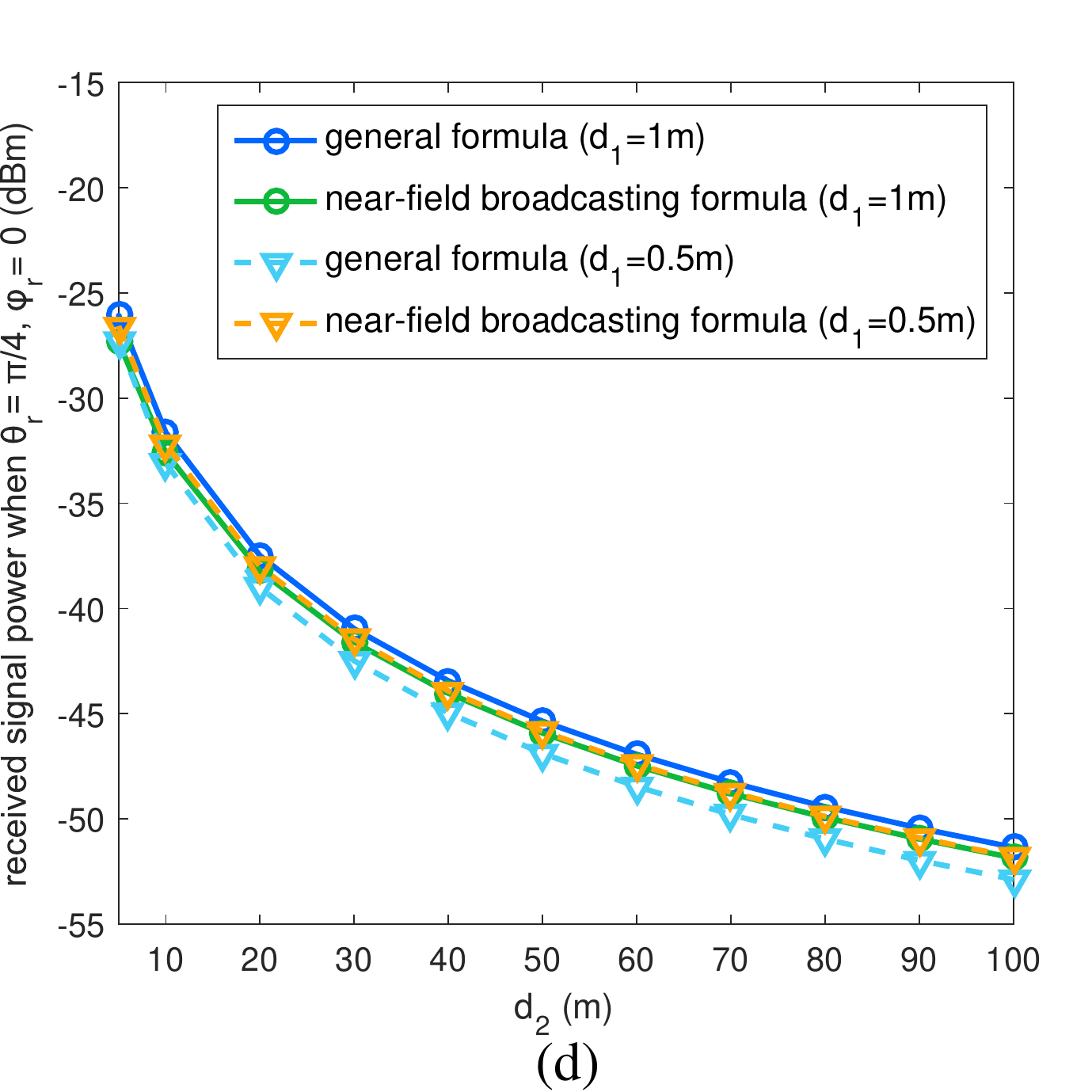}
    \vspace{-0.4cm}
	\caption{. Simulation results of RIS-assisted near-field broadcasting through intelligent reflection. (a) Received signal power distribution when $d_1$ = 2 m and $d_{2} = 100 m$ with large RIS1. (b) Received signal power versus $d_1$ and $d_2$ when $d_1$ = 1 m and 2 m with large RIS1. (c) Received signal power distribution when $d_1$ = 1 m and $d_{2} = 100 m$  with large RIS2. (d) Received signal power versus $d_1$ and $d_2$ when $d_1$ = 0.5 m and 1 m  with large RIS2.}
	\label{abnormalonebignearfield}
\end{figure}

${P_t} = 0\ dBm$, ${\theta _{t}} = \frac{\pi }{4}$, ${\varphi _t} = \pi$. The large RIS1 and X-band antennas are employed in the simulation. ${\phi _{n,m}}$ are designed according to (\ref{sr1}) for desired broadcasting direction of ${\theta _{des}} = \frac{\pi }{4}$ and ${\varphi _{des}} = \frac{\pi }{4}$ . Fig. \ref{abnormalonebignearfield}(a) shows the received signal power distribution in various directions when $d_1$ = 2 m and $d_2$ = 100 m. The large RIS1 successfully broadcasts the reflected signal into the desired direction. Fig. \ref{abnormalonebignearfield}(b) illustrates that the near-field broadcasting formula (\ref{s19}) fits well with the general formula (\ref{s14}), which validates the free-space path loss in (\ref{s25}) for RIS-assisted near-field intelligent broadcasting scenario.
\vspace{-0.5cm}
\vspace{-0.9cm}
The large RIS2 is employed in the simulation when ${\theta _{t}} = 0$. ${\phi _{n,m}}$ are designed to steer the reflected signal towards two desired directions: ${\theta _{des}} = \frac{\pi }{4}, {\varphi _{des}} = 0$ and ${\theta _{des}} = \frac{\pi }{4}, {\varphi _{des}} = \pi$. ${\phi _{n,m}} = 0$ when $mod(m,4)= 0\ or\ 1$ and ${\phi _{n,m}} = \pi$ when $mod(m,4)= 2\ or\ 3$ \cite{MetaInfo}. Fig. \ref{abnormalonebignearfield}(c) shows the received signal power distribution when $d_1$ = 1 m and $d_2$ = 100 m. It can be observed that the large RIS2 successfully shapes the reflected signal towards two desired directions at the same time. As shown in Fig. \ref{abnormalonebignearfield}(d), the near-field broadcasting formula (\ref{s19}) fits well with the general formula (\ref{s14}), which reveals the free-space path loss also follows (\ref{s25}) under this scenario.
\vspace{-1.3cm}
\subsection{Path Loss Simulation Summary}\label{SimulationSummary}
\vspace{-0.2cm}
The numerical simulations have verified the free-space path loss models of RIS-assisted wireless communications summarized in Table \ref{pathlosssummary} via the general formula (\ref{s14}). In addition, the boundary of the far field and the near field of the RIS is redefined in (\ref{s28}), which can be used to determine whether the transmitter/receiver is in the near or far field of the RIS.
\vspace{-0.4cm}
\section{Validation of Path Loss Models via Experimental Measurements}\label{Measurement}
\vspace{-0.15cm}
Experimental measurements are carried out to further validate the proposed free-space path loss models for RIS-assisted systems. We set up a path loss measurement system in a microwave anechoic chamber and utilize three different metasurfaces to act as RISs in different scenarios.
\vspace{-0.7cm}
\subsection{Measurement Setup}\label{MeasurementSetup}
\begin{figure}[]
	\centering
	\includegraphics[height=2.15in]{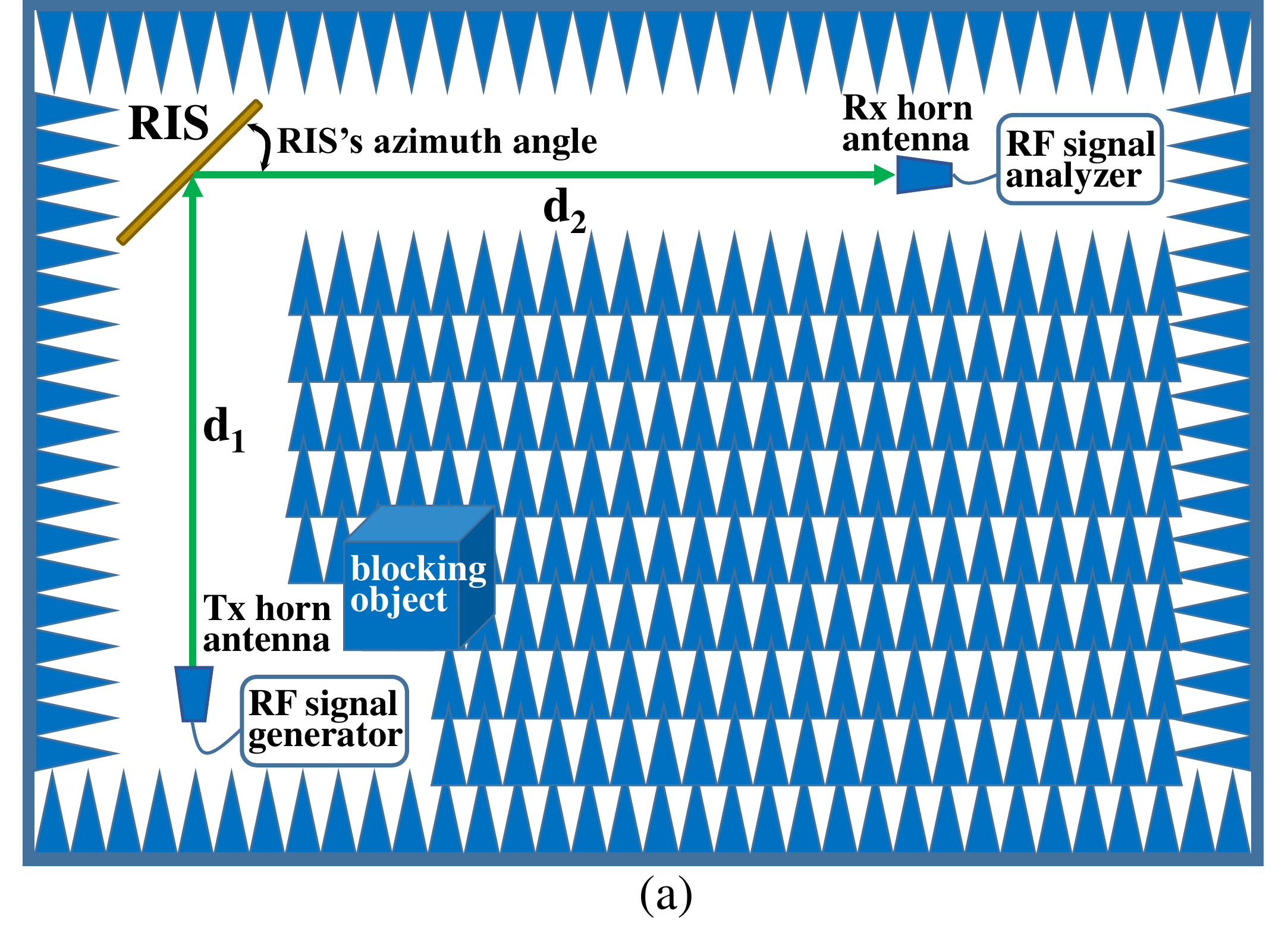}
    \hspace{0.4cm}
	\includegraphics[height=2.15in]{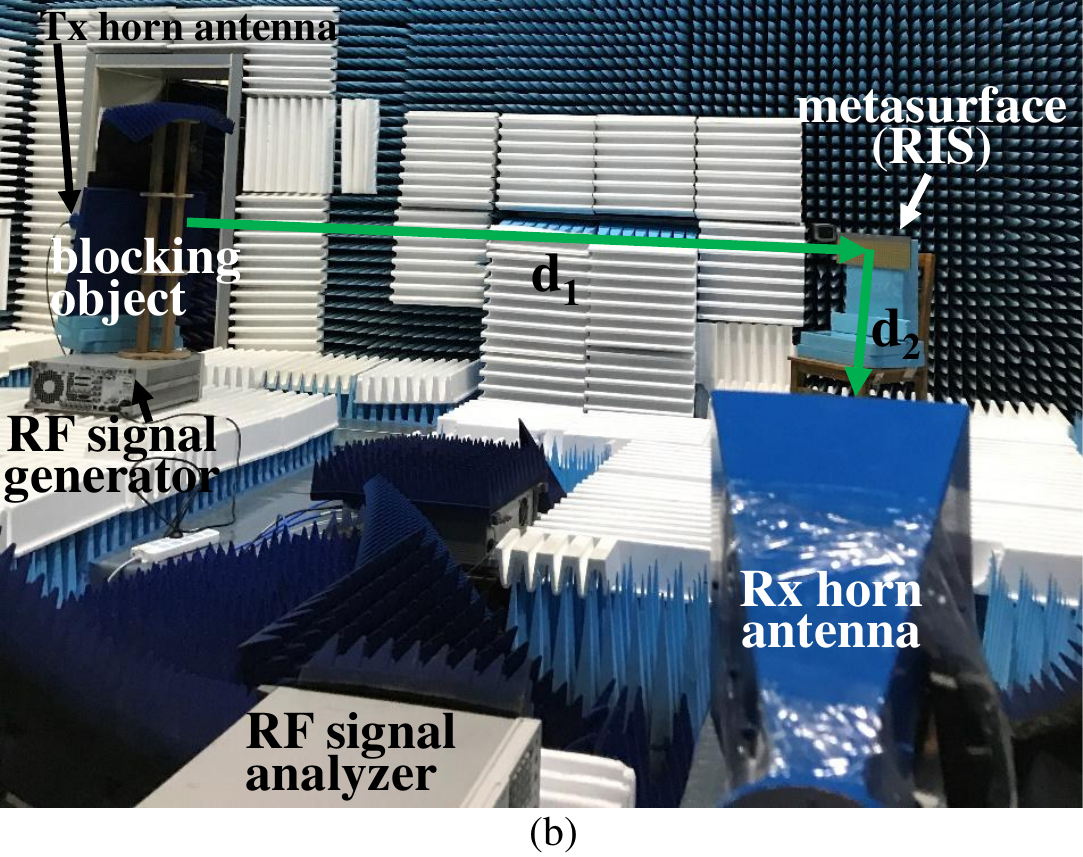}
    \vspace{-0.4cm}
	\caption{. Free-space path loss measurement system of RIS-assisted wireless communications: (a) diagram. (b) photo.}
    \vspace{-1.15cm}
	\label{measurementsetup}
\end{figure}

\begin{figure}[H]
	\centering
	\includegraphics[height=2.7in]{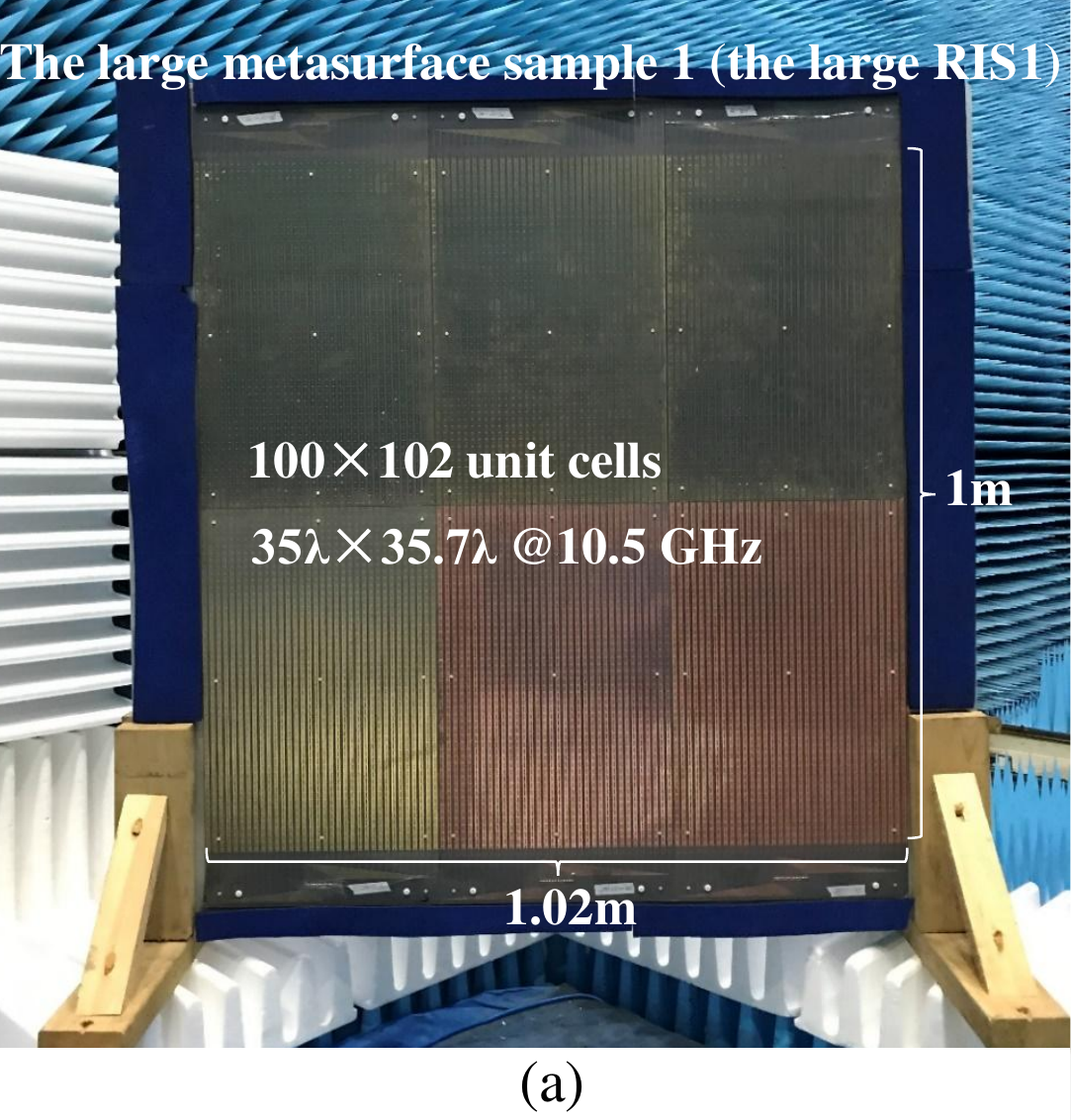}
    \hspace{0.1cm}
    \vspace{0.1cm}
	\includegraphics[height=2.7in]{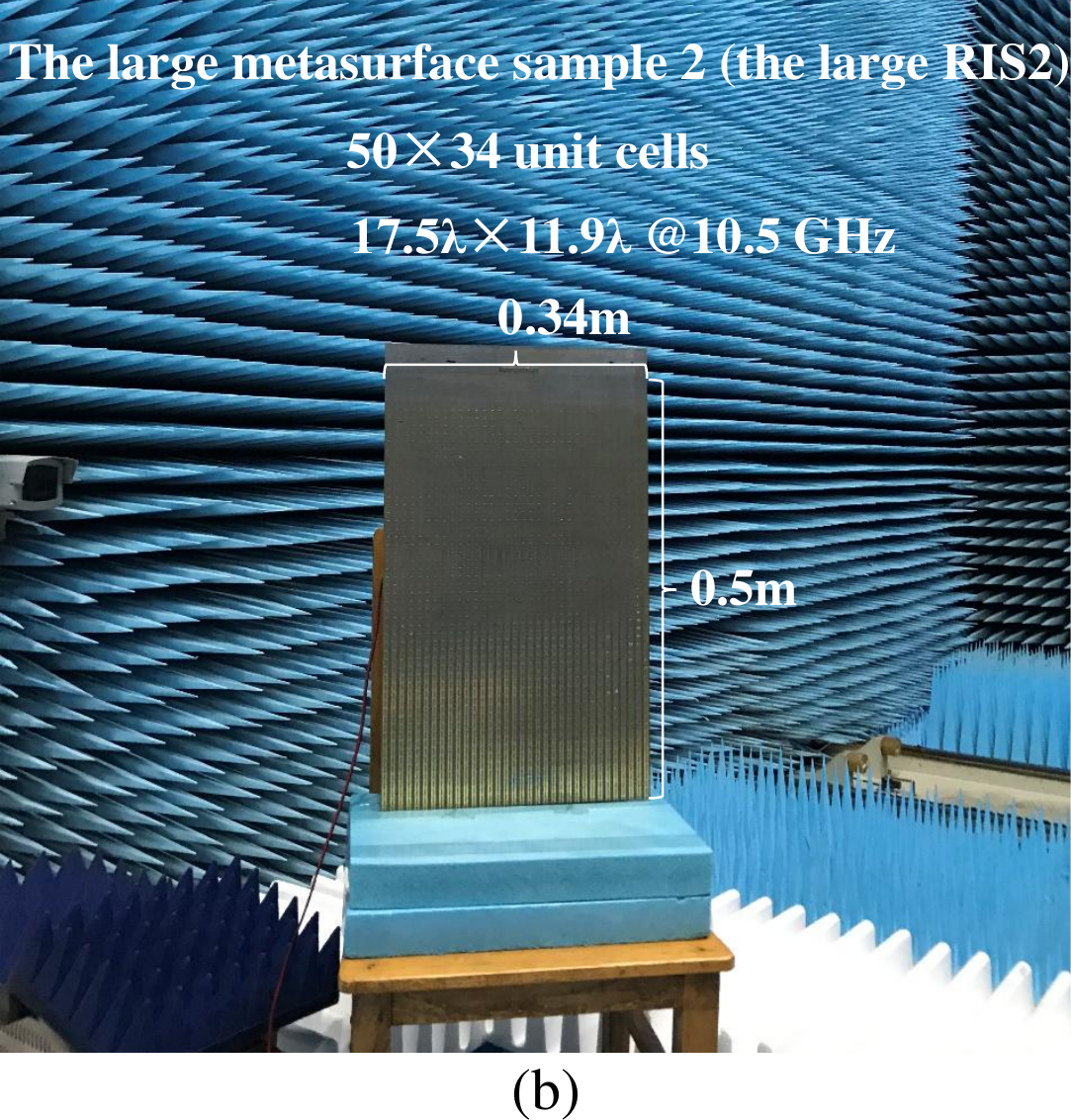}
    \includegraphics[height=2in]{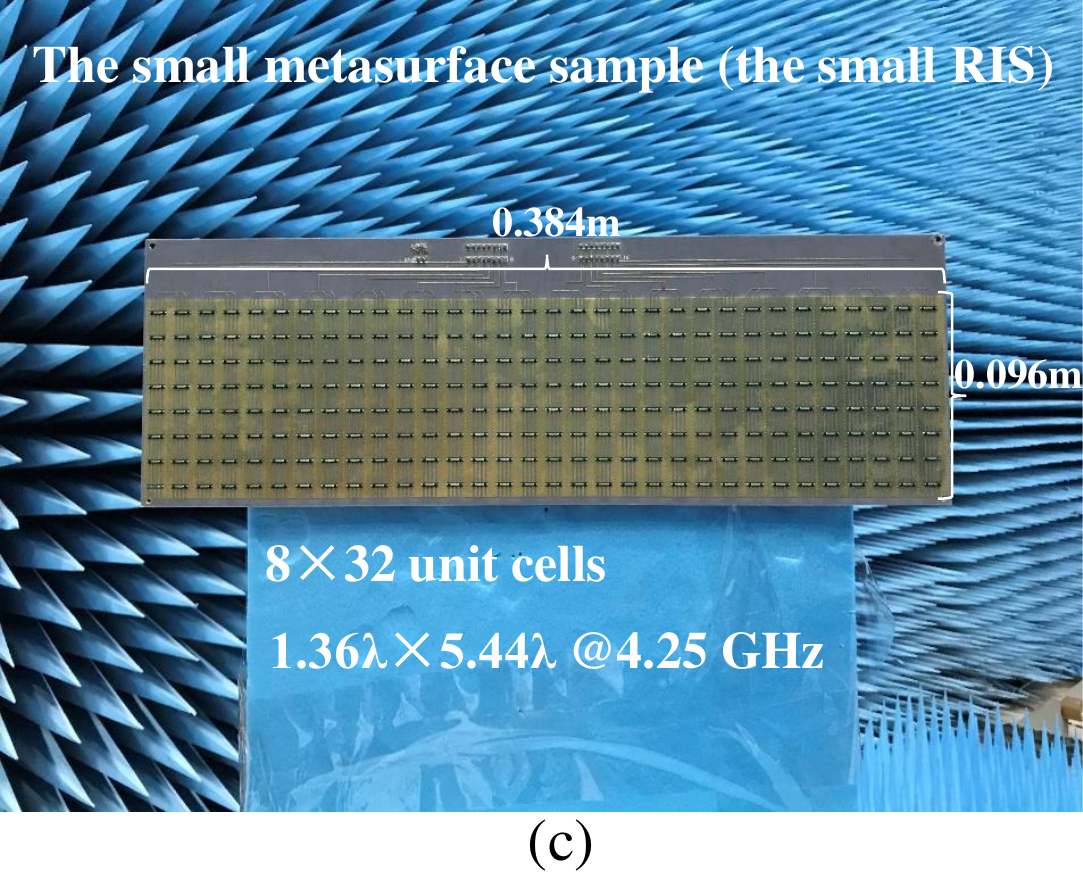}
    \hspace{0.1cm}
	\includegraphics[height=2in]{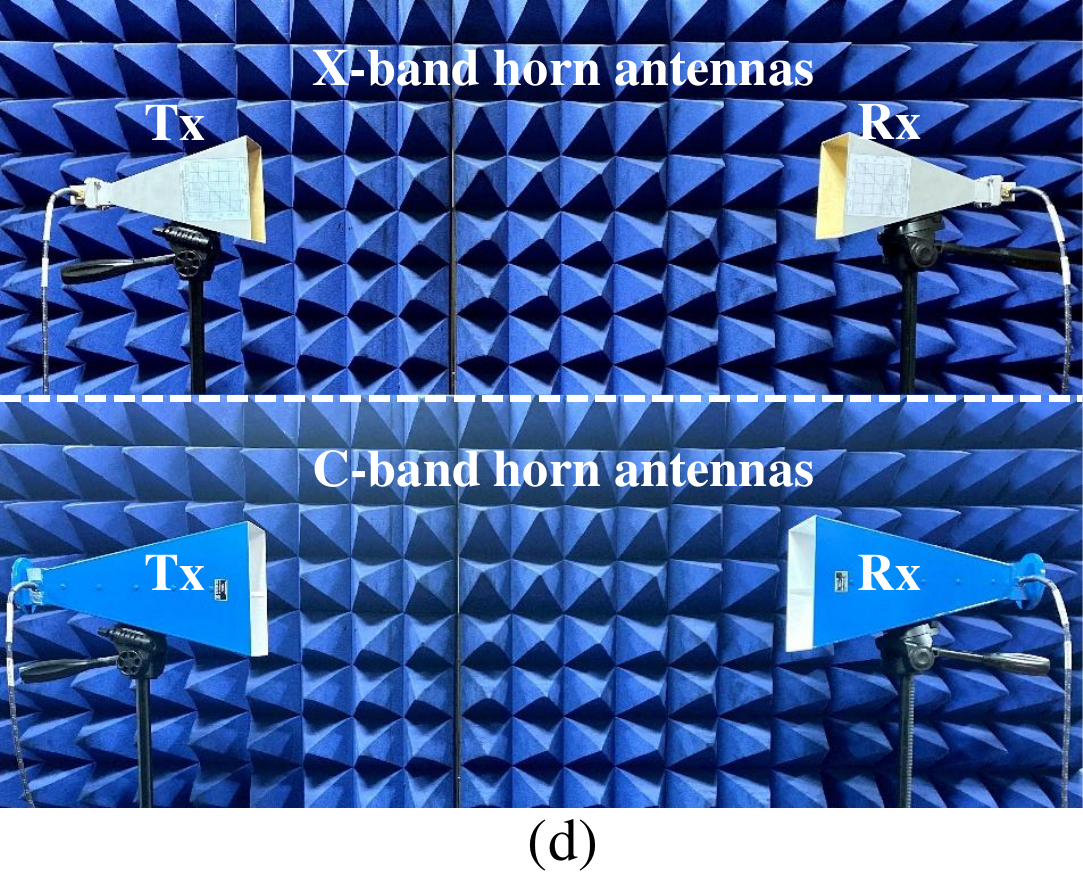}
    \vspace{-0.35cm}
	\caption{. Photographs of the metasurfaces and horn antennas utilized for the free-space path loss measurements of RIS-assisted wireless communications. (a) the large RIS1. (b) the large RIS2. (c) the small RIS. (d) horn antennas.}
	\label{metasurfaces}
\end{figure}

\vspace{-0.15cm}
Fig. \ref{measurementsetup} illustrates our free-space path loss measurement system for RIS-assisted wireless communications, which is composed of a metasurface, RF signal generator (Agilent E8267D), Tx horn antenna, RF signal analyzer (Agilent N9010A), Rx horn antenna and accessories such as cables and blocking object (electromagnetic wave-absorbing materials). During the experimental measurement, the polarization directions of the metasurface, Tx horn antenna, and Rx horn antenna are well matched, which are all horizontally polarized. The RF signal generator provides the RF signal with a constant power to the Tx horn antenna, which illuminates the metasurface through distance $d_1$. The signal reflected by the metasurface propagates over distance $d_2$, and is received by the Rx horn antenna and the RF signal analyzer, which gives the measurement result of the received signal power. Because multipath propagation is greatly suppressed in the microwave anechoic chamber and the blocking object cuts off the direct path between the Tx antenna and the Rx antenna, the received signal mainly comes from the reflection of the RIS, which ensures the feasibility of the free-space path loss measurements of RIS-assisted wireless communications. Since $d_1$ and $d_2$ need to be flexibly changed during the measurement to study the relationship of the path loss as a function of $d_1$ and $d_2$, we utilize the two aisles perpendicular to each other in the anechoic chamber and place the metasurface at the junction of the two aisles. The angles between the metasurface and the two aisles are both $45^{\circ}$ (i.e., ideally, RIS's azimuth angle in Fig. \ref{measurementsetup}(a) equals $45^{\circ}$), which indicates that the metasurface performs specular reflection for most of our measurements. In addition, we also measured an intelligent reflection scenario, in which the incident wave is perpendicular to the metasurface and the reflected wave is manipulated into two directions.
Fig. \ref{metasurfaces} shows the metasurfaces and horn antennas utilized in the free-space path loss measurements in different scenarios. Their main parameters are consistent with those used for the simulations described in Table \ref{parasummary} at the beginning of Section IV.
\vspace{-0.6cm}
\subsection{Measurement Results}\label{MeasurementResults}
\vspace{-0.15cm}
We utilize the two large RISs and the small RIS to measure the free-space path loss in the RIS-assisted broadcasting and beamforming scenarios, respectively. The ${L_{bound}}$ (boundary between the far field and the near field) of the large RIS1, the large RIS2 and the small RIS is 28.77 m, 4.8 m and 0.866 m according to (\ref{s28}), respectively. This makes it possible to measure the path loss in the near field case of the two large RISs and in the far field case of the small RIS, as the area of the microwave anechoic chamber is about $6\ m\times 5\ m$.
\subsubsection{Specular Reflection of Large RIS1 in the Near Field Broadcasting Case}
\begin{figure}[H]
	\centering
	\includegraphics[height=2.6in]{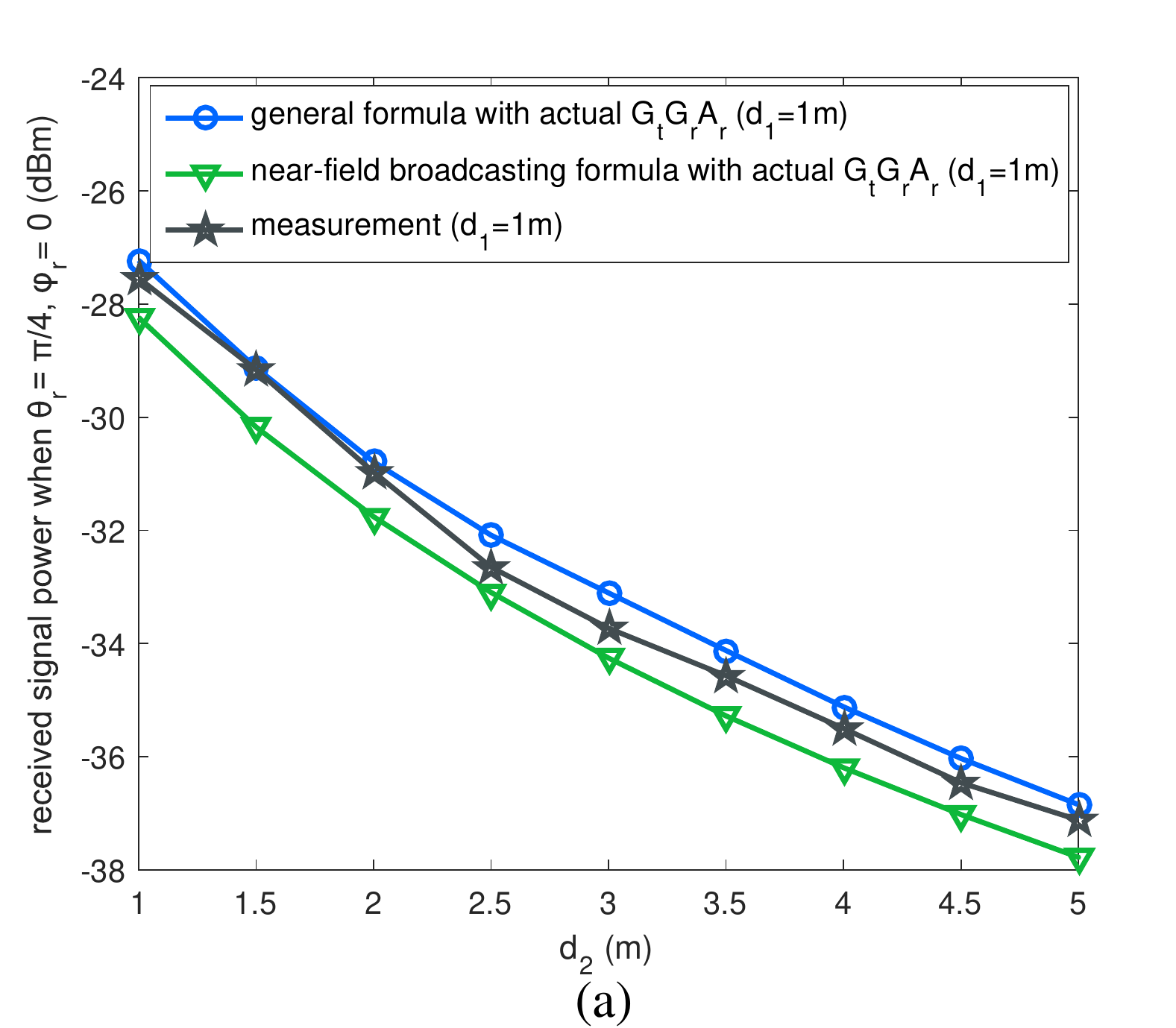}
    \hspace{0.5cm}
    \includegraphics[height=2.6in]{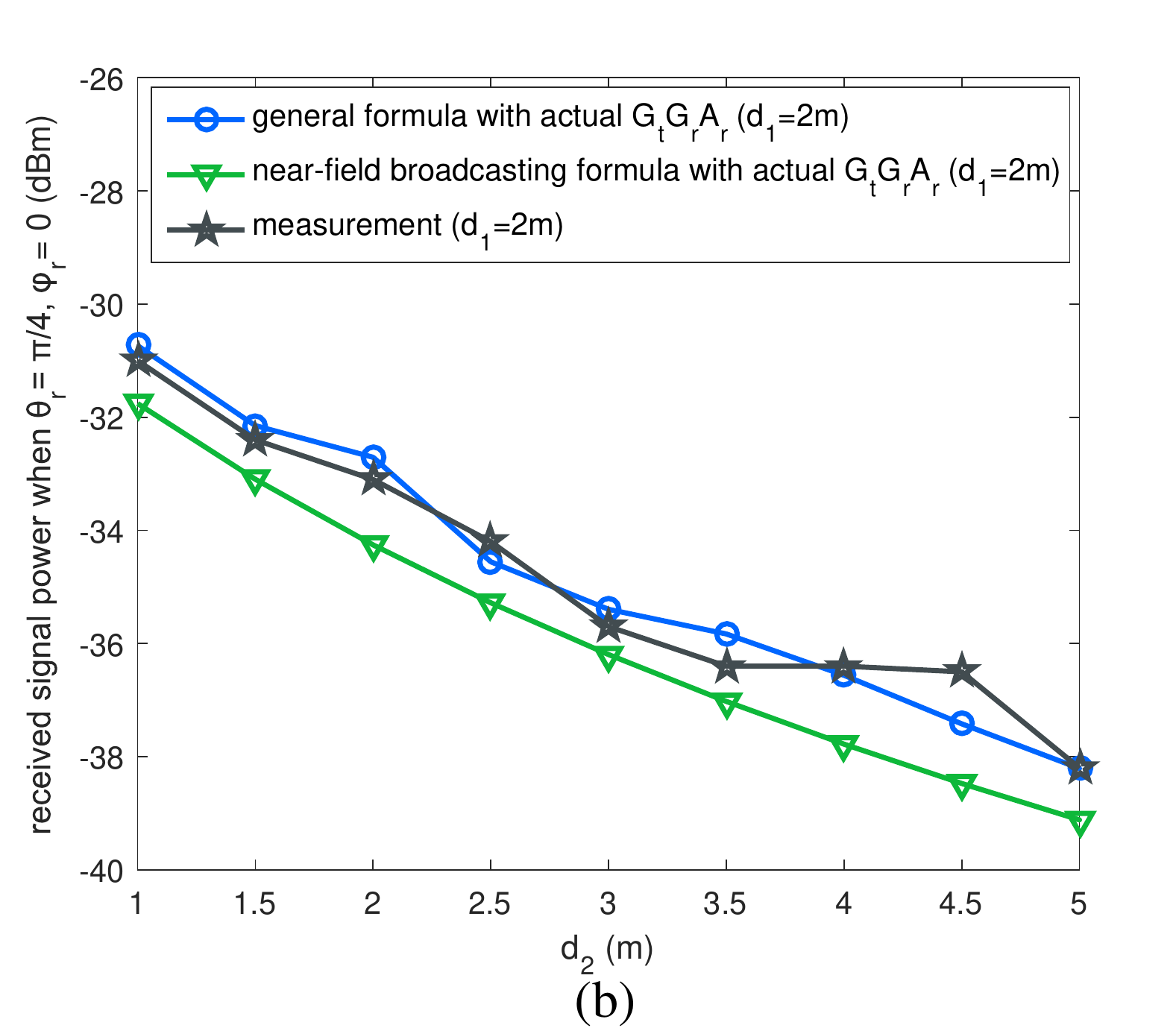}
    \vspace{-0.35cm}
	\caption{. Measurement results of specular reflection through the large RIS1 in the near field broadcasting case. (a) received signal power versus $d_2$ when $d_1$ = 1m. (b) received signal power versus $d_2$ when $d_1$ = 2m.}
	\label{sixbignearfieldmeasure}
\end{figure}

Let all the unit cells of the large RIS1 have the same reflection coefficient by applying the same control voltage (0 V) to them. The transmit power is fixed to $0\ dBm$ and the large RIS1 specularly reflects the incident signal as depicted in Fig.\ref{measurementsetup}(a). Fig.\ref{sixbignearfieldmeasure} illustrates the measured received signal power versus $d_1$ and $d_2$, from which we can observe that the measurement results are in good agreement with the theoretical general formula (\ref{s14}) and the proposed near-field broadcasting formula (\ref{s19}). The measurement results practically validate that the free-space path loss of RIS-assisted wireless communication follows (\ref{s25}) under the scenario of specular reflection through a large RIS in the near field broadcasting case. The path loss is proportional to $(d_1+d_2)^2$. It should be noted that in order to fairly compare the measurement results with the theoretical models, we measured the actual value of $G_tG_rA_rG_{lineloss}$ ($G_{lineloss}$ denotes the cable loss, which is introduced by the RF cables utilized in the measurements) and substituted it into the proposed theoretical formulas.
\vspace{-0.25cm}
\vspace{-1.45cm}
\vspace{-0.8cm}
\subsubsection{Intelligent Reflection of Large RIS2 in the Near Field Broadcasting Case}
\begin{figure}[H]
	\centering
	\includegraphics[height=2.5in]{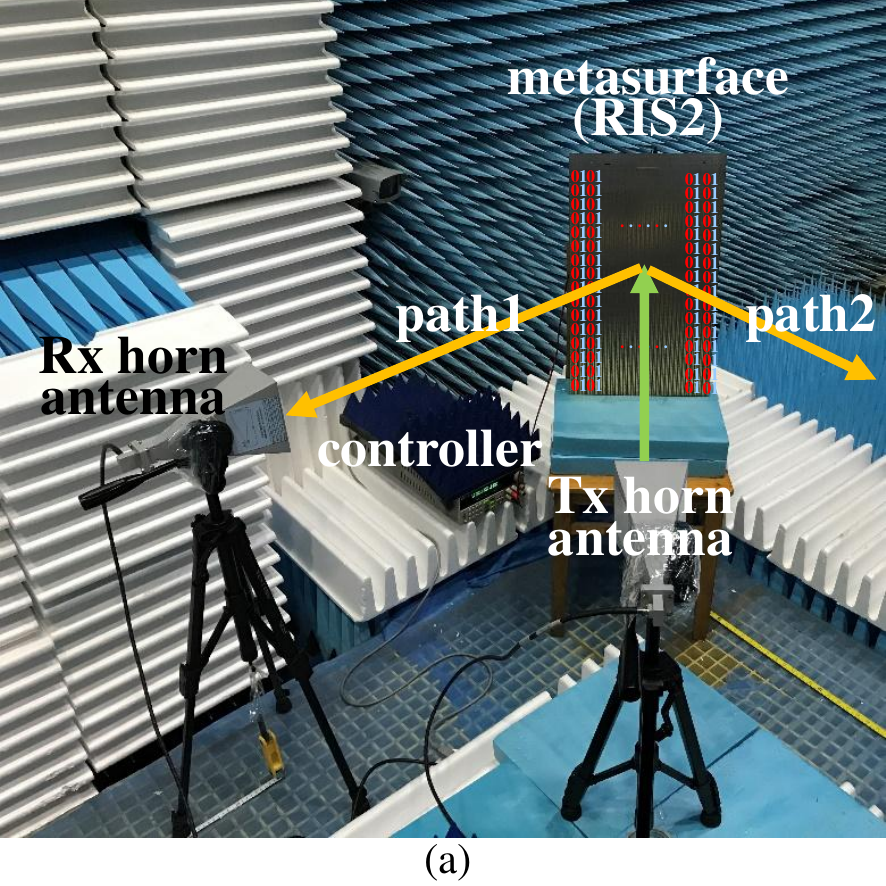}
    \hspace{0.8cm}
	\includegraphics[height=2.6in]{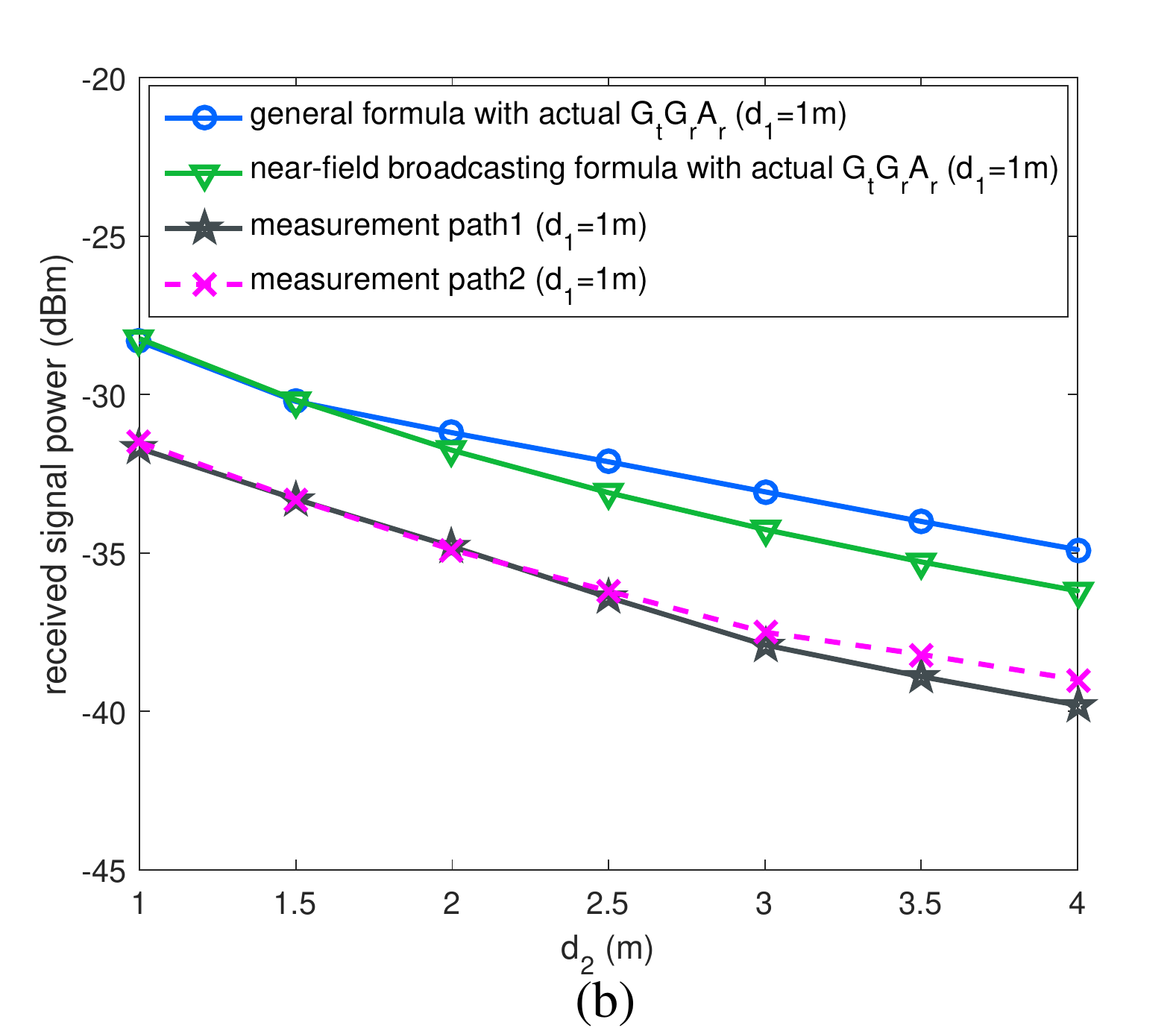}
    \vspace{-0.35cm}
	\caption{. Measurement photograph and results of intelligent reflection through the large RIS2 in the near field broadcasting case. (a) Photograph. (b) Received signal power along the two paths versus $d_2$ when $d_1$ = 1m.}
	\label{onebignearfieldmeasure}
\end{figure}

The incident wave is perpendicular to the large RIS2 and the reflected wave is reflected towards two paths (${\theta _{r}} = \frac{\pi }{4}$, ${\varphi _r} = \pi$ and ${\theta _{r}} = \frac{\pi }{4}$, ${\varphi _r} = 0$) simultaneously, as sketched in Fig. \ref{onebignearfieldmeasure}(a).
Fig. \ref{onebignearfieldmeasure}(b) illustrates the measured received signal power along the two paths versus $d_2$ when $d_1$ = 1m. We can see that the trend of the two measurement curves are exactly the same as that of the proposed theoretical curves, and the difference between them is only about 3 dB\footnote{The slight position deviations of the RIS, Tx horn antenna, and Rx horn antenna will lead to the differences between theory and the measurements. Meanwhile, these slight position deviations are hard to avoid in practice. The deviation of the relative position causes the center direction of the reflected beam of the RIS to deviate from the measurement path, which results in lower measurement results. For example, just a very few degrees' deviation on RIS's azimuth angle can cause 2$\sim$3 dB decrease in the measured received power, which can be verified through simulations based on the general formula (\ref{s14}) in Theorem 1.}. The measurement results validate that the free-space path loss of RIS-assisted wireless communication follows (\ref{s25}) under the scenario of intelligent reflection through a large RIS in the near field broadcasting case.
\vspace{-0.45cm}
\vspace{-0.8cm}
\subsubsection{Specular Reflection of Small RIS in the Far Field Beamforming Case}
\begin{figure}[H]
	\centering
	\includegraphics[height=2.6in]{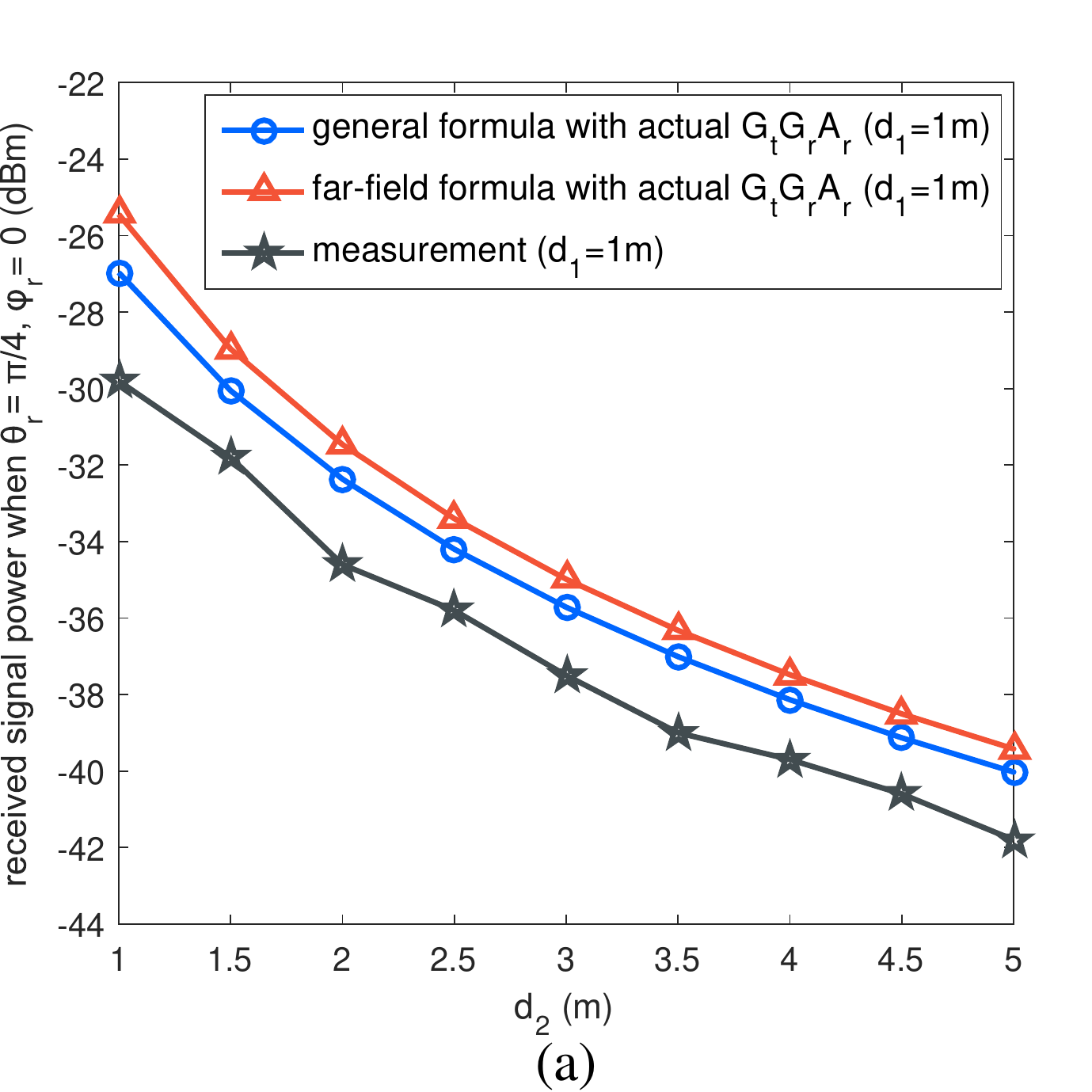}
    \hspace{1cm}
    \includegraphics[height=2.6in]{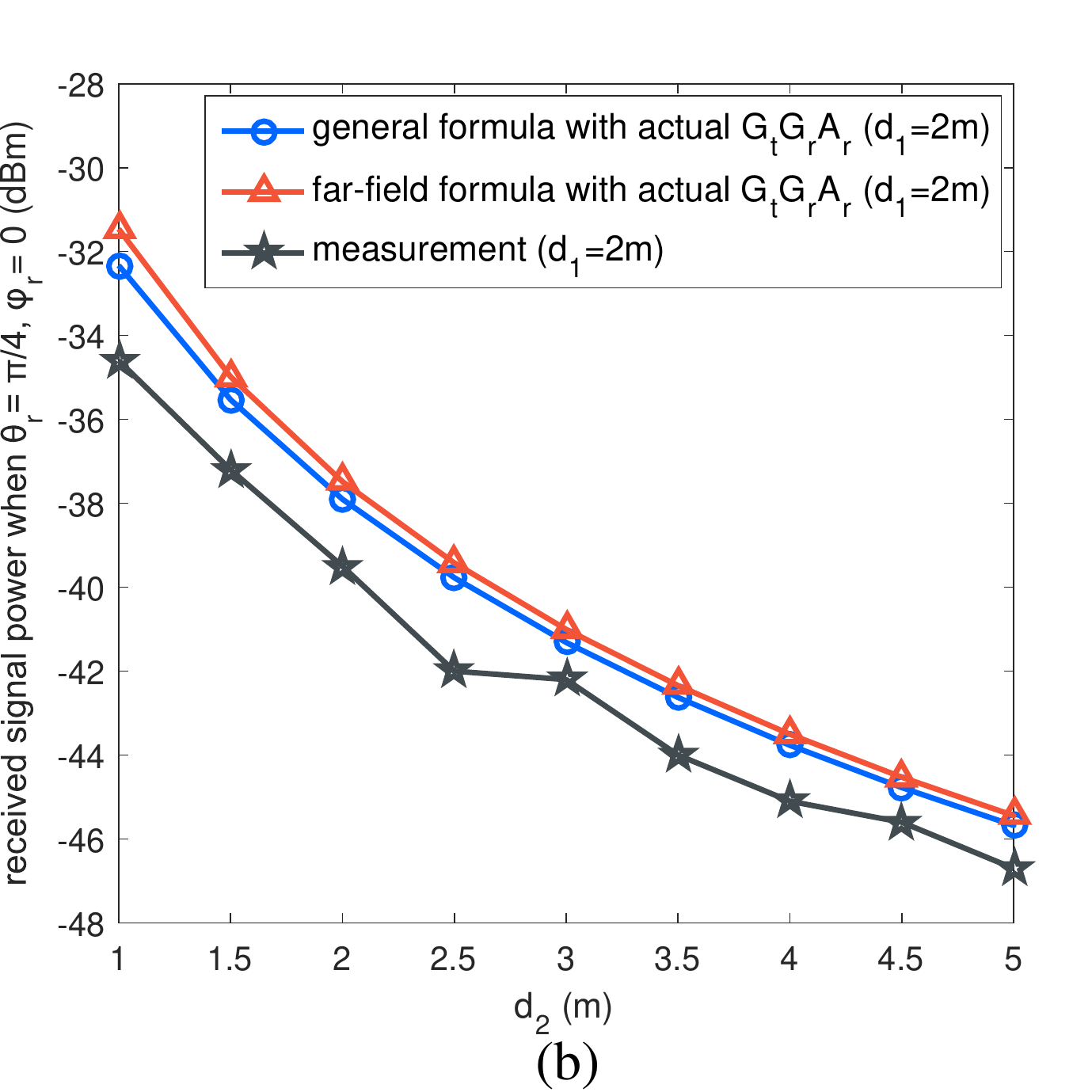}
    \vspace{-0.35cm}
	\caption{. Measurement results of specular reflection through the small RIS in the far field beamforming case. (a) Received signal power versus $d_2$ when $d_1$ = 1 m. (b) Received signal power versus $d_2$ when $d_1$ = 2 m.}
	\label{smallfarfieldmeasure}
\end{figure}

Let all the unit cells of the small RIS have the same reflection coefficient by applying the same control voltage (0 V) to them. The transmit power is fixed to $0\ dBm$ and the small RIS specularly reflects the incident signal to the direction of $\theta _{r} = \frac{\pi }{4}$ and $\varphi _{r} = 0$, along which we measured the received signal power as depicted in Fig.\ref{measurementsetup}. Fig.\ref{smallfarfieldmeasure} illustrates the measured received signal power versus $d_1$ and $d_2$. We observe that the measurement results are in good agreement with the theoretical general formula (\ref{s14}) and the proposed far-field formula (\ref{s15}). The difference between them is only about 2 dB (see footnote7). The measurement results practically validate that the free-space path loss of RIS-assisted wireless communication follows (\ref{s23}) under the scenario of specular reflection through RIS in the far field beamforming case. The free-space path loss is proportional to $(d_1d_2)^2$ when the RIS-assisted wireless communications aim to achieve beamforming in the far field case.
\vspace{-0.65cm}
\vspace{-0.75cm}
\subsubsection{On the Sensitivity of Phase Regulation to the Incident Angle}
\begin{figure}[H]
	\centering
	\includegraphics[height=2.45in]{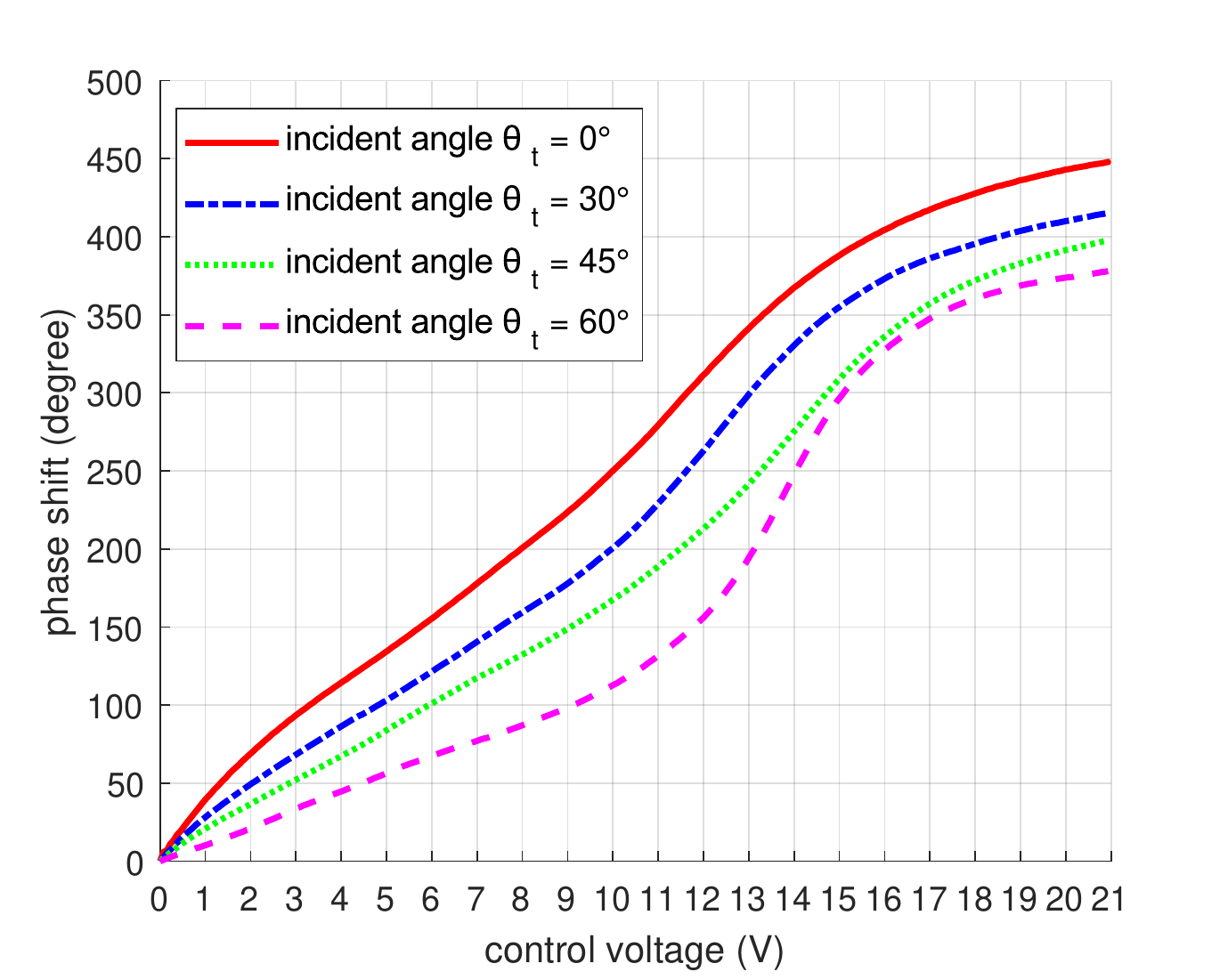}
    \vspace{-0.35cm}
	\caption{. Relationship between the control voltage and the phase shift of the unit cells of the used small RIS.}
	\label{anglesensitivity}
\end{figure}

The general formula (\ref{s14}) reveals that RIS-assisted wireless communication systems are transmitter-receiver reciprocal, which is based on the assumption that $\Gamma_{n,m}$ is not sensitive to the changes of the incident angle $\theta_t$. If $\Gamma_{n,m}$ is sensitive to $\theta_t$ and only sensitive to $\theta_t$, especially if its phase ${\phi _{n,m}}$ is sensitive to $\theta_t$ and only sensitive to $\theta_t$, the channel reciprocity of RIS-assisted wireless communications will not hold anymore. We measured the relationship between the control voltage and the phase shift of the unit cells of our small RIS under different incident angles as plotted in Fig. \ref{anglesensitivity}. It can be observed that the used RIS is highly sensitive to the incident angle. (More comprehensive measurements are underway.)
\vspace{-0.6cm}
\vspace{-0.8cm}
\subsubsection{Power Consumption of the RISs}
The utilized small RIS belongs to the family of varactor-diode-based programmable metasurfaces, while the two large RISs belong to the family of PIN-diode-based programmable metasurfaces. The power consumption of varactor-diode-based RISs is almost zero, because the current in the varactor diodes of the unit cells is negligible when it is working. Therefore, the power comsumption of varactor-diode-based RISs is $P_{RIS}^{varactor}{\approx}0$. The power consumption of PIN-diode-based RISs is related to the states of their unit cells. The phase shift of each unit cell can be adjusted between 0 and $\pi$ by electrically controlling the ``on'' and ``off'' state of the PIN diode. In the power consumption experiment, 0 V voltage signal is provided to all unit cells of the large RIS2 by a DC voltage source instrument. In this case, all diodes are in the ``off'' state, and the current output of the DC voltage source is negligible. On the contrary, when 0.7 V voltage signal is provided, all diodes are in the ``on'' state. In this case, the current output of the DC voltage source is 0.8 A, which indicates that the power consumption of each unit cell in the ``on'' state is 0.33 mW (0.7 V*0.8 A/(50*34 unit cells)=0.33 mW/unit cell). Therefore, the power comsumption of our designed PIN-diode-based RIS is $P_{RIS}^{PIN}{=}0.33{N_{{\text{on}}}}\ mW$, where ${N_{{\text{on}}}}$ is the number of unit cells in the "on" state. It should be noted that the above calculations account only for the RISs themselves. In practice, the controller which provides control signals to the RIS also consumes energy, and its power consumption is related to the controller's circuit design and the number of output control signals. The power consumption of the controller we designed for the utilized small RIS is about 0.72 W, and that for the large RIS2 is about 10 W.
\vspace{-0.5cm}
\section{Conclusion}\label{Conclusion}
In this paper, we have theoretically developed free-space path loss models for RIS-assisted wireless communications based on the electromagnetic and physics properties of the RISs. The general formula that yields the free-space path loss of RIS-assisted wireless communications in different scenarios was first proposed. Subsequently, three more insightful free-space path loss models have been derived, namely the far-field formula, the near-field beamforming formula and the near-field broadcasting formula, to characterize the free-space path loss of RIS-assisted beamforming and broadcasting, respectively. In addition, the boundary of the far field and the near field regimes of RIS-assisted wireless communications has been discussed according to the numerical simulations. Moreover, free-space path loss measurements of RIS-assisted wireless communications have been conducted in a microwave anechoic chamber by considering different scenarios. The measurement results match quite well with the modeling results, which validates the proposed path loss models. This work can help researchers understand the basic large-scale fading of RIS-assisted wireless communication systems, which is useful for link budget calculation and performance analysis. More fading factors, such as small-scale fading, deserve further studies in the future as an important generalization of this paper.
\vspace{-0.55cm}
\begin{appendices}
\section*{Appendix A-Proof of Theorem 1}\label{A}
\vspace{-0.2cm}
The power of the incident signal into unit cell $U_{n,m}$ can be expressed as
\begin{equation}\label{s3}
P_{n,m}^{in} = \frac{{{G_t}{P_t}}}{{4\pi r{{_{n,m}^t}^2}}}{F^{tx}}\left( {\theta _{n,m}^{tx},\varphi _{n,m}^{tx}} \right)F\left( {\theta _{n,m}^t,\varphi _{n,m}^t} \right){d_x}{d_y},
\end{equation}
and the electric field of the incident signal into $U_{n,m}$ is given by
\begin{equation}\label{s4}
E_{n,m}^{in} = \sqrt {\frac{{2{Z_0}P_{n,m}^{in}}}{{{d_x}{d_y}}}}\ {e^{\frac{{ - j2\pi r_{n,m}^t}}{\lambda }}},
\end{equation}
where $Z_0$ is the characteristic impedance of the air, and $r_{n,m}^t$ can be written as
\begin{equation}\label{s5}
r_{n,m}^t = \sqrt {{{\left( {{x_t}{-}\left({m{-}\frac{1}{2}} \right){d_x}} \right)}^2}{+}{{\left( {{y_t}{-}\left( {n{-}\frac{1}{2}}\right){d_y}} \right)}^2}{+}{{ {{z_t}} }^2}}.
\end{equation}

According to the law of energy conservation, for the unit cell $U_{n,m}$, the power of the incident signal times the square of the reflection coefficient is equal to the total power of the reflected signal, thus we have
\begin{equation}\label{s6}
P_{n,m}^{in}{|}\Gamma _{n,m}^2{|} = P_{n,m}^{reflect},
\end{equation}
where $P_{n,m}^{reflect}$ is the total reflected signal power of the unit cell $U_{n,m}$, and the reflection coefficient can be written as
\begin{equation}\label{s7}
{\Gamma _{n,m}} = {A_{n,m}}{e^{j{\phi _{n,m}}}},
\end{equation}
where ${A_{n,m}}$ and ${\phi _{n,m}}$ represent the controllable amplitude and phase shift of $U_{n,m}$, respectively.

The power of the reflected signal received by the receiver from $U_{n,m}$ can be expressed as
\begin{equation}\label{s8}
P_{n,m}^{r} = \frac{{{G}{P_{n,m}^{reflect}}}}{{4\pi r{{_{n,m}^r}^2}}}F\left( {\theta _{n,m}^r,\varphi _{n,m}^r} \right){F^{rx}}\left( {\theta _{n,m}^{rx},\varphi _{n,m}^{rx}} \right)A_r,
\end{equation}
where $A_r$ represents the aperture of the receiving antenna, and $r_{n,m}^r$ can be written as
\begin{equation}\label{s9}
r_{n,m}^r = \sqrt {{{\left( {{x_r}{-}\left( {m{-}\frac{1}{2}} \right){d_x}} \right)}^2}{+}{{\left( {{y_r}{-}\left( {n{-}\frac{1}{2}} \right){d_y}} \right)}^2}{+}{{ {{z_r}} }^2}}.
\end{equation}

By combining (\ref{s3}), (\ref{s6}) and (\ref{s8}), the electric field of the reflected signal received by the receiver from $U_{n,m}$ is obtained as

\begin{equation}\label{s10}
E_{n,m}^r{=}\sqrt {\frac{{2{Z_0}P_{n,m}^r}}{{{A_r}}}} {e^{ - j(\frac{{2\pi r_{n,m}^t}}{\lambda }{+}\frac{{2\pi r_{n,m}^r}}{\lambda }{{-}{\phi _{n,m}}})}}
 {=} \sqrt {\frac{{Z_0}{P_t}{G_t}G{d_x}{d_y}{{F_{n,m}^{combine}}}}{{8{\pi ^2}}}} \frac{{{\Gamma _{n,m}}}}{{r_{n,m}^tr_{n,m}^r}}{e^{\frac{{ - j2\pi (r_{n,m}^t + r_{n,m}^r)}}{\lambda }}},
\end{equation}
where $-(\frac{{2\pi r_{n,m}^t}}{\lambda }{+}\frac{{2\pi r_{n,m}^r}}{\lambda }{-}{\phi _{n,m}})$ is the phase alteration caused by the propagation and the reflection coefficient of $U_{n,m}$.
The total electric field of the received signal is the superposition of the electric fields reflected by all unit cells towards the receiver, which can be written as \cite{Appendix1,Appendix2,Appendix3}
\begin{equation}\label{s11}
{E^r} = \sum\limits_{m = 1 - \frac{M}{2}}^{\frac{M}{2}} {\sum\limits_{n = 1 - \frac{N}{2}}^{\frac{N}{2}} {E_{n,m}^r} }.
\end{equation}

The received signal power of the receiver is
\begin{equation}\label{s12}
{P_{r}} = \frac{{{{\left| {{E^r}} \right|}^2}}}{{2{Z_0}}}{A_r},
\end{equation}
where the aperture of the receiving antenna can be written as
\begin{equation}\label{s13}
{A_r} = \frac{{{G_r}{\lambda ^2}}}{{4\pi }}.
\end{equation}

We obtain Theorem 1 by substituting (\ref{s10}), (\ref{s11}), and (\ref{s13}) into (\ref{s12}).
\vspace{-0.2cm}
\section*{Appendix B-Proof of Proposition 1}\label{B}
\vspace{-0.2cm}
The position of the transmitter in Fig. \ref{systemmodel} is
\begin{equation}\label{s29}
({x_t},{y_t},{z_t}) = ({d_1}\sin {\theta _{t}}\cos {\varphi _{t}},{d_1}\sin {\theta _{t}}\sin {\varphi _{t}},{d_1}\cos {\theta _{t}}).
\end{equation}

By combining (\ref{s5}) and (\ref{s29}), when the transmitter is in the far field of the RIS, $r_{n,m}^t$ can be further expressed as
\begin{equation}\label{s30}
\begin{aligned}
r_{n,m}^t &= \sqrt {{{\left( {{d_1}\sin {\theta _{t}}\cos {\varphi _{t}}{-}(m{-}\frac{1}{2}){d_x}} \right)}^2}{+}{{\left( {{d_1}\sin {\theta _{t}}\sin {\varphi _{t}}{-}(n{-}\frac{1}{2}){d_y}} \right)}^2}{+}{{\left( {{d_1}\cos {\theta _{t}}} \right)}^2}} \\
 &\approx {d_1}{-}\sin {\theta _{t}}\cos {\varphi _{t}}(m{-}\frac{1}{2}){d_x}{-}\sin {\theta _{t}}\sin {\varphi _{t}}(n{-}\frac{1}{2}){d_y},
\end{aligned}
\end{equation}

Similarly, the position of the receiver in Fig. \ref{systemmodel} is
\begin{equation}\label{s31}
({x_r},{y_r},{z_r}) = ({d_2}\sin \theta_{r} \cos \varphi_{r} ,{d_2}\sin \theta_{r} \sin \varphi_{r} ,{d_2}\cos \theta_{r} ).
\end{equation}

By combining (\ref{s9}) and (\ref{s31}), when the receiver is in the far field of the RIS, $r_{n,m}^r$ can be further expressed as
\begin{equation}\label{s32}
\begin{aligned}
r_{n,m}^r &= \sqrt {{{\left( {{d_2}\sin \theta_{r} \cos \varphi_{r}{-}(m{-}\frac{1}{2}){d_x}} \right)}^2}{+}{{\left( {{d_2}\sin \theta_{r} \sin \varphi_{r}{-}(n{-}\frac{1}{2}){d_y}} \right)}^2}{+}{{\left( {{d_2}\cos \theta_{r} } \right)}^2}} \\
 &\approx {d_2}{-}\sin \theta_{r} \cos \varphi_{r} (m{-}\frac{1}{2}){d_x}{-}\sin \theta_{r} \sin \varphi_{r} (n{-}\frac{1}{2}){d_y},
\end{aligned}
\end{equation}

In the far field case, ${(\theta _{n,m}^t,\varphi _{n,m}^t)}$ and ${(\theta _{n,m}^r,\varphi _{n,m}^r)}$ can be approximated as ${(\theta_t,\varphi_t)}$ and ${(\theta_r,\varphi_r)}$, respectively. In addition, as assuming that the direction of peak radiation of both the transmitting antenna and receiving antenna point to the center of the RIS, we have ${F^{tx}}(\theta _{n,m}^{tx},\varphi _{n,m}^{tx}) \approx  1$ and ${F^{rx}}(\theta _{n,m}^{rx},\varphi _{n,m}^{rx}) \approx 1$ in the far field case.

By substituting (\ref{s30}) and (\ref{s32}) into the general formula (\ref{s14}), the received signal power can be further written as
\begin{equation}\label{s33}
\begin{aligned}
{P_r} &\approx {P_t}\frac{{{G_t}{G_r}G{d_x}{d_y}{\lambda ^2}}}{{64{\pi ^3}}}{\left| {\sum\limits_{m = 1 - \frac{M}{2}}^{\frac{M}{2}} {\sum\limits_{n = 1 - \frac{N}{2}}^{\frac{N}{2}} {\sqrt {F({\theta _t},{\varphi _t})F({\theta _r},{\varphi _r})}\ {\Gamma _{n,m}}\frac{{{e^{\frac{{ - j2\pi \alpha }}{\lambda }}}}}{{{d_1}{d_2}}}} } } \right|^2}\\
 &= {P_t}\frac{{{G_t}{G_r}G{d_x}{d_y}{\lambda ^2}F({\theta _t},{\varphi _t})F({\theta _r},{\varphi _r})}}{{64{\pi ^3}{d_1}^2{d_2}^2}}{\left| {\sum\limits_{m = 1 - \frac{M}{2}}^{\frac{M}{2}} {\sum\limits_{n = 1 - \frac{N}{2}}^{\frac{N}{2}} {{\Gamma _{n,m}}\ {e^{\frac{{j2\pi ({d_1} + {d_2} - \alpha )}}{\lambda }}}} } } \right|^2},
\end{aligned}
\end{equation}
where $\alpha{=}{d_1}{-}\sin {\theta _t}\cos {\varphi _t}(m{-}\frac{1}{2}){d_x}{-}\sin {\theta _t}\sin {\varphi _t}(n{-}\frac{1}{2}){d_y}{+}{d_2}{-}\sin {\theta _r}\cos {\varphi _r}(m{- }\frac{1}{2}){d_x}{-}\\\sin {\theta _r}\sin {\varphi _r}(n{-}\frac{1}{2}){d_y}$.

As all the unit cells of RIS share the same reflection coefficient $A{e^{j{\phi}}}$, (\ref{s33}) can be written as
\begin{equation}\label{s34}
\begin{aligned}
{P_r} = {P_t}\frac{{{G_t}{G_r}G{d_x}{d_y}{\lambda ^2}F({\theta _t},{\varphi _t})F({\theta _r},{\varphi _r}){A^2}}}{{64{\pi ^3}{d_1}^2{d_2}^2}}{\left| \beta  \right|^2},
\end{aligned}
\end{equation}
where
\begin{equation}\label{s35}
\beta  = \sum\limits_{m = 1 - \frac{M}{2}}^{\frac{M}{2}} {\sum\limits_{n = 1 - \frac{N}{2}}^{\frac{N}{2}} {{e^{\frac{{j2\pi ((\sin {\theta _t}\cos {\varphi _t} + \sin {\theta _r}\cos {\varphi _r})(m - \frac{1}{2}){d_x} + (\sin {\theta _t}\sin {\varphi _t} + \sin {\theta _r}\sin {\varphi _r})(n - \frac{1}{2}){d_y})}}{\lambda }}}} }.
\end{equation}

Define $u = \frac{{2\pi }}{\lambda }(\sin {\theta _{t}}\cos {\varphi _{t}} + \sin \theta_{r} \cos \varphi_{r} ){d_x}$ and $v = \frac{{2\pi }}{\lambda }(\sin {\theta _{t}}\sin {\varphi _{t}} + \sin \theta_{r} \sin \varphi_{r} ){d_y}$, then (\ref{s35}) can be rewritten as
\begin{equation}\label{s36}
\beta  = \sum\limits_{m = 1 - \frac{M}{2}}^{\frac{M}{2}} {\sum\limits_{n = 1 - \frac{N}{2}}^{\frac{N}{2}} {{e^{j(m - \frac{1}{2})u}}} } {e^{j(n - \frac{1}{2})v}} = \sum\limits_{m = 1 - \frac{M}{2}}^{\frac{M}{2}} {{e^{j(m - \frac{1}{2})u}}\sum\limits_{n = 1 - \frac{N}{2}}^{\frac{N}{2}} {{e^{j(n - \frac{1}{2})v}}} }
\end{equation}

By applying the formula of the sum of the geometric progression, (\ref{s36}) can be rewritten as
\begin{equation}\label{s37}
\begin{aligned}
\beta  &= \frac{{{e^{ - j\frac{{Mu}}{2}}} - {e^{j\frac{{Mu}}{2}}}}}{{{e^{ - j\frac{u}{2}}} - {e^{j\frac{u}{2}}}}}\frac{{{e^{ - j\frac{{Nv}}{2}}} - {e^{j\frac{{Nv}}{2}}}}}{{{e^{ - j\frac{v}{2}}} - {e^{j\frac{v}{2}}}}} = \frac{{\sin \left( {\frac{{Mu}}{2}} \right)}}{{\sin \left( {\frac{u}{2}} \right)}}\frac{{\sin \left( {\frac{{Nv}}{2}} \right)}}{{\sin \left( {\frac{v}{2}} \right)}} = MN\frac{{\sinc\left( {\frac{{Mu}}{2}} \right)}}{{\sinc(\frac{u}{2})}}\frac{{\sinc\left( {\frac{{Nv}}{2}} \right)}}{{\sinc\left( {\frac{v}{2}} \right)}}\\
 &= MN\frac{{\sinc\left( {\frac{{M\pi }}{\lambda }(\sin {\theta _{t}}\cos {\varphi _{t}}{+}\sin \theta_{r} \cos \varphi_{r} ){d_x}} \right)}}{{\sinc(\frac{\pi }{\lambda }(\sin {\theta _{t}}\cos {\varphi _{t}}{+}\sin \theta_{r} \cos \varphi_{r} ){d_x})}}\frac{{\sinc\left( {\frac{{N\pi }}{\lambda }(\sin {\theta _{t}}\sin {\varphi _{t}}{+}\sin \theta_{r} \sin \varphi_{r} ){d_y}} \right)}}{{\sinc\left( {\frac{\pi }{\lambda }(\sin {\theta _{t}}\sin {\varphi _{t}}{+}\sin \theta_{r} \sin \varphi_{r} ){d_y}} \right)}}.
\end{aligned}
\end{equation}

By plugging (\ref{s37}) into (\ref{s34}), the received signal power in the far field case can be obtained as stated in (\ref{s15}). When $\sin {\theta _{t}}\cos {\varphi _{t}} + \sin \theta_{r} \cos \varphi_{r}  = 0$ and $\sin {\theta _{t}}\sin {\varphi _{t}} + \sin \theta_{r} \sin \varphi_{r}  = 0$, that is, $\theta_{r}  = {\theta _{t}}$ and $\varphi_{r}  = {\varphi _{t}} + {\pi}$, (\ref{s15}) is maximized as stated in (\ref{s16}).
\section*{Appendix C-Proof of Proposition 2}\label{C}
For the intelligent reflection in the far field case, by substituting ${\Gamma _{n,m}} = {A}{e^{j{\phi _{n,m}}}}$ into (\ref{s33}), we have
\begin{equation}\label{s38}
\begin{aligned}
{P_r} = {P_t}\frac{{{G_t}{G_r}G{d_x}{d_y}{\lambda ^2}F({\theta _t},{\varphi _t})F({\theta _r},{\varphi _r}){A^2}}}{{64{\pi ^3}{d_1}^2{d_2}^2}}{\left| {\tilde \beta } \right|^2},
\end{aligned}
\end{equation}
where
\begin{equation}\label{s39}
\tilde \beta  = \sum\limits_{m = 1 - \frac{M}{2}}^{\frac{M}{2}} {\sum\limits_{n = 1 - \frac{N}{2}}^{\frac{N}{2}} {{e^{\frac{{j2\pi ((\sin {\theta _t}\cos {\varphi _t} + \sin {\theta _r}\cos {\varphi _r})(m - \frac{1}{2}){d_x} + (\sin {\theta _t}\sin {\varphi _t} + \sin {\theta _r}\sin {\varphi _r})(n - \frac{1}{2}){d_y} + \frac{{\lambda {\phi _{n,m}}}}{{2\pi }})}}{\lambda }}}} }.
\end{equation}

Define $\tilde u = \frac{{2\pi }}{\lambda }(\sin {\theta _{t}}\cos {\varphi _{t}} + \sin \theta_{r} \cos \varphi_{r}  + {\delta _1}){d_x}$ and $\tilde v = \frac{{2\pi }}{\lambda }(\sin {\theta _{t}}\sin {\varphi _{t}} + \sin \theta_{r} \sin \varphi_{r}  + {\delta _2}){d_y}$, where ${\delta _1}\left( {m - \frac{1}{2}} \right){d_x} + {\delta _2}\left( {n - \frac{1}{2}} \right){d_y} = \frac{{\lambda {\phi _{n,m}}}}{{2\pi }}$, (\ref{s39}) can be rewritten as
\begin{equation}\label{s40}
\tilde\beta  = \sum\limits_{m = 1 - \frac{M}{2}}^{\frac{M}{2}} {\sum\limits_{n = 1 - \frac{N}{2}}^{\frac{N}{2}} {{e^{j(m - \frac{1}{2})\tilde u}}} } {e^{j(n - \frac{1}{2})\tilde v}} = \sum\limits_{m = 1 - \frac{M}{2}}^{\frac{M}{2}} {{e^{j(m - \frac{1}{2})\tilde u}}\sum\limits_{n = 1 - \frac{N}{2}}^{\frac{N}{2}} {{e^{j(n - \frac{1}{2})\tilde v}}} }
\end{equation}

Similarly, by applying the formula of the sum of the geometric progression, the received signal power can be obtained as (\ref{s17}), which is maximized when $\sin {\theta _{t}}\cos {\varphi _{t}} + \sin \theta_{r} \cos \varphi_{r}  + {\delta _1}= 0$ and $\sin {\theta _{t}}\sin {\varphi _{t}} + \sin \theta_{r} \sin \varphi_{r} + {\delta _2}= 0$.

\section*{Appendix D-Proof of Proposition 3}\label{D}
By substituting ${\Gamma _{n,m}} = {A}{e^{j{\phi _{n,m}}}}$ into the general formula (\ref{s14}), (\ref{s20}) can be simply obtained. For the desired receiver at $\left( {{x_{r}},{y_{r}},{z_{r}}} \right)$, (\ref{s20}) is maximized as (\ref{s21}) when $2\pi (r_{n,m}^t + r_{n,m}^r) - \lambda {\phi _{n,m}} = 0$, which can be further expressed as (\ref{s22}).
\end{appendices}

\end{document}